\documentclass[12pt,a4paper]{article}
\usepackage[utf8]{inputenc}
\usepackage{scalerel,stackengine,amsmath}
\usepackage{amsfonts,amssymb}
\usepackage{ifthen}
\usepackage{tikz}
\usepackage{xspace}
\usepackage{physics,bm}
\usepackage{booktabs}
\usepackage{hyphenat}
\usepackage{enumitem}
\usepackage{dsfont}
\usepackage[left=2cm,right=2cm,top=2.5cm,bottom=2.5cm]{geometry}
\usepackage{cite}
\usepackage{slashed}
\usepackage{hyperref}
\usepackage{graphicx}
\usepackage{caption}
\usepackage{float}
\usepackage{subcaption}
\usepackage[labelformat=simple]{subcaption}

\usepackage{authblk}
\hypersetup{
	unicode=true,          % non-Latin characters in Acrobat's bookmarks
	pdftoolbar=true,        % show Acrobat's toolbar?
	pdfmenubar=true,        % show Acrobat's menu?
	pdffitwindow=false,     % window fit to page when opened
	pdfstartview={FitH},    % fits the width of the page to the window
	pdftitle={CriticalUnstableQubits},    % title
	pdfauthor={},     % author
	pdfsubject={},   % subject of the document
	pdfcreator={},   % creator of the document
	pdfproducer={Producer}, % producer of the document
    pdfkeywords={Bloch Sphere} {Critical Unstable Qubits} {B0B0-Meson System},
    pdfnewwindow=true,      % links in new PDF window
	colorlinks=true,       % false: boxed links; true: colored links
	linkcolor=blue,          % color of internal links (change box color with linkbordercolor)
	citecolor=red,        % color of links to bibliography
	filecolor=black,      % color of file links
	urlcolor=blue           % color of external links
}

\newcommand{\de}{{\rm d}}
\newcommand{\be}{\mathbf{b}}

\newcommand{\vect}{\widehat{\mathbf{t}}}
\newcommand{\vecn}{\widehat{\mathbf{n}}}
\newcommand{\vecm}{\widehat{\mathbf{m}}}

\newcommand{\bvec}[1]{\vectorbold{#1}\xspace}

\newcommand{\vecb}{\bvec b\xspace}

\newcommand{\lrb}[1]{\left( #1 \right)}
\newcommand{\lrsb}[1]{\left[ #1 \right]}
\newcommand{\lrcb}[1]{\left\{ #1 \right\}}
\newcommand{\lrBigb}[1]{\Big( #1 \Big)}

\newcommand{\lrBigcb}[1]{\Big\{ #1 \Big\}}

\renewcommand{\Re}[1]{\operatorname{Re}\lrsb{#1}\xspace}
\renewcommand{\Im}[1]{\operatorname{Im}\lrsb{#1}\xspace}

\renewcommand{\Tr}[1]{\textrm{Tr}\lrsb{#1}\xspace}

\newcounter{NumArgs}

\newcommand{\eqs}[1]{\setcounter{NumArgs}{0}\foreach\i in{#1}{\stepcounter{NumArgs}}%
	\ifthenelse{\equal{\theNumArgs}{1}}{(\ref{#1})}%
	{\ifthenelse{\equal{\theNumArgs}{2}}%
		{\foreach\i[count=\q]in{#1}{\ifthenelse{\equal{\q}{\theNumArgs}}{and (\ref{\i})}{(\ref{\i})~}}}%
		{\foreach\i[count=\q]in{#1}{\ifthenelse{\equal{\q}{\theNumArgs}}{and (\ref{\i})}{(\ref{\i}),~}}}}}

\newcommand{\refs}[1]{\setcounter{NumArgs}{0}\foreach\i in{#1}{\stepcounter{NumArgs}}%
	\ifthenelse{\equal{\theNumArgs}{1}}{(\ref{#1})}%
	{\ifthenelse{\equal{\theNumArgs}{2}}%
		{\foreach\i[count=\q]in{#1}{\ifthenelse{\equal{\q}{\theNumArgs}}{and (\ref{\i})}{(\ref{\i})~}}}%
		{\foreach\i[count=\q]in{#1}{\ifthenelse{\equal{\q}{\theNumArgs}}{and (\ref{\i})}{(\ref{\i}),~}}}}}

\newcommand{\Figs}[1]{\setcounter{NumArgs}{0}\foreach\i in{#1}{\stepcounter{NumArgs}}%
	\ifthenelse{\equal{\theNumArgs}{1}}{Figure~\ref{#1}}%
	{\ifthenelse{\equal{\theNumArgs}{2}}%
		{Figures~\foreach\i[count=\q]in{#1}{\ifthenelse{\equal{\q}{\theNumArgs}}{and \ref{\i}}{\ref{\i}~}}}%
		{Figures~\foreach\i[count=\q]in{#1}{\ifthenelse{\equal{\q}{\theNumArgs}}{and \ref{\i}}{\ref{\i},~}}}}}

 \renewcommand{\theequation}{\arabic{section}.\arabic{equation}}

\usepackage{indentfirst}

\definecolor{ao}{rgb}{0.0, 0.0, 1.0}
\newenvironment{bl}[1]{{\color{ao}  #1}}

\newcommand*{\email}[1]{%
	\footnotesize\href{mailto:#1}{\bl{#1}}
}

\title{\textbf{Critical Unstable Qubits:\\ an Application to 
{\boldmath $B^0\bar{B}^0$}-Meson System}}

\author[]{Dimitrios Karamitros${}^{\,a,b\,}$\footnote{\email{dimitrios.d.karamitros@jyu.fi}} }
\author[]{Thomas McKelvey${}^{\,c\,}$\footnote{\email{thomas.mckelvey@manchester.ac.uk}} } 
\author[]{Apostolos Pilaftsis${}^{\,c,d\,}$\footnote{\email{apostolos.pilaftsis@manchester.ac.uk}} }

\affil[]{
\normalsize
\textit{\hspace{1.4cm}${}^a$Helsinki Institute of Physics, University of Helsinki, Helsinki,\newline P.O. Box 64, FIN-00014, Finland} 
 
\textit{\hspace{1.4cm}${}^b$Department of Physics, University of Jyv\"askyl\"a, Jyv\"askyl\"a,\newline  P.O.Box 35 (YFL), FIN-40014, Finland} 

\textit{\hspace{1.4cm}${}^c$Department
 of Physics and Astronomy, University of Manchester,\newline Manchester, M13 9PL, United Kingdom}

\textit{\hspace{1.4cm}${}^d$PRISMA Cluster of Excellence \& Mainz Institute for Theoretical Physics,\newline Johannes Gutenberg University, 55099 Mainz, Germany}
}

\date{\empty}

\begin{document}
\setcounter{page}{1}
	
{\vspace*{-3cm}
	\begin{flushright}
		MITP-25-020\\
		April 2025
	\end{flushright}
\vspace*{-1cm}
}

{\let\newpage\relax\maketitle}

\maketitle

\flushbottom
\vspace{-1cm}
\begin{abstract}

\noindent 
We extend our previous work on a novel class of unstable qubits which we have identified recently and called them Critical Unstable Qubits (CUQs). The characteristic property of CUQs is that the energy-level and decay-width  vectors, ${\bf E}$ and ${\bf \Gamma}$, are orthogonal to one another, and the key parameter $r = |{\bf \Gamma}|/|2{\bf E}|$ is less than 1. Most remarkably, CUQs exhibit two atypical behaviours: (i) they display coherence-decoherence oscillations in a co-decaying frame of the system described by a unit Bloch vector ${\bf b}$, and (ii)~the unit Bloch vector ${\bf b}$ describing a pure CUQ sweeps out unequal areas during equal intervals of time, while rotating about the vector ${\bf E}$. The latter {\em anharmonic} phenomenon emerges beyond the usual oscillatory pattern due to the energy-level difference of the two-level quantum system, which governs an ordinary qubit. By making use of a Fourier series decomposition, we define anharmonicity observables that quantify the degree of {\em non}-sinusoidal oscillation of a CUQ. We apply the results of our formalism to the $B^0\bar{B}^0$-meson system and derive, for the first time, generic upper limits on these new observables.

\end{abstract}

\newpage
\tableofcontents
\newpage

\section{Introduction}\label{sec:Intro}
\setcounter{equation}{0}

The time evolution of Quantum Mechanical~(QM) systems has been a topic of immense research for several decades now~\cite{Feynman:1986vej,Barenco:1995na,Kabir_1996}, either within the context of meson--antimeson systems~\cite{Lee:1957qq,Pontecorvo:1957cp}, neutrino-flavour oscillations~\cite{Pontecorvo:1967fh,Mikheyev:1985zog,Wolfenstein:1977ue}, or through axion-photon mixing~\cite{Raffelt:1987im}. 
In addition to quantum sensing techniques based on optical and cold-atom interferometers~\cite{Alonso:2022oot}, significant progress has recently been made in the construction of large-scale quantum computing facilities. With the aid of these facilities, algorithms for quantum simulations can be deployed to investigate particle-physics systems that exhibit mixing, oscillation, and decay~\cite{Pilaftsis:1997dr} in a more systematic way. Several critical phenomena~\cite{Karamitros:2022oew,Cao:2023syu} found in new physics models can now be explored and experimentally tested, since the model parameters of such systems can be tuned accordingly\footnote{We caution the reader that the critical phenomena we will be studying here should not be confused with phase transitions which could occur near or at quantum critical points in entangled spin-chain models~\cite{Vidal:2002rm}.}. These quantum systems could thus provide a perfect environment for studying the transport equations that occur in describing the dynamics of the early Universe~\cite{BhupalDev:2014pfm,Karamitros:2023tqr}.

Two-level quantum systems were first studied by Rabi~\cite{Rabi:1937dgo} in order to describe his nominal oscillatory phenomenon of spin precession of fermions in the presence of external magnetic fields. In the context of Quantum Information Theory, such two-level quantum systems are known as~{\it qubits}, whose quantum coherence properties can be accurately described by employing the Bloch sphere formalism~\cite{PhysRev.70.460,PhysRevA.51.2738}.

If a two-level quantum system turns out to be unstable like the kaon system, one has to consider decay-width effects in addition to oscillations. These effects have been taken into account by Lee, Oehme and Yang~\cite{Lee:1957qq} who made use of the famous effective Weisskopf and Wigner~(WW) approximation~\cite{Weisskopf1930}.  In the WW approximation, the dynamics of an unstable quantum system, such as an unstable qubit, may be conveniently described by making use of an effective non-Hermitian Hamiltonian, $\textrm{H}_\textrm{eff}$.
The time evolution of an unstable quantum state, $\ket{\Psi}$, is then governed by an effective Schr\"{o}dinger equation,
\begin{equation}
  \label{eq:Schrodinger_def}
    i\partial_t \ket{\Psi} = \textrm{H}_\textrm{eff} \ket{\Psi}\;.  
\end{equation}
This equation plays a central role in computing observable quantities in the time evolution of unstable qubits in particle physics, such as the $K^0\bar{K}^0$-, $B^0\bar{B}^0$- and $D^0\bar{D}^0$-meson systems~\cite{ParticleDataGroup:2024cfk}.

Despite~\eqref{eq:Schrodinger_def} being formulated more than six decades ago~\cite{Lee:1957qq}, the complete set of exact analytic solutions to this equation and their physical significance were found to be inadequate and only advanced recently~\cite{Karamitros:2022oew}. Although it was known that an initially mixed unstable qubit is expected to approach a pure QM state aligned to its long-lived eigenstate of~${\rm H}_{\rm eff}$ at sufficiently large times, the time evolution of an unstable qubit becomes rather complex if the two energy eigenstates happen to have equal lifetimes. In particular, we identified in~\cite{Karamitros:2022oew} a new class of {\em critical} unstable two-level systems, which we called Critical Unstable Qubits (CUQs). The characteristic feature of CUQs is that the energy-level and decay-width  vectors, ${\bf E}$ and ${\bf\Gamma}$, are orthogonal to one another, and their norm ratio, $r = |{\bf \Gamma}|/(2|{\bf E}|)$ is less than~1\footnote{Note that the limiting case $r = 1$ was originally studied in~\cite{Pilaftsis:1997dr}, and corresponds to a non-diagonalisable~$\textrm{H}_{\rm eff}$ which describes a two-level mixing scenario with maximal resonant CP violation.}. Specifically, we found that the CUQs display a number of unusual quantum properties. Most strikingly, if a CUQ ensemble is prepared to be initially at a fully mixed state characterised by a Bloch vector ${\bf b} = {\bf 0}$ of the normalised density matrix $\widehat{\rho}\equiv\rho / \text{Tr}\,\rho$, its time evolution will exhibit coherence--decoherence oscillations in the {\em co-decaying} frame defined by~$\widehat{\rho}$. Equally remarkably, if a pure QM state given by a normalised unit Bloch vector, with $|{\bf b}| = 1$ and ${\bf b} \nparallel {\bf E}$, then the unit vector ${\bf b}(t)$ will rotate about the direction defined by the energy-level vector~${\bf E}$, and it will sweep out {\em unequal} areas in equal amounts of time. We must note that this phenomenon differs from the usual sinusoidal oscillatory pattern that would result from the energy-level difference of a typical two-level quantum system. 

It is therefore important to observe that CUQs exhibit unusual quantum behaviours that have no analogue not only in classical, but even in ordinary quantum systems. Their simulation would require the use of quantum computers~\cite{DiMolfetta:2016gzc,Arguelles:2019phs,Jha:2021itm,Chen:2021cyw}, which could provide an independent confirmation of the many unexpected predictions. Such a study goes beyond the scope of the present article, which focuses on the $B^0\bar{B}^0$ system. In this work, we will use a Fourier series decomposition to define new anharmonicity observables which would enable us to quantify the degree of {\em non}-sinusoidal oscillation of a CUQ. The results of our formalism will be applied to the $B^0\bar{B}^0$-meson system, and we will derive, for the first time, generic upper limits on these new observables.

The paper is structured as follows. After this introductory section, Section~\ref{sec:density_matrix} lays out the effective Hamiltonian and the density matrix approach to qubit systems. We outline the key conditions defining a CUQ and derive an expression for the evolution of its co-decaying Bloch vector~${\bf b}(t)$. In Section~\ref{sec:Fourier}, we use the analytic expression for the time evolution of CUQ oscillations to compute their Fourier coefficients. Additionally, we give estimates for the reliability of the Fourier series and introduce anharmonicity factors which may be used to estimate defining parameters of the effective Hamiltonian. In Section~\ref{sec:ApplBmeson}, considering that the $B^0_{\rm d}\bar{B}_{\rm d}^0$-meson system is a CUQ, 
we compute the Fourier coefficients of its oscillation pattern deduced from current experimental data. We also estimate the precision required for future experiments to detect anharmonicities in $B$-meson oscillations. Section~\ref{sec:Concl} summarises the main results of our work, and proposes potential future research directions. Finally, all technical details relevant to our work are given in Appendices~\ref{App:Geometry} and~\ref{App:Errors}. 

\vfill\eject

\section{The Density Matrix}\label{sec:density_matrix}
\setcounter{equation}{0}

The density matrix formalism~\cite{Shankar:102017,Sakurai:1341875} gives a complete description for both pure and mixed quantum states. This formalism makes use of the so-called density matrix operator~$\rho$, which is defined through a collection of normalised eigenstates of the Hamiltonian $\ket{\Psi^k}$ and is given by
\begin{equation}\label{eq:DenMat}
    \rho \: = \: \sum_{k=1}^N \, w_k \, \ket{\Psi^k}\!\bra{\Psi^k} \, ,
\end{equation}
where $w_k \in \lrsb{0,1}$ are positively valued weights equal to the probability of observing the quantum state~$\ket{\Psi^k}$. As mentioned above, the density matrix formalism has the additional feature that enables one to distinguish between pure and mixed quantum states. In this formalism, a pure state is defined by a density matrix consisting of a single quantum state, 
\begin{equation}
    \rho_{\rm pure} \: = \: \ket{\Psi}\!\bra{\Psi} \, ,
\end{equation}
which manifestly obeys the idempotency property,
\begin{equation}
    \rho^2 \: = \: \rho \, .
\end{equation}
In contrast, a mixed state is simply defined as a quantum state which is not pure, $\rho^2 \neq \rho$.

In addition to the advantages outlined above, the density-matrix formalism permits us to study the correlations between different states and their associated probabilities. As can be deduced from its definition given in~\eqref{eq:DenMat}, the diagonal elements of the density matrix~$\rho$ describe probabilities, while the off-diagonal elements quantify the quantum correlations of the system. As a consequence of unitarity, the trace of $\rho$ should evaluate to unity,
\begin{equation}
    {\rm Tr} \, \rho \: = \: \sum_{\alpha, \, k} w_k \left|\left\langle \Psi^k| \alpha \right\rangle \right|^2 \: = \: \sum_{\alpha} \, \mathbb{P}(\alpha) \: = \: 1 \, .
\end{equation}
In the above, $\ket{\alpha}$ is a basis state of the Hilbert space, and $\mathbb{P}(\alpha)$ is the probability of the system being in the state described by $\ket{\alpha}$.

The density matrix has been well studied within the context of unitarity preserving systems described by Hermitian Hamiltonians~\cite{PoincareH1889,PhysRev.70.460}. However, its extension to effective Hamiltonians which may violate unitarity can lead to statistically inconsistent results, since one is considering a particular sub-matrix of a larger density matrix~\cite{Kabir_1996,Bernabeu:2003ym}. In essence, one only considers some part, $\mathcal{H}_P$, of a wider Hilbert space, $\mathcal{H}_T = \mathcal{H}_P \otimes \mathcal{H}_D$, with the remaining states contained in $\mathcal{H}_D$ neglected. Such unitarity-violating effective Hamiltonians may be parametrised in terms of two Hermitian matrices, ${\rm E}$ and $\Gamma$, describing {\em dispersive} and {\em dissipative} quantum effects, respectively. The effective Hamiltonian then takes the form,
\begin{equation}
    {\rm H_{eff}} \: = \: {\rm E}\, -\, \frac{i}{2}\Gamma \; .
\end{equation}
In our analysis, we assume that the density matrix is constructed from eigenstates of the Hermitian\- part of the effective Hamiltonian, ${\rm E}$, with the anti-Hermitian part, $i\Gamma$, describing unitarity-violating processes such as decays. Here, we note in passing that similar effective Hamiltonians have been studied in a much wider contexts, such as open quantum systems and trace-preserving Markovian master equations~\cite{cmp/1103899849, REDFIELD19651}. 

Assuming a Schr\"{o}dinger-like evolution for the eigenstates, one can derive the evolution equation for the density matrix,
    \begin{equation}
        \frac{{\rm d} \rho}{{\rm d}t} \: = \: -i \lrsb{{\rm E}\, , \, \rho} - \frac{1}{2} \lrcb{\Gamma \, , \, \rho} \, .
    \end{equation}
By taking the trace on both sides of this last equation, it is not difficult to find that 
\begin{equation}
        \frac{{\rm d \, Tr}\rho}{{\rm d}t} \: = \: - {\rm Tr} \lrb{\Gamma \rho} \: \neq \: 0 \, ,
\end{equation}
which indicates a leakage of the probability from the quantum system defined in the subspace~$\mathcal{H}_P$. 
In addition, we observe that the inclusion of non-trivial decay terms prevents the factorisation of the evolution equation for the trace, ${\rm Tr}\rho$, and as such, ${\rm Tr} \rho (t)$ no longer shows up the usual exponential fall-off behaviour. On the other hand, the apparent violation of unitarity is readily remedied by normalising the density matrix to its trace,
\begin{equation}\label{eq:NormDenMatDef}
    \widehat{\rho} \: = \: \frac{\rho}{{\rm Tr} \, \rho} \ .
\end{equation}
By careful consideration of the states, we see that $\widehat{\rho}$ is an operator which will be acting on a normalised Hilbert subspace $\mathcal{H}_P$ pertinent to the two-level quantum system of interest to us. 
The so-normalised matrix,~$\widehat{\rho}$, was called the {\em co-decaying} density matrix in~\cite{Karamitros:2022oew}, and obeys the traceless evolution equation,
\begin{equation}\label{eq:NormDenEq}
    \frac{{\rm d} \widehat{\rho}}{{\rm d}t} \: = \: -i \lrsb{{\rm E}\, , \, \widehat{\rho}} - \frac{1}{2} \lrcb{\Gamma \, , \, \widehat{\rho}} + {\rm Tr} \lrb{\Gamma \widehat{\rho}} \, \widehat{\rho} \, ,
\end{equation}
given that ${\rm Tr} \, \widehat{\rho} = 1$. Evidently, the modified time-evolution equation stated in~\eqref{eq:NormDenEq} implies conservation of probability. Nevertheless, due to the non-exponential decay of the trace, the latter comes at the cost of linearity. This can be seen from the RHS in~(\ref{eq:NormDenEq}), where its final term is quadratic in $\widehat{\rho}$. Notice that in the simple case, with $\Gamma\propto \mathds{1}$, the two decay-width terms cancel against each other on the RHS of~\eqref{eq:NormDenEq}, and we return to the standard paradigm of Rabi oscillations which dictates the time evolution of ordinary quantum systems. In~\cite{Karamitros:2022oew}, the inclusion of the aforementioned non-linear term is shown to give rise to novel phenomena, such as coherence-decoherence oscillations, and, as we will see in the subsequent sections, to anharmonic oscillations between energy levels which deviate from the usual picture of sinusoidal oscillations.

\vfill\eject

\subsection{Two-Level Quantum Systems}\label{sec:2Level}

We will now employ the Bloch-sphere formalism to analyse the time evolution of two-level quantum systems, also known as {\em qubits}. From~(\ref{eq:DenMat}), it is clear that the density matrix $\rho$, and hence its normalised version $\widehat{\rho}$ (with $\text{Tr}\,\widehat{\rho} = 1$), are Hermitian. Hence, the co-decaying density matrix~$\widehat{\rho}$ can be elegantly represented through a vector, $\vecb\in \mathbb{R}^3$, the so-called Bloch vector, once~$\widehat{\rho}$ is decomposed in terms of the 2D identity matrix~$\mathds{1}$ and the Pauli vector $\boldsymbol{\sigma} = (\sigma_1\, , \sigma_2\, , \sigma_3)$,
\begin{equation}
    \widehat{\rho} \: = \: \frac{1}{2}\, \Big(\mathds{1} + \vecb \cdot\boldsymbol{\sigma} \Big) \, .
\end{equation}
This matrix representation was initially used by Poincar\'{e} and Bloch in studies on light~\cite{PoincareH1889,stokes_2009} and nuclei~\cite{PhysRev.70.460}, respectively. By construction, $\widehat{\rho}$ satisfies the unitary and Hermiticity conditions of the density matrix, and it has the unique feature that the coherence of the system may be characterised through the magnitude of the Bloch vector, $\mathbf{b}$. By squaring  $\mathbf{b}$, one can see that the system is pure, when $|\vecb|=1$, and mixed, if $|\vecb|<1$. In the case of the effective Hamiltonians we are considering, we refer to  $\mathbf{b}$ as the {\em co-decaying} Bloch vector, so as to  explicitly indicate that the system is normalised according to~(\ref{eq:NormDenMatDef}). 

In line with the representation of the density matrix on the Pauli basis, we may similarly express the Hermitian energy and decay matrices as
\begin{equation}
    {\rm E} \: = \: {\rm E}_\mu \sigma^\mu \: = \: {\rm E}^0 \mathds{1}_2 - \mathbf{E}\cdot\boldsymbol{\sigma} \, ,
\end{equation}
\begin{equation}
    \Gamma \: = \: \Gamma_\mu \sigma^\mu \: = \: \Gamma^0 \mathds{1}_2 - \boldsymbol{\Gamma}\cdot\boldsymbol{\sigma} \, .
\end{equation}
Substituting the above two expansions in the evolution equation~\eqref{eq:NormDenEq}, we derived in~\cite{Karamitros:2022oew} a non-linear, first-order differential equation which describes the motion of ${\bf b}(t)$,
\begin{equation}
  \label{eq:dbdt}
    \frac{{\rm d}\vecb}{{\rm d}t} \: = \: -2\mathbf{E}\times\vecb + \boldsymbol{\Gamma} - \lrb{\boldsymbol{\Gamma}\cdot\vecb}\vecb \, .
\end{equation}
Following the parameterisation laid out in~\cite{Karamitros:2022oew}, we introduce the dimensionless parameters
\begin{equation}
    \tau \: = \: |\boldsymbol{\Gamma}|\,t \, , \qquad r \: = \: \frac{|\boldsymbol{\Gamma}|}{2|\mathbf{E}|}\, ,
\end{equation}
as well as the unit vectors $\mathbf{e}=\mathbf{E}/|\mathbf{E}|$ and $\boldsymbol{\gamma}=\boldsymbol{\Gamma}/|\boldsymbol{\Gamma}|$. With the help of all these parameters, the evolution equation of  $\mathbf{b}$ in~\eqref{eq:dbdt} may be recast into a fully dimensionless differential equation,
\begin{equation}
  \label{eq:dbdtau}
    \frac{{\rm d}\vecb}{{\rm d}\tau} \: = \: -\frac{1}{r}\,\mathbf{e}\times\vecb\, +\, \boldsymbol{\gamma}\, -\, \lrb{\boldsymbol{\gamma}\cdot\vecb}\vecb \; ,
\end{equation}
which only depends on two unit vectors, $\mathbf{e}$ and $\boldsymbol{\gamma}$, and a single real-valued parameter, $r$.
Here, it is important to stress that \eqref{eq:dbdtau} is our {\em master evolution equation} for studying the motion of~${\bf b}(\tau )$ for unstable qubits.  

In the co-decaying frame introduced above, a CUQ is identified as a two-level quantum system whose co-decaying Bloch vector, ${\bf b}(\tau)$, does not approach a specific stationary point in the 3D space. In~general, we expect that a decaying system will relax in the state with longest lifetime. But this frequently assumed picture is not always true. From the parameter space, it can be seen that there exist two-level quantum systems that could exhibit a critical behaviour, if $\mathbf{e} \cdot \boldsymbol{\gamma} = 0$. For such systems, the asymptotic value of $\mathbf{b}$ is given by~\cite{Karamitros:2022oew}
\begin{equation}
    \mathbf{b}(\tau \to \infty)\: =\: \dfrac{\sqrt{r^2-1}}{r} \boldsymbol{\gamma}\, 
    -\, \dfrac{1}{r} \mathbf{e} \times \boldsymbol{\gamma}\; .
    \label{eq:asymptotic_b}
\end{equation}
Observe that there is {\em no} admissible asymptotically fixed solution for a real-valued vector, like~$\mathbf{b}$, when~${r<1}$. Consequently, CUQs are characterised by the two conditions: $\mathbf{e} \cdot \boldsymbol{\gamma} = 0$ and $r<1$. For~this special class of qubits, we will find that energy-level oscillations will proceed indefinitely and have unique features, such as anharmonicities in the oscillations between energy levels, while all trajectories of ${\bf b}(\tau )$ will be entirely contained onto a plane.

\begin{figure}[t!]
    \centering
    \includegraphics[width=0.7\linewidth]{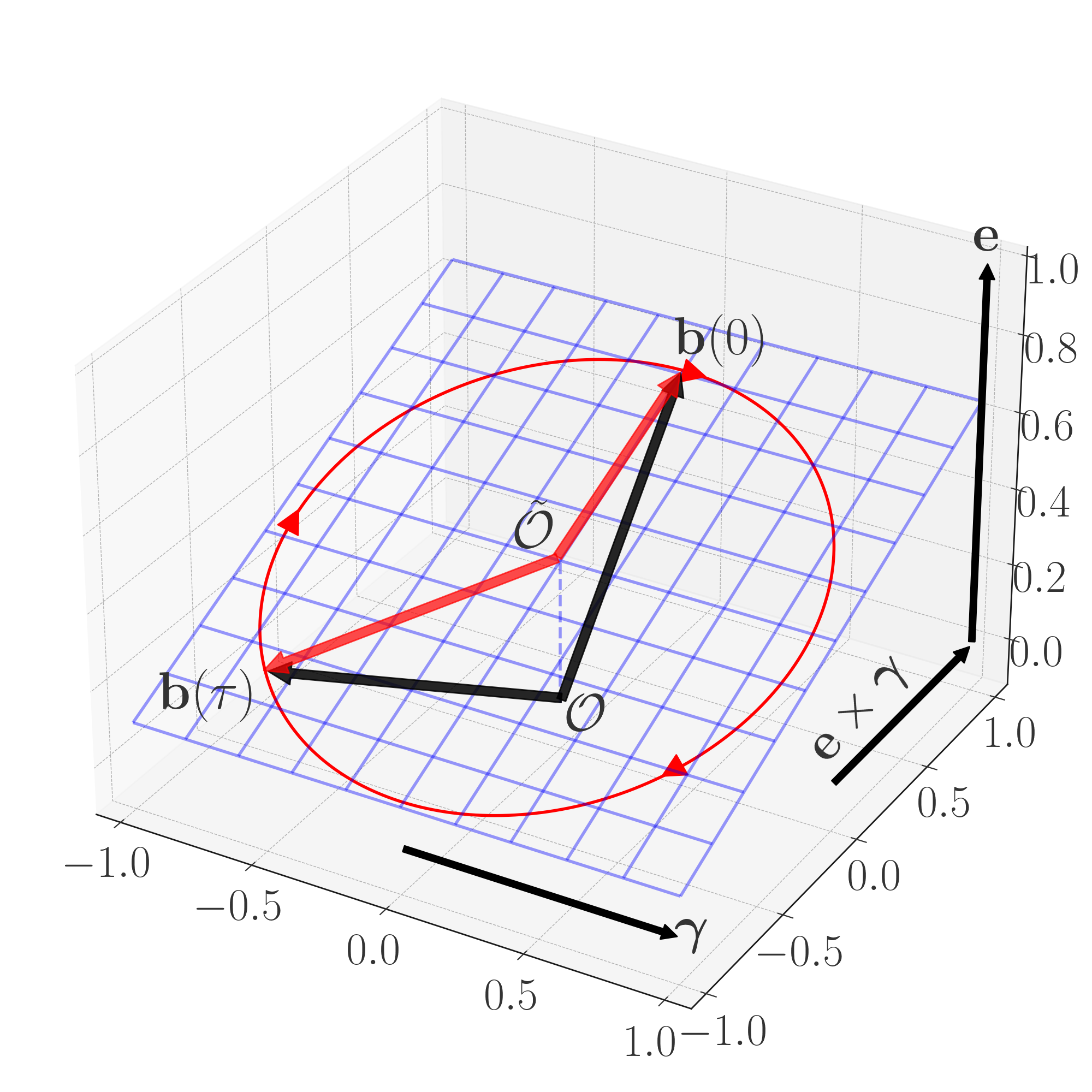}
    \caption{The plane of oscillation for the co-decaying Bloch vector, $\mathbf{b}$, over a full period. The system is assumed to be critical with $\mathbf{e}\cdot\boldsymbol{\gamma}=0$, and $r=0.7$. The initial condition is taken to be $\mathbf{b}(0)=0.6\,\mathbf{e} + 0.8\, \mathbf{e}\times\boldsymbol{\gamma}$. The blue plane is constructed using the Frenet-Serret procedure, and contains the trajectory of the $\mathbf{b}$. The red curve indicates the trajectory traced by $\mathbf{b}$ with the red arrows indicating the direction of motion. The two example Bloch vectors: $\mathbf{b}(0)$ and $\mathbf{b}(\tau)$ with $\tau=\frac{3}{4}\widehat{P}$, are drawn from the true origin, and the projections of $\mathbf{b}$ onto the plane are displayed as red arrows drawn from the origin of the plane $\tilde{\mathcal{O}}$.}
    \label{fig:BlochPlane}
\end{figure}

For illustration, in Figure~\ref{fig:BlochPlane} we give a typical example that describes the time evolution of the co-decaying Bloch vector, $\mathbf{b}(\tau )$, for a CUQ. After solving~\eqref{eq:dbdtau}, we find that the time evolution of a~CUQ, as described by $\mathbf{b}(\tau )$, follows a closed path (red curve), namely that of a circle $S^1$, and as such exhibits a periodic behaviour with period $\widehat{\rm P}$. As argued in~\cite{Karamitros:2022oew}, the trajectory of~$\mathbf{b}(\tau)$ for a CUQ is contained in a plane that can be identified by the Frenet-Serret procedure denoted blue in Figure~\ref{fig:BlochPlane}. Moreover, the black arrows denote the basis of the space, as well as $\mathbf{b}(\tau )$ at $\tau=0$ and $\tau=\frac{3}{4}\widehat{\rm P}$ as spanned from the true origin, $\mathcal{O}$. 
The red arrows describe the projection of the co-decaying Bloch vector, $\mathbf{b}$, onto the plane, with the base at the origin of the plane, $\tilde{O}$. Upon performing a geometric analysis of this plane, one finds that the plane is always parallel to the decay vector $\boldsymbol{\gamma}$. Consequently, $\boldsymbol{\gamma}$ forms one of the basis vectors of the plane, with a particular complementary vector, $\boldsymbol{v}$, which satisfies $\boldsymbol{v}\cdot\boldsymbol{\gamma} = 0$ and thus uniquely defines the plane. An analytic expression for $\boldsymbol{v}$ is given in~\eqref{eq:PerpVec} of Appendix~\ref{App:Geometry}. The vector, $\boldsymbol{v}$, approaches $\boldsymbol{v}=\mathbf{e}\times\boldsymbol{\gamma}$, when the initial condition is perpendicular to~$\mathbf{e}$.
\begin{figure}[t!]
    \centering
    \begin{subfigure}{0.49\linewidth}
    \centering
    \includegraphics[width=\linewidth]{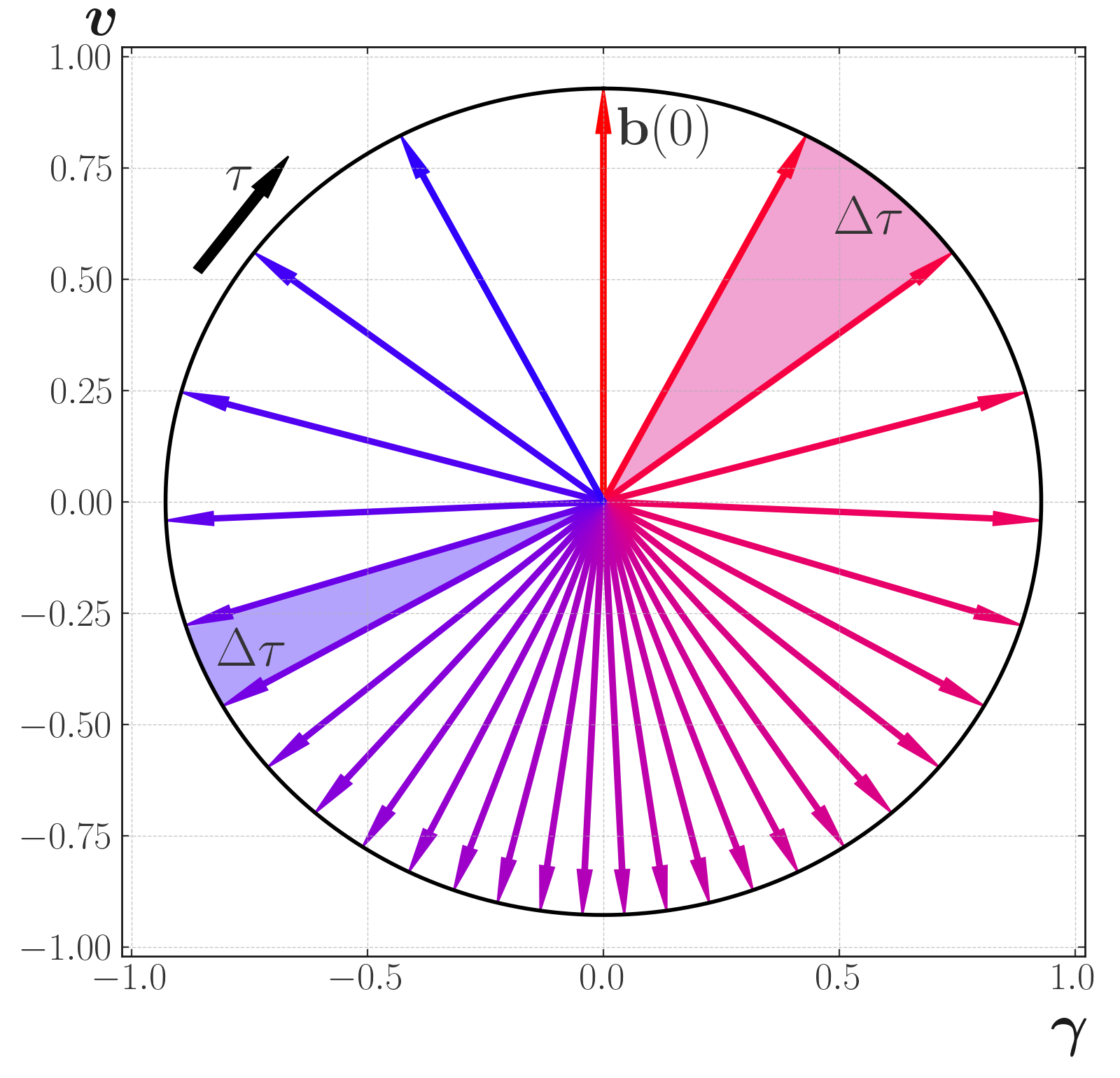}
    \caption{\empty}
    \label{fig:Arrow_1}
    \end{subfigure}
    \begin{subfigure}{0.49\linewidth}
    \centering
    \includegraphics[width=\linewidth]{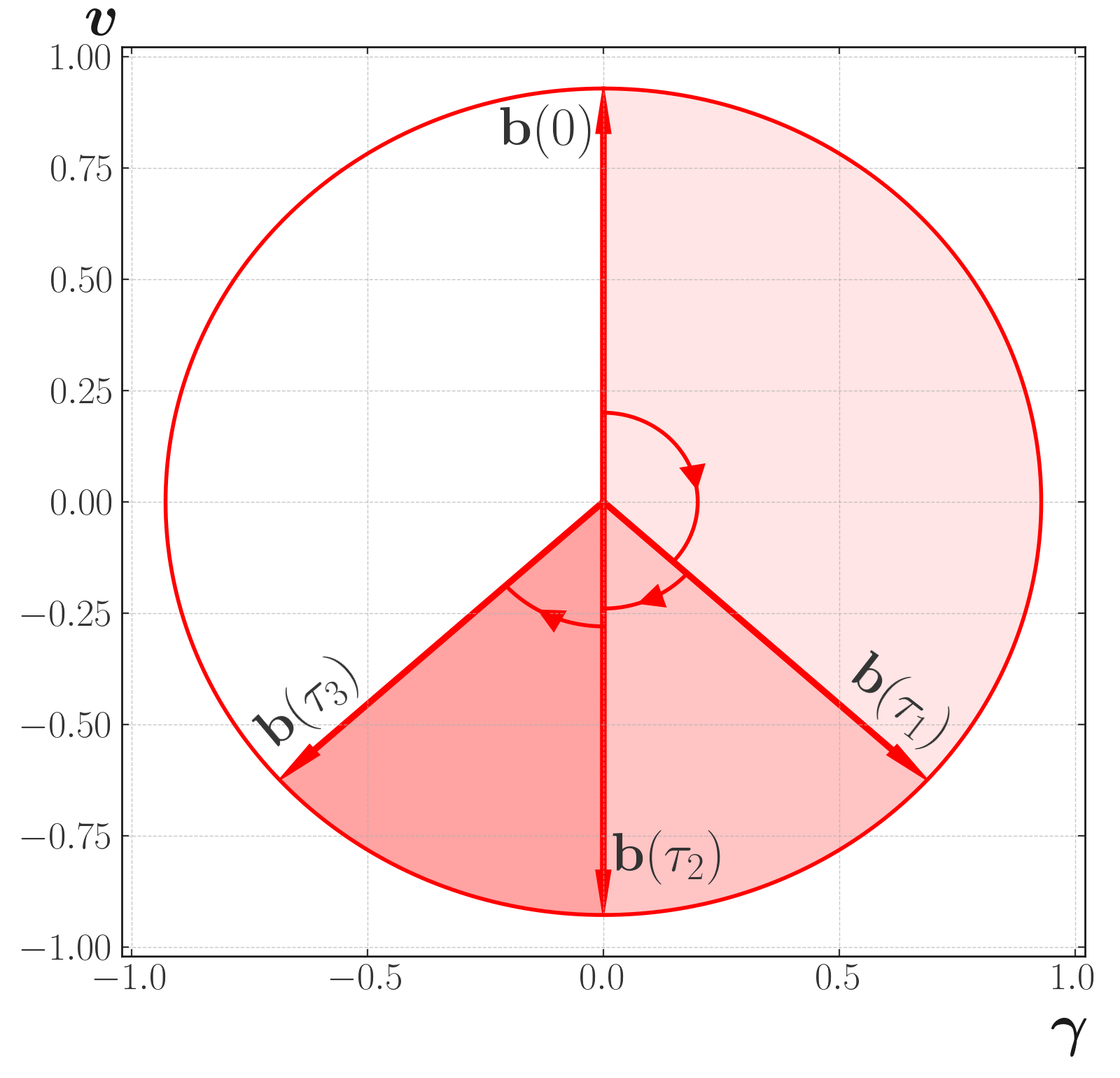}
    \caption{\empty}
    \label{fig:Arrow_2}
    \end{subfigure}
    \caption{Evolution of the projection of the co-decaying Bloch vector, $\mathbf{b}$, onto the plane in a CUQ scenario for the same input parameters as given in Figure~\ref{fig:BlochPlane}. In the left panel ($a$), each arrow is equally $\tau-$spaced, with red hues indicating the evolution at and immediately proceeding $\tau=0$ and blue hues indicating the evolution in the approach to $\tau=\widehat{\rm P}$. The right panel ($b$) shows the evolution of $\mathbf{b}$ split into quartiles of the period, with $\tau_1 = \frac{1}{4}\widehat{\rm P}$, $\tau_2 = \frac{1}{2}\widehat{\rm P}$, and $\tau_3 = \frac{3}{4}\widehat{\rm P}$. Observe that for a CUQ, the fundamental domain of its non-sinusoidal oscillation is isomorphic to $S^1/\mathbb{Z}_2$, instead of  $S^1/(\mathbb{Z}_2\times\mathbb{Z}_2)$ which is for an ordinary qubit.}
    \label{fig:Arrow_figs}
\end{figure}

Figure~\ref{fig:Arrow_figs} further exemplifies the results presented in Figure~\ref{fig:BlochPlane}, although it explicitly shows the motion of the projection of the Bloch vector onto the plane. As before, the initial condition of $\mathbf{b}$ is taken to be: $\vecb(0)=0.6\, \mathbf{e}+0.8\,\mathbf{e}\times\boldsymbol{\gamma}$, with $r=0.7$. As can be seen in Figure~\ref{fig:Arrow_figs}, the characteristic features of CUQs still persist, even when we consider more general initial conditions. In particular, we see that the inhomogeneous oscillation of~$\mathbf{b}(\tau )$ is found to occur. This observation therefore provides a foundation on which we may build a novel method for the analysis of CUQ oscillations, by means of a Fourier series decomposition, in order to study the anharmonicity of the signal.

We should remark that the fundamental domain of a typical qubit oscillation is $S^1/(\mathbb{Z}_2\times\mathbb{Z}_2)$. This means that only a quarter period is needed to infer the full oscillation profile which is mapped to the circle $S^1$. However, in Figure~\ref{fig:Arrow_figs}, we observe that for CUQs, the above symmetry is reduced, as a consequence of non-trivial off-diagonal decay-width effects. For~CUQs, the fundamental domain of the oscillation is enlarged to: $S^1/\mathbb{Z}_2$. In this case, full knowledge of the half-period would be required to detail the oscillation profile over the whole period~$S^1$.

Since the analysis required to identify the oscillation plane generated by the motion of the co-decaying Bloch vector~${\bf b}(\tau)$ is technical, the full computation of its construction, including analytic expressions for the basis of the plane, is given in Appendix~\ref{App:Geometry}. In particular, the calculation demonstrates clearly how the phenomena described in~\cite{Karamitros:2022oew} can be generalised to more generic initial conditions, with anharmonic oscillation of $\mathbf{b}$ present in the projection onto the plane. Additionally, Appendix~\ref{App:Geometry} demonstrates a clear correspondence between setups with generic initial conditions and setups constrained to the $\lrcb{\boldsymbol{\gamma}, \mathbf{e}\times\boldsymbol{\gamma}}$ plane. As a consequence, it is sufficient to carry out the analysis on the $\lrcb{\boldsymbol{\gamma}, \mathbf{e}\times\boldsymbol{\gamma}}$ plane, with the results of this analysis then appropriately remapped. However, there are some additional considerations that one must make when considering more general initial conditions, such as an effective value of~$r$, henceforth referred to as $\tilde{r}$. For the sake of brevity, the necessary rescaling of $\tilde{r}$ to find the true value of $r$ which characterises the effective Hamiltonian is stated in the main body without justification, but a full exposition of the mathematical details is given in Appendix~\ref{App:Geometry}.

\subsection{Critical Evolution of a Pure State Qubit}\label{sec:CritScen}

In this subsection, we outline the derivation of an  equation that explicitly describes the $\tau$-dependence of the angle, $\varphi (\tau )$, between the co-decaying Bloch vector, ${\bf b}(\tau)$, and the fixed unit decay vector, $\boldsymbol{\gamma}$. For simplicity, we consider the most straightforward circumstance, where $\mathbf{b}(\tau )$ is always restricted to lying on the $\lrcb{\boldsymbol{\gamma}, \mathbf{e}\times\boldsymbol{\gamma}}$ plane. However, the calculation applies to more general CUQ scenarios. In Appendix~\ref{App:Geometry}, we give a detailed demonstration of the equivalence of all systems characterised by $\mathbf{e}\cdot\boldsymbol{\gamma}=0$ and explicitly show that the unit decay vector, $\boldsymbol{\gamma}$, remains a basis of the plane in the most general CUQ setup. In light of this geometric property, we~calculate the Fourier series of the projection of ${\bf b}(\tau)$ along $\boldsymbol{\gamma}$. For completeness, the projection ${\bf b}(\tau)$ along the remaining axis is also given towards the end of the section, which, for the simple case we consider here, is $\mathbf{e}\times\boldsymbol{\gamma}$.

It was shown in~\cite{Karamitros:2022oew} that the torsion of the trajectory of the co-decaying Bloch vector, $\mathbf{b}(\tau)$, vanishes whenever $\mathbf{e}\cdot\boldsymbol{\gamma}=0$. As a consequence, in CUQs, $\mathbf{b}(\tau)$ is necessarily contained to a plane. By choosing to study the motion of $\mathbf{b}(\tau)$ in this plane, we may reduce the vector equation to a scalar equation by projecting along $\boldsymbol{\gamma}$, without losing information. In this way, we find the evolution equation,
\begin{equation}
   \label{eq:dbgammadtau}
    \frac{{\rm d} (\mathbf{b}\cdot\boldsymbol{\gamma})}{{\rm d}\tau} \: = \: - \frac{1}{r}\lrb{\mathbf{e}\times \mathbf{b}}\cdot\boldsymbol{\gamma} + 1 - (\mathbf{b} \cdot \boldsymbol{\gamma})^2 \, .
\end{equation}
Assuming a pure state, $|\mathbf{b}|=1$, the projection along $\boldsymbol{\gamma}$ may be written as $\mathbf{b}\cdot\boldsymbol{\gamma}=\cos\, \varphi$. For~later convenience, we redefine the angle $\varphi$ by shifting its value by $\pi/2$, such that~$\varphi = \theta + \frac{\pi}{2}$. After making this shift, the projection, $\mathbf{b}\cdot\boldsymbol{\gamma}$, becomes an odd function of $\theta$, since~
\begin{equation}
  \label{eq:bgamma}
\mathbf{b}\cdot\boldsymbol{\gamma}\: =\: -\,\sin \theta\,.
\end{equation}
Also, we employ the triple product identity to obtain the projection along $\mathbf{e}\times\boldsymbol{\gamma}$,
\begin{equation}
  \label{eq:ebgamma}
    \lrb{\mathbf{e}\times \mathbf{b}}\cdot\boldsymbol{\gamma} \: = \: -\lrb{\mathbf{e}\times \boldsymbol{\gamma}}\cdot\mathbf{b} \: = \: -\cos \, \theta \, .
\end{equation}
Substituting~\eqref{eq:bgamma} and~\eqref{eq:ebgamma}  in~\eqref{eq:dbgammadtau}, we find a rather simple expression for the angular velocity, ${\rm d}\theta (\tau)/{\rm d}\tau$, given by 
\begin{equation}
  \label{eq:dthetadtau}
    \frac{{\rm d} \theta}{{\rm d} \tau} \: = \: - \frac{1}{r}\, -\, \cos \, \theta \; .
\end{equation}

Since CUQ oscillations are periodic as we discussed in previous subsection, the choice of initial condition is somewhat arbitrary, as long as $|\vecb|=1$. Therefore, we will take an initial condition that gives us useful symmetries and set the initial state of ${\bf b}(\tau )$ to be equal to $\mathbf{e}\times \boldsymbol{\gamma}$, i.e.~$\theta(\tau=0) = 0$.  
Given that the differential equation in~\eqref{eq:dthetadtau} is separable, its analytic solution can be obtained as follows:
\begin{equation}
  \label{eq:tautheta}
       \tau \: = \: - \int_0^\theta \, \frac{r}{1+r\cos \, x}\: {\rm d} x \: = \: - \frac{2r}{\sqrt{1-r^2}} \tan^{-1}\lrb{\sqrt{\frac{1-r}{1+r}} \, \tan \, \frac{\theta}{2}} \,. 
\end{equation}
In the co-decaying frame, the oscillation period $\widehat{\rm P}$ of a CUQ may be found by integrating~\eqref{eq:dthetadtau} over the full range of $\theta \in (-\pi, \pi)$,
\begin{equation}\label{eq:DLPeriod}
    \widehat{\rm P}\: =\: \int_{-\pi}^\pi \, \frac{r}{1+r\cos \, x}\: {\rm d} x \: = \: \frac{2\pi r}{\sqrt{1-r^2}}\ .
\end{equation}
Given~$\widehat{\rm P}$,  we may define the dimensionless angular frequency
\begin{equation}
    \widehat{\omega} \: = \: \frac{2\pi}{\widehat{\rm P}} \: = \: \frac{\sqrt{1-r^2}}{r} \, .
\end{equation}
Taking this last expression into account, we find an expression for the angle $\theta$ in terms of~$\tau$,
\begin{equation}\label{eq:AngSol}
    \tan \frac{\theta}{2} \: = \: -\,\sqrt{\frac{1+r}{1-r}} \, \tan \frac{\widehat{\omega} \tau}{2} \, .
\end{equation}
In the limit $r \to 0$, we can see that~\eqref{eq:AngSol} returns the expected result corresponding to Rabi oscillations: $\theta (\tau ) = -\widehat{\omega} \tau$, where the lingering pre-factor $-1$ indicates that ${\bf b}(\tau )$ rotates in the plane in a clockwise direction. Restoring the units of the period and angular frequency is easily done by appropriately rescaling by the magnitude of the decay vector $\mathbf{\Gamma}$:
\begin{equation}
    {\rm P} \: = \: \frac{\pi}{|\mathbf{E}|\sqrt{1-r^2}} \ , \qquad \omega \: = \: 2|\mathbf{E}|\,\sqrt{1-r^2}\ .
\end{equation}
Notice that as $r$ approaches $1$, these expressions imply an oscillation period ${\rm P}$ that tends to infinity. Thus, in the extreme case $r=1$,  the co-decaying Bloch vector $\mathbf{b}$ of CUQ no longer oscillates, independently of any initial condition for $\mathbf{b}(0)$ at $t=0$. This non-oscillatory feature must be anticipated for such extremal CUQs, since the two energy eigenstates of the two-level quantum system become exactly degenerate in this case. In~\cite{Pilaftsis:1997dr}, it was shown that such critical non-oscillatory two-level quantum systems maximise the phenomenon of resonant CP~violation through mixing of states.

\bigskip

\section{Fourier Coefficients for Critical Unstable Qubits}\label{sec:Fourier}
\setcounter{equation}{0}

Having stated the basic equations governing the time evolution of the co-decaying Bloch vector, $\mathbf{b}(\tau )$, in the previous section [cf.~\eqref{eq:dbdtau} and~\eqref{eq:AngSol}], we will now consider observables that would allow us to physically distinguish CUQs from ordinary unstable qubits. In general, observable quantities are often proportional to the projection of ${\bf b}(\tau )$  along a particular axis, e.g.~along the unit decay vector~$\boldsymbol{\gamma}$ [cf.~\eqref{eq:AngSol}]. Although the expression in~\eqref{eq:AngSol} could be used to fit a signal and extract estimates for $r$,  it may still be difficult to reach a precision high enough to determine these values if $r \ll 1$. However, as discussed in the previous section, CUQs have the unique feature that their oscillation is \textit{anharmonic} rather than harmonic. Therefore, we~propose searching for Fourier modes beyond the principal harmonic mode as an alternative and more efficient method of extracting the key model parameter~$r$. With this in mind, we consider in the following two subsections a Fourier series decomposition of the expected signal arising from~a CUQ.

\subsection{Evaluation of Fourier Coefficients}

As a reference model, we assume a system which is initially prepared so that $\be(0) = \mathbf{e}\times \boldsymbol{\gamma}$. Additionally, since $\widehat{\omega}\tau = \omega t$ is a dimensionless quantity, we choose to evaluate the Fourier coefficients in terms of the dimensionless quantities, $r$, $\widehat{\omega}$, and~$\tau$. By choosing to study the evolution of $\vecb(\tau)$ with this initial condition, we are able to fully exploit the symmetries present in the evolution and express $\be \cdot \boldsymbol{\gamma}$ as an exclusively odd-term Fourier series:
\begin{equation}
  \label{eq:bdotgamma}
    \be (\tau) \cdot \boldsymbol{\gamma} \: = \: \sum_{n=1}^\infty \, c_n \, \sin \lrb{n\widehat{\omega} \tau} \, , 
\end{equation}
with
\begin{equation}\label{eq:FourierIntegral}
    c_n \: = \: -\frac{2}{\widehat{\rm P}} \int_{-\widehat{\rm P}/2}^{\widehat{\rm P}/2} \, \sin(n\widehat{\omega} \tau) \, \sin \theta \: \de \tau \, .
\end{equation}
By making use of~\eqref{eq:AngSol}, it is not difficult to find an expression for $\be \cdot \boldsymbol{\gamma} \, = \, - \sin \theta$, in terms of~$\tau$, by inverting the tangent half-angle transformations, 
\begin{subequations}
    \begin{align}
    \sin \theta \: &= \: \frac{2 \tan \frac{\theta}{2}}{1+\tan^2\frac{\theta}{2}} \: = \: -\sqrt{1-r^2}\frac{\sin \lrb{\widehat{\omega} \tau}}{1-r\cos\lrb{\widehat{\omega}\tau}} \ , \label{eq:sinEvol}\\
    \cos \theta \: &= \: \frac{1-\tan^2 \frac{\theta}{2}}{1+\tan^2 \frac{\theta}{2}} \: = \: \frac{\cos \lrb{\widehat{\omega}\tau} - r}{1 - r\cos\lrb{\widehat{\omega} \tau}} \, , \label{eq:cosEvol}
\end{align}
\end{subequations}
where the last formula for $\cos\theta$ was included for completeness. After inserting the expression for $\sin\theta$ in~\eqref{eq:sinEvol} into~\eqref{eq:FourierIntegral}, the Fourier coefficients may be computed as follows:
\begin{align}
    c_n \: &= \: \frac{2\sqrt{1-r^2}}{\widehat{\rm P}}  \int_{-\widehat{\rm P}/2}^{\widehat{\rm P}/2} \, \sin(n\widehat{\omega} \tau) \, \frac{\sin \lrb{\widehat{\omega} \tau}}{1-r\cos\lrb{\widehat{\omega}\tau}} \, \de \tau \,\nonumber\\
    &=\: \frac{\sqrt{1-r^2}}{\pi}  \int_{-\pi}^{\pi} \, \sin(n\phi) \, \frac{\sin \phi}{1-r\cos\phi} \, \de \phi
    \label{eq:cn}
\end{align}
Making use of the substitution $z=\exp\lrsb{i\phi}$, as well as exploiting the symmetry properties of complex integrals, allows one to express the Fourier coefficients $c_n$ as a closed contour integral bounded by the unit circle, $|z|=1$,
\begin{equation}
    c_n \: = \: \frac{\sqrt{1-r^2}}{i\pi r} \oint_{|z|=1} \, \frac{z^{n-1}\,(z^2-1)}{z^2-\frac{2}{r}z+1} \, \de z \, .
\end{equation}
This integral can be easily evaluated using the residue theorem, upon careful consideration of the relevant poles. In the region $|z|<1$, the numerator of the fraction in the integrand is analytic in $z$ for any $n\geq 1$, while its denominator is a quadratic equation in $z$ and has two roots given by 
\begin{equation}
     z_\pm \: = \: \frac{1}{r}\, \pm\, \frac{1}{r}\sqrt{1-r^2} \; .
\end{equation}
As we are interested in identifying CUQ oscillations, we assume that $0<r<1$. Hence, only the root, $z_- < 1$, will contribute to the integral. Taking this fact into account, the Fourier series coefficients are found to be
\begin{align}\label{eq:FourierCoeffC}
    c_n \: = \: 2 \frac{\sqrt{1-r^2}}{r}\,\bigg(\frac{1}{r}\,-\,\frac{\sqrt{1-r^2}}{r}\bigg)^n \, .
\end{align}
We should clarify here that there is no divergence 
in $c_n$ for very small values of $r$. To see this, we utilise
the MacLaurin series expansion for the expression of interest,
\begin{equation}
    \frac{1}{r}-\frac{\sqrt{1-r^2}}{r} \ \simeq \ \frac{1}{r} - \bigg(\frac{1}{r} - \frac{r}{2} + \mathcal{O}(r^3)\bigg) \ \simeq \  \frac{r}{2} + \mathcal{O}(r^3)\, .
\end{equation}
As can be seen in the above, the offending $1/r$ terms cancel  as $r\to 0^+$, thereby rendering the expression for $c_n$ finite. Thus, we find that the leading term of $c_n$ is of order $r^{n-1}$. 
In fact, in the limit $r\to 0^+$, we recover 
the typical Rabi oscillations for which only the principal harmonic survives, i.e.
\begin{equation}
    \lim_{r\to 0^+} \, c_1\: =\: 1 \, , \qquad \lim_{r\to 0} \, c_{n \, \geq \, 2}\: =\: 0 \, .
\end{equation}

To evaluate the Fourier coefficients for the projection of $\be (\tau )$ along $\mathbf{e}\times\boldsymbol{\gamma}$, we follow a nearly identical approach.\footnote{We utilize the same variable change as in~\eqref{eq:cn} to bring the coefficients in a form that can be evaluated by integrating over the unit circle.} In this case, we use the expression for $\cos\theta$ given in~\eqref{eq:cosEvol} and expand over even Fourier modes as follows:
\begin{equation}
  \label{eq:bexgamma}
    \be (\tau)\cdot(\mathbf{e}\times\boldsymbol{\gamma}) \: = \: d_0\, +\, \sum_{n=1}^\infty \, d_n \cos(n\widehat{\omega} \tau) \, ,
\end{equation}
where
\begin{subequations}
\begin{align}
  \label{eq:d0}
    d_0 \: &= \: \frac{i}{2\pi r}\oint_{|z|=1}  \, \frac{1}{z}\,\frac{z^2-2rz+1}{z^2-\frac{2}{r}z+1} \, \de z  \: = \: -\frac{1}{r}\,+\,\frac{\sqrt{1-r^2}}{r}\ ,
    \\
  \label{eq:dn}
    d_{n\geq 1} \: &= \: \frac{i}{\pi r}\oint_{|z|=1}  \, z^{n-1}\,\frac{z^2-2rz+1}{z^2-\frac{2}{r}z+1} \, \de z \: = \:  2 \frac{\sqrt{1-r^2}}{r}\,\bigg(\frac{1}{r}\,-\,\frac{\sqrt{1-r^2}}{r}\bigg)^n \, .
\end{align}
\end{subequations}
In the case of $d_0$, we should take care of an additional pole at $z=0$. As far as the calculation of~$d_{n\ge 1}$ is concerned, we observe that despite the different analytic forms of their integrals, the Fourier coefficients of the even series are identical to those found in the odd series, i.e.~$d_n = c_n$.

It is instructive to verify that for the Fourier coefficients~$d_n$ given in~\eqref{eq:d0} and~\eqref{eq:dn}, setting $\tau=0$ in~\eqref{eq:bexgamma} returns the initial condition, i.e.~$\be (0)\cdot(\mathbf{e}\times\boldsymbol{\gamma}) = 1$, for the reference model under study where $\be (0) = \mathbf{e}\times\boldsymbol{\gamma}$. For the odd Fourier series in~\eqref{eq:bdotgamma}, the initial condition, $\be (0)\cdot \boldsymbol{\gamma} = 0$, is satisfied automatically, so no extra information can be inferred in this case. However, for the even Fourier series in~\eqref{eq:bexgamma}, setting $\tau = 0$ implies the constraint, 
\begin{equation}
    \sum_{n=0}^\infty \, d_n \: = \: d_0 - 2 \frac{\sqrt{1-r^2}}{r}\frac{d_0}{1+d_0} \: = \: 1 \, .
\end{equation}
Solving this last constraint for $d_0$ reproduces its expression in~\eqref{eq:d0}, as it should be.

\begin{figure}[t!]
    \centering
    \begin{subfigure}{0.49\linewidth}
    \centering
    \includegraphics[width=\linewidth]{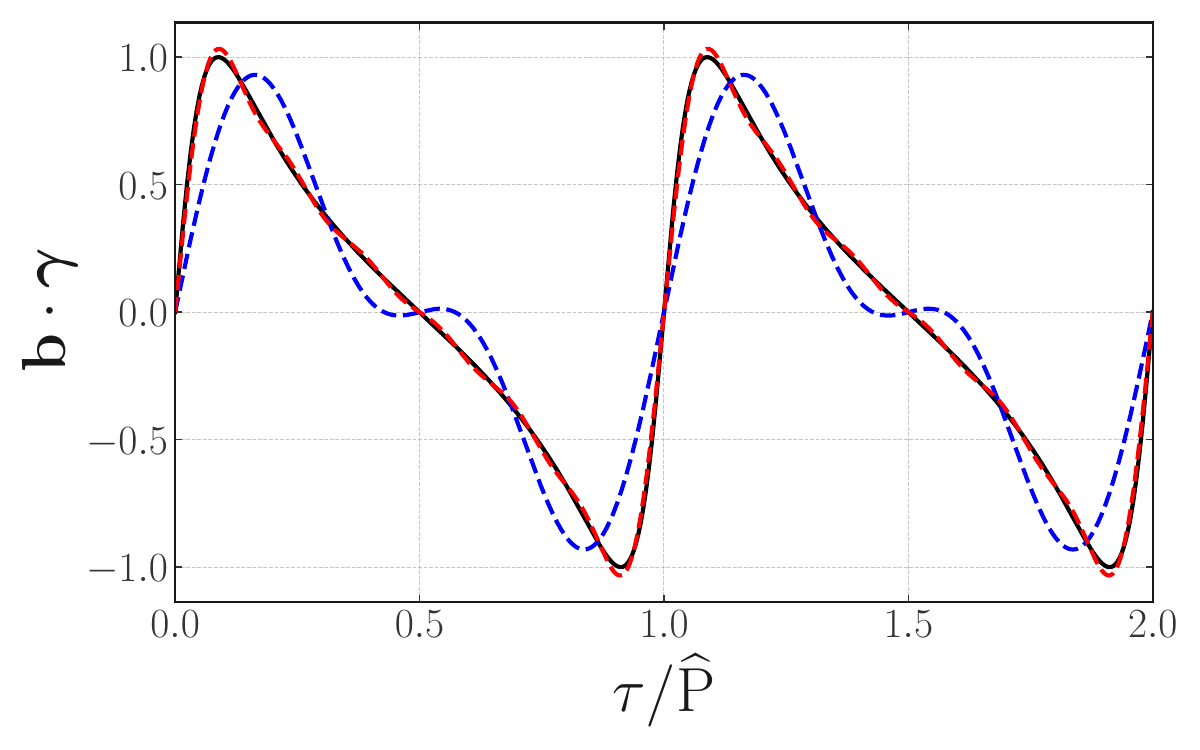}
    \caption{\empty}
    \label{fig:FourierE}
    \end{subfigure}
    \begin{subfigure}{0.49\linewidth}
    \centering
    \includegraphics[width=\linewidth]{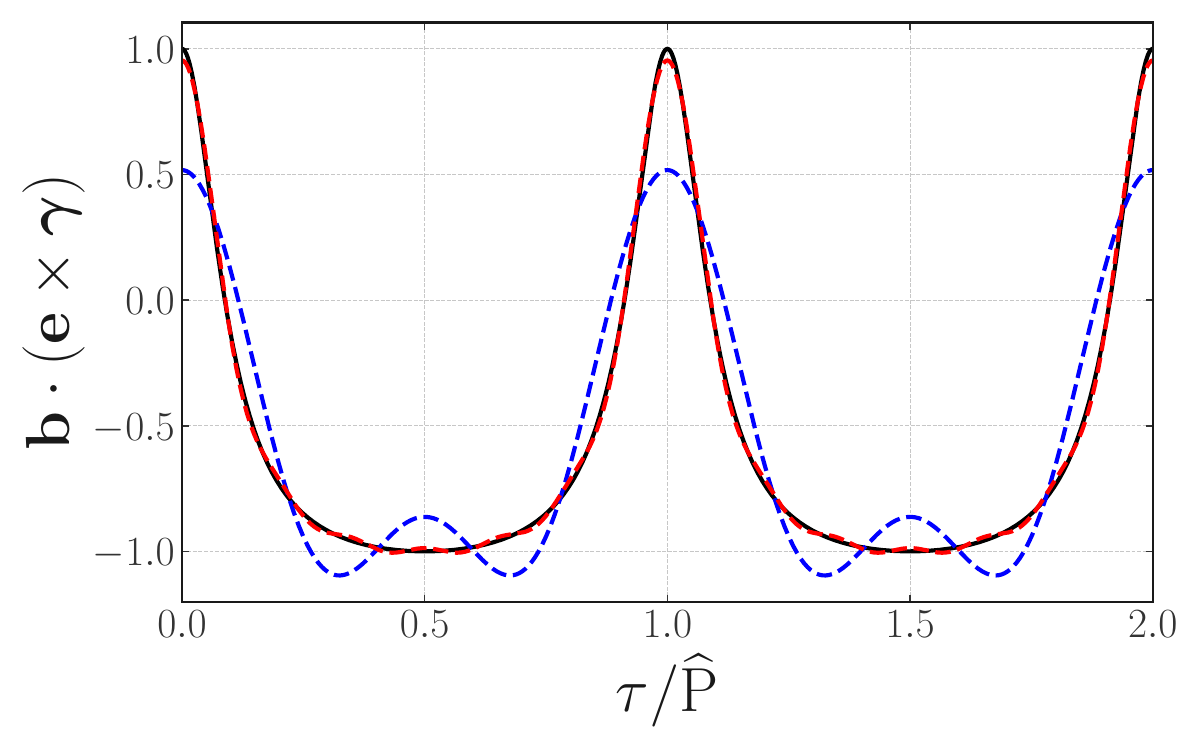}
    \caption{\empty}
    \label{fig:FourierEGam}
    \end{subfigure}
    \caption{Fourier series approximations for a CUQ with $r=0.85$ and $\be(0)=\mathbf{e}\times\boldsymbol{\gamma}$. The black line is the exact evolution of ${\bf b}(\tau)$, the blue dashed line is the Fourier series including terms up to and including $n=2$, and the red dashed line is the Fourier series including terms up to and including $n=6$. The left panel ($a$) shows the projection onto $\boldsymbol{\gamma}$, and the right panel ($b$) shows the projection onto $\mathbf{e}\times\boldsymbol{\gamma}$.}
    \label{fig:Fourier_Proj}
\end{figure}

In Figure~\ref{fig:Fourier_Proj}, we show only the projections of ${\bf b}(\tau )$ along the directions $\boldsymbol{\gamma}$ and $\mathbf{e}\times\boldsymbol{\gamma}$. According to our reference model, we~assume the initial condition $\vecb(0) = \mathbf{e}\times\boldsymbol{\gamma}$ and also take $r=0.85$. The black lines show the exact evolution of CUQ oscillations, the blue dashed lines show the Fourier series up to and including the $n=2$ Fourier coefficient, and the red dashed lines show the Fourier series up to and including the $n=6$ Fourier coefficient. As shown in~\cite{Karamitros:2022oew}, CUQs exhibit an inhomogeneous rotation of ${\bf b}(\tau )$, which slows down and freezes as $r$ approaches~1. Moreover, in Figure~\ref{fig:Fourier_Proj}, we observe that the oscillation profile of ${\bf b}(\tau )$ strongly depends on the particular axis onto which $\vecb (\tau )$ is projected, as could happen in realistic experiments. In~particular, we see that projections along $\boldsymbol{\gamma}$ have a triangular wave form, whereas the projection along $\mathbf{e}\times\boldsymbol{\gamma}$ features sharp peaks and wide troughs. Nonetheless, we have verified that the two different oscillation profiles displayed in the two panels of Figure~\ref{fig:Fourier_Proj} preserve the condition expected from a pure qubit, $|\vecb(\tau)|=1$, throughout their evolution.

With the aid of the analytic results derived in this section, we can approximate the odd and even Fourier series projections of~$\be (\tau)$ given in~\eqref{eq:bdotgamma} and~\eqref{eq:bexgamma} up to any given order $N$, by including all Fourier modes $n \leq N$. In this respect, it is worth noting that the Fourier coefficients, $c_n$ and $d_n$, follow a sequence of geometric progression. Based on this property, we~define the {\em anharmonicity factors},
\begin{equation}
    \mathcal{C}_n \: \equiv \: \frac{c_{n+1}}{c_n}\, , \qquad  \mathcal{D}_{n-1} \: \equiv \: \frac{d_{n}}{d_{n-1}} \, ,
\end{equation}
with $n\geq 1$. These factors are independent of~$n$, and as such, they can be regarded as physical observables, as their measurements allow us to determine the key model parameter~$r$ from oscillation data.
For instance,  for the odd Fourier series given in~\eqref{eq:bdotgamma}, we find 
\begin{equation}
    r \: = \: \frac{2\, \mathcal{C}_n}{\mathcal{C}_n^2+1}\: \overset{\mathcal{C}_n\ll1}{\approx}\: 2\, \mathcal{C}_n\, +\, \mathcal{O}(\mathcal{C}_n^{3})\, ,
\end{equation}
for all $n\geq 1$. By analogy, for the even Fourier series in~\eqref{eq:bexgamma}, we may deduce, independently of~$n$, the value of $r$ from the the following relations:
\begin{subequations}
\begin{align}
     \label{eq:Fourier_Ratios}
    r \: &= \: \frac{1}{\sqrt{1+\frac{1}{4}\mathcal{D}_0^2}} \: \overset{\mathcal{D}_0\gg 1}{\approx} \: \frac{2}{\mathcal{D}_0} + \mathcal{O}(\mathcal{D}_0^{-3})\, ,\\ r \: &= \: \frac{2\, \mathcal{D}_n}{\mathcal{D}_n^2+1} \: \overset{\mathcal{D}_n\ll1}{\approx} \: 2\, \mathcal{D}_n + \mathcal{O}(\mathcal{D}_n^{3})\,, \qquad \mbox{for $n\geq 1$.}
\end{align}
\end{subequations}
A significant benefit of estimating $r$ by taking ratios of the Fourier coefficients is that some of the amplitude dependence drops out. However, there are some non-trivial aspects to be considered. For pure state oscillations where the projection of ${\bf b}(\tau)$ has an amplitude $R<1$, the actual value of $r$ extracted from the anharmonicity factors will only be an effective value, which is denoted as $\tilde{r}$ in Appendix~\ref{App:Geometry}. In order to map the latter onto the true value of $r$, we employ the relation,
\begin{equation}\label{eq:r_effective}
    r \: = \: \frac{\tilde{r}}{\sqrt{R^2 + \tilde{r}^2 \lrb{1-R^2}}}\ . 
\end{equation}
Note that for $\tilde{r}\ll 1$, the value $r$ is obtained by the simple rescaling: $r= \tilde{r}/R\: +\: \mathcal{O}(\tilde{r}^3)$.  

The origin of the correction to $r$ given in~\eqref{eq:r_effective} is rooted in the geometry of general CUQs. It~is the fact that $\be (\tau )$ is not really oscillating about the energy vector, $\mathbf{e}$, but truly about a linear combination of $\mathbf{e}$ and $\boldsymbol{\gamma}\times\be(0)$ [cf.~\eqref{eq:Plane_Norm}]. Such a generalisation is discussed in detail in Appendix~\ref{App:Geometry}, and the relevant formulae are derived by utilising the so-called Frenet-Serret~frame.  An outline of the adjustments necessary to map the general CUQ onto our simpler reference CUQ model are also presented in Appendix~\ref{App:Geometry}.

%\vfill\eject

\subsection{Accuracy of the Fourier Series}

\begin{figure}
    \centering
    \begin{subfigure}{0.49\linewidth}
    \centering
    \includegraphics[width=\linewidth]{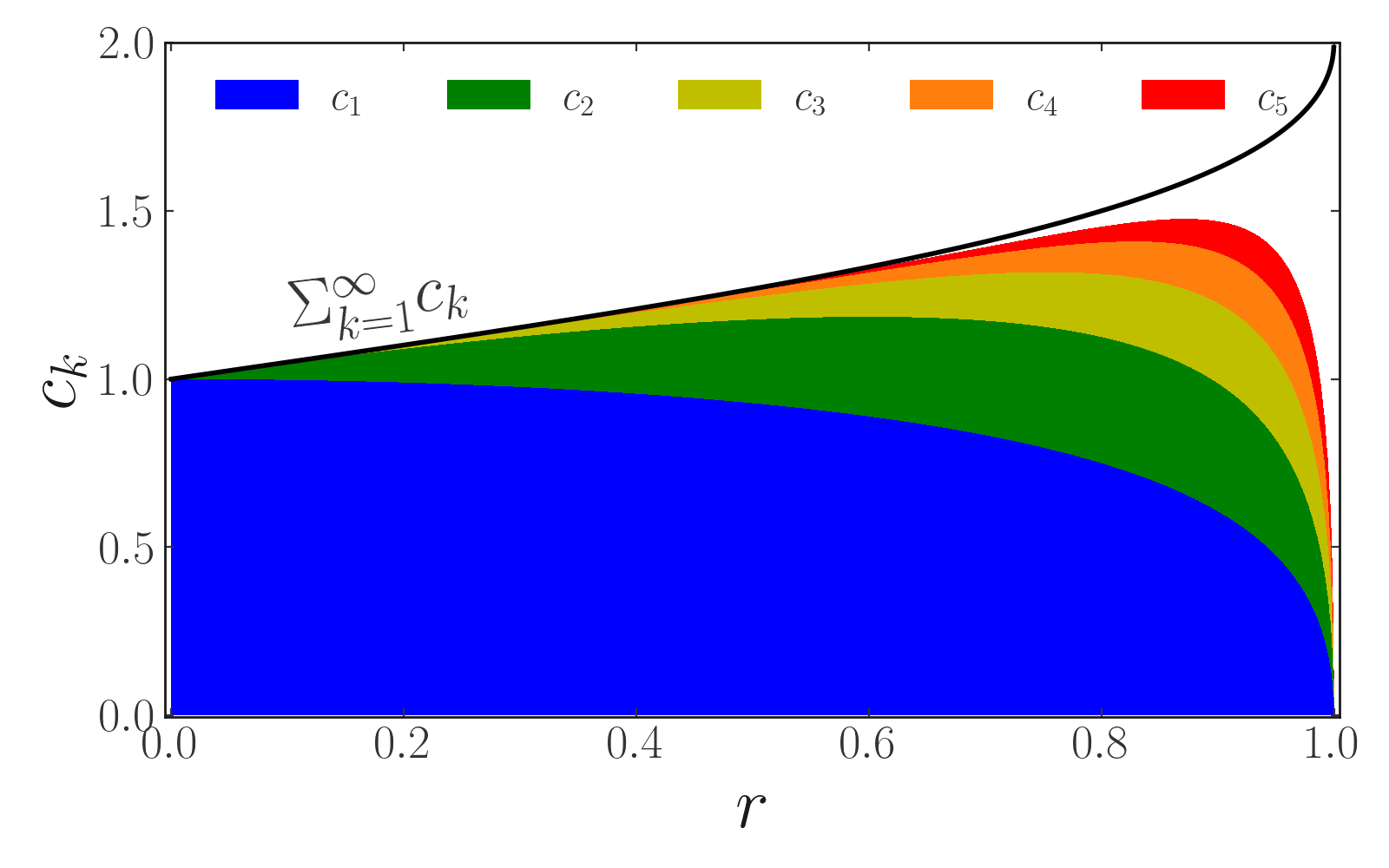}
    \caption{\empty}
    \label{fig:FourierCoeffs}
    \end{subfigure}
    \begin{subfigure}{0.49\linewidth}
    \centering
    \includegraphics[width=\linewidth]{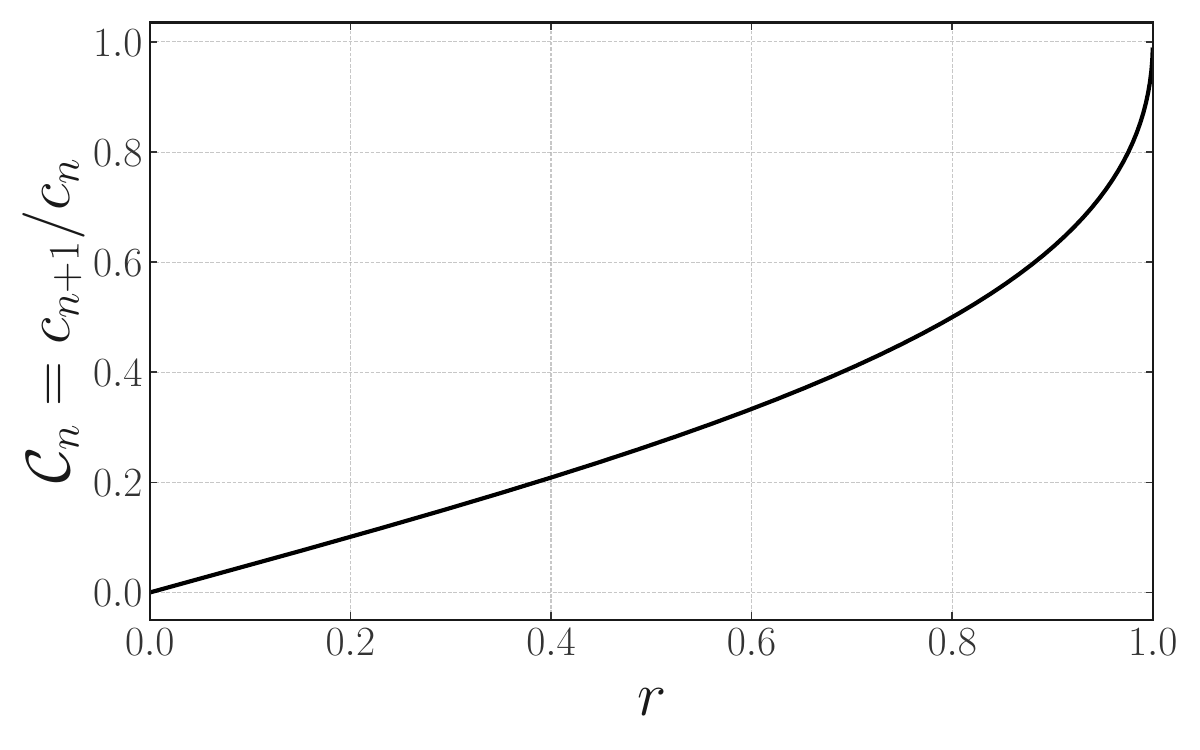}
    \caption{\empty}
    \label{fig:FourierRatio}
    \end{subfigure}
    \caption{Estimation of the dominant contributions to the CUQ Fourier series. The left panel ($a$) shows the cumulative total of the leading Fourier coefficients, with the black line giving the infinite sum of Fourier coefficients, $c_n$. The right panel ($b$) shows the anharmonicity factor $\mathcal{C}_n$, as a function of $r$.}
    \label{fig:coefficients}
\end{figure}

We may achieve a high degree of accuracy in the evolution of the co-decaying Bloch vector, ${\bf b}(\tau )$, if its Fourier series representation is truncated {\em above} a sufficiently high Fourier mode~$N$. As~we will see in this subsection, as few as $N=5$ Fourier modes suffice to match signals that deviate significantly from the principal harmonic~${n=1}$. Of course, for values of $r\ll1$, we may only need the leading terms to match the signal accurately. However, the inclusion of additional modes may be used, via the anharmonicity factors $\mathcal{C}_n$ and $\mathcal{D}_n$, to obtain independent estimations for the key model parameter,~$r$.

In Figure~\ref{fig:coefficients} we show the importance of Fourier coefficients, both relative to one another, and relative to the infinite sum over all coefficients. Figure~\ref{fig:FourierCoeffs} shows the relative size of each additional Fourier coefficient on the total Fourier series, as a function of $r$, by displaying the running total of Fourier coefficients. Additionally, the black line indicates the sum over all Fourier coefficients. For values of $r$ close to zero, we see that the principal harmonic carries the bulk information of the oscillation profile. However, as $r$ increases, higher harmonics begin to have non-negligible contributions to the oscillation and additional Fourier modes may be required to fully represent the anharmonicities present in the oscillation. Moreover, we see that for $r\ll 1$, the lowest order Fourier modes dominate over higher modes, and consequently, one expects a good fit with few Fourier modes. 

In the limit $r\to 1$, Figure~\ref{fig:coefficients} shows a divergence between the black line and the first five harmonics, as each individual Fourier coefficient shrinks in size and becomes of similar magnitude. In this extreme region, it becomes necessary to include even higher Fourier modes to match the true oscillation profile. This is similarly seen in Figure~\ref{fig:FourierRatio}, where we show the size of the anharmonicity factor $\mathcal{C}_n$. For $r\ll1$, it is clear that the anharmonicity factors approach zero, indicating that few Fourier coefficients will be necessary for an accurate representation of the oscillation signal. In contrast, for $r\to 1$, many Fourier modes are necessary, since each new mode will be of similar magnitude. Hence, higher Fourier modes contribute similarly to the expected leading modes, and one needs to fit a large number of Fourier coefficients to achieve a high degree of precision.

\begin{figure}[t!]
    \centering
    \begin{subfigure}{0.49\linewidth}
    \centering
    \includegraphics[width=\linewidth]{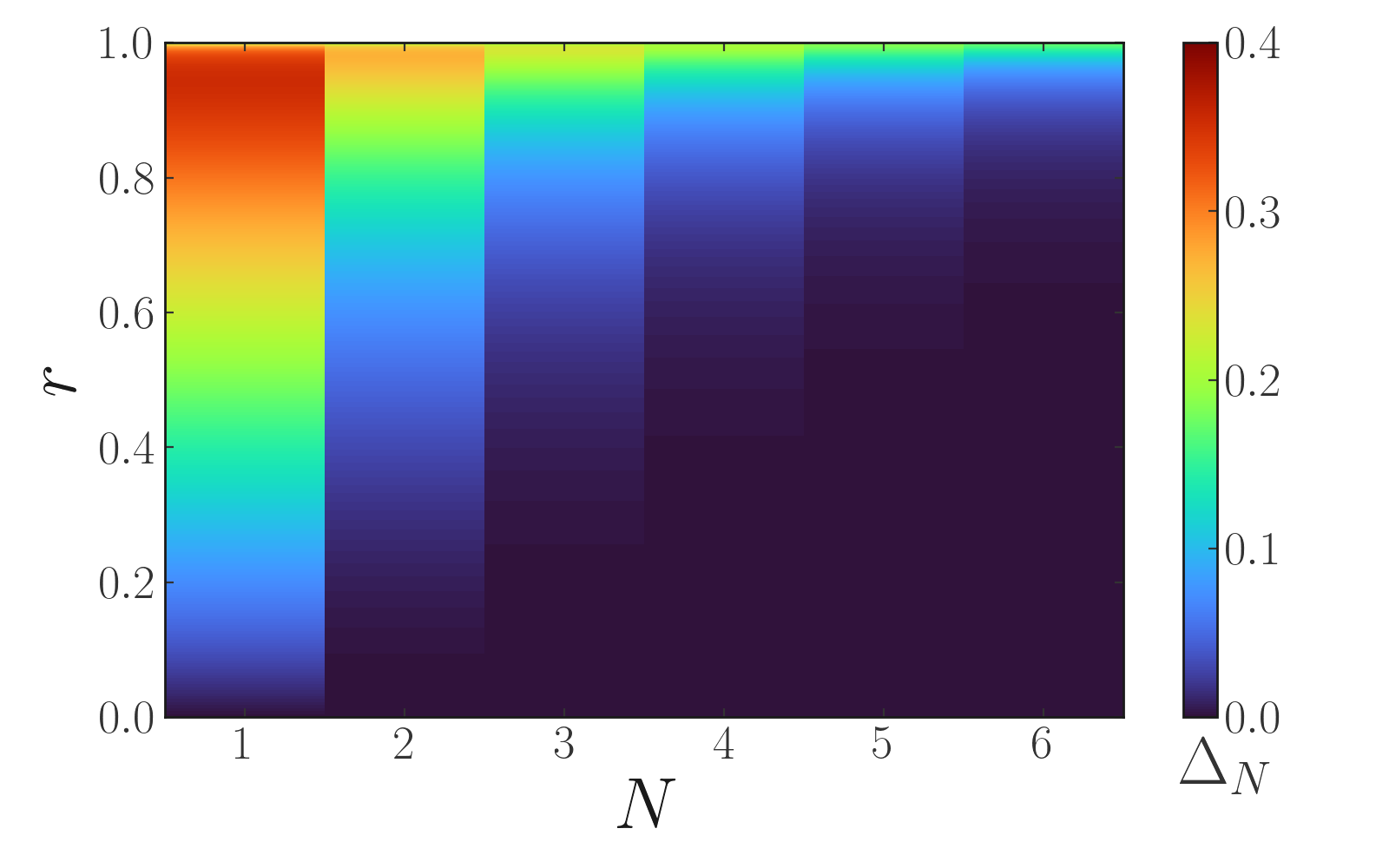}
    \caption{\empty}
    \label{fig:FourierErrorMap}
    \end{subfigure}
    \begin{subfigure}{0.49\linewidth}
    \centering
    \includegraphics[width=\linewidth]{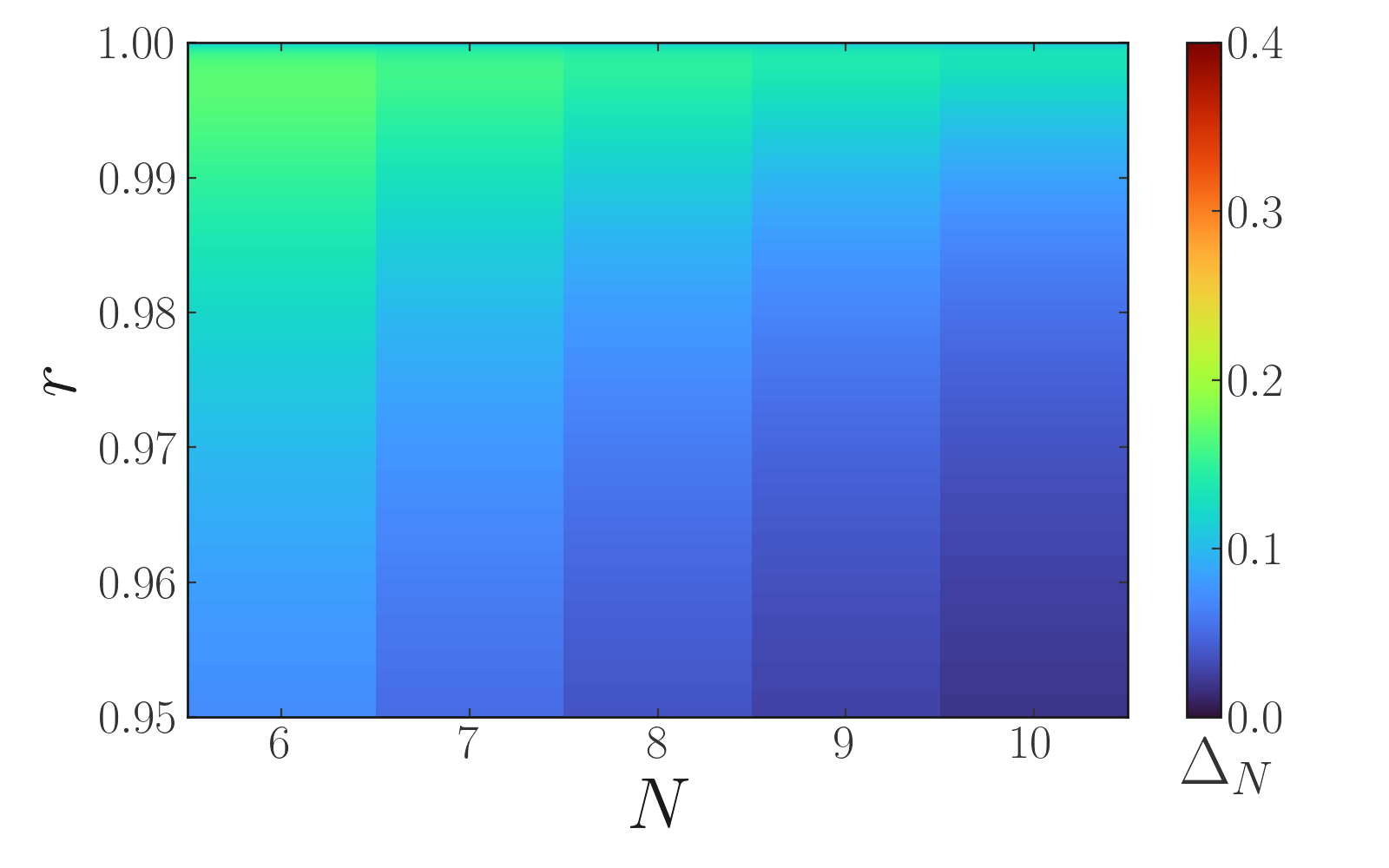}
    \caption{\empty}
    \label{fig:FourierErrorMapTight}
    \end{subfigure}
    \caption{Estimation of the root-mean-squared error, $\Delta_N$, of the $N$-term Fourier series around the exact oscillation profile. The left panel ($a$) shows the leading order Fourier series, $N\leq 6$ for $0<r<1$, and the right panel ($b$) focuses in on the higher order contributions $6 \leq N \leq 10$ for values of $r$ close to unity where we expect to require additional Fourier modes to achieve high precision.}
    \label{fig:Fourier_figs}
\end{figure}

We would like to assess how well the Fourier series can reliably reproduce the oscillation profile of a~CUQ. Typically, when dealing with experimental datasets, these evaluations can be performed by calculating the value of $\chi^2$ statistics. However, arbitrarily chosen partitions of the oscillation period would lead to inaccurate and misleading estimations of $\chi^2$. 
This problem can be avoided by calculating the variance of the $N^{\text{th}}$-mode truncated Fourier series, which we denote as~$\Delta^2_N$, around the expected signal by integrating over the full oscillation period.  Equivalently, we may compute $\Delta^2_N$ as an integral of $\phi=\widehat{\omega}\tau$ over $\lrb{-\pi,\pi}$,
\begin{equation}
    \Delta_N^2 \: = \: \frac{1}{2\pi} \int_{-\pi}^{\pi} \lrsb{\sum_{k=1}^N c_k \sin\lrb{k\phi} - \lrb{-\sin\theta(\phi)}}^2 \: {\rm d}\phi \, .
\end{equation}
We note that since the exact analytic oscillation profile and its Fourier series representation are both odd functions, the average of these two expressions vanishes individually, when they are integrated over the full period. This result indicates that the Fourier series gives an unbiased estimation regardless of the number of harmonics, $N$, included. Careful evaluation of the integral gives an expression for $\Delta_N^2$ which is solely dependent on the model parameter~$r$. Taking into account the geometric progression form of the Fourier coefficients $c_n$ in~\eqref{eq:cn}, we may evaluate the variance~$\Delta_N^2$ as follows:
\begin{equation}\label{eq:FSVar}
    \Delta_N^2 \: = \: \frac{1-r^2}{\sqrt{1-r^2} - 1 + r^2} - 2\frac{1-r^2}{r^2} - \frac{1}{2}\sum_{k=1}^N c_k^2 \ = \ \frac{r^2}{4}\frac{c_{N+1}^2(r)}{\sqrt{1-r^2} - 1 + r^2}\ .
\end{equation}
From the very last expression, we may verify that in the limit $N\to \infty$, the variance vanishes due to the convergence of the Fourier series, $\Delta_N^2 \to 0$, indicating perfect matching between the $N^{\text{th}}$-mode truncated Fourier series  and the exact oscillation profile. Similarly, in the proximity to 
the point $r = 0$, we have that $\Delta_N^2 \simeq c_{N+1}^2/2$. In this small $r$ region, each of the Fourier coefficients with $N\geq 2$ is vanishingly small, and so $\Delta_N^2 \to 0$ for all $N$ as $r\to 0$. On the other hand, we notice that the denominator of $\Delta_N^2$, given in~\eqref{eq:FSVar}, leads to a divergence as $r \to 1$, signifying a reduction in the accuracy of the Fourier series. To improve the accuracy in this case,
additional Fourier modes must be included to properly capture the exact oscillation profile. Given the analytic expressions for $c_{N+1}$ and $\Delta_N$ that can be inferred from~\eqref{eq:cn} and~\eqref{eq:FSVar}, respectively, it~is straightforward to estimate the number of Fourier modes, $N$, needed to achieve the desirable precision for a given initial estimate of~$r$.

Figure~\ref{fig:FourierErrorMap} shows the value of the root-mean-squared deviation, i.e.~$\Delta_N$, for $0<r<1$ and a $N^{\text{th}}$-mode truncated Fourier series, with $1\leq N \leq 6$. Blue hues indicate parametric regions where $\Delta_N$ is small, displaying good agreement between the Fourier series and the oscillation profile. Red hues indicate where the deviation between the Fourier series and oscillation is at its greatest. Colour bars are provided as a reference for intermediate shades. Figure~\ref{fig:FourierErrorMap} shows~$\Delta_N$ for the leading contributions to the approximate Fourier series truncated at $N \leq 6$, for $0<r<1$. In this region, we see that when $r$ is small, there is a high degree of matching between the approximate Fourier series and the exact analytic profile, even with a relatively low number of Fourier modes included. However, for $r$ values close to 1, as expected, we find that the root-mean-squared deviation is large unless significantly more Fourier modes are included. Figure~\ref{fig:FourierErrorMapTight} focuses on the narrow interval, $0.95 \leq r < 1$, and shows $\Delta_N$ for $6\leq N \leq 10$. Finally, we find that the closer $r$ is to unity, the greater the deviation is, and hence more Fourier modes are needed to accurately represent the full profile in such cases.

%\vfill\eject

%%%

\section{Application to {\boldmath $B^0\bar{B}^0$}-Meson Oscillations} \label{sec:ApplBmeson}
\setcounter{equation}{0}

In this section, we apply the approximate Fourier series approach developed in the previous section to studying anharmonicities in the oscillation profile of a meson--antimeson system. For~definiteness, we will focus on the $B^0\bar{B}^0$-meson system. 

\subsection{Bloch Sphere Parameters for the {\em B}--Meson System}

To implement our approach more efficiently, we must first translate the relevant $B$-meson observables, which are derived from the effective Hamiltonian and were extensively discussed in the literature~\cite{Kabir:1989dd,Grimus:1988,Paschos:1989ur,Buras:1990fn,Neubert:1993mb,Nierste:2004uz}, into the Bloch sphere parameters introduced in Subsection~\ref{sec:2Level}. In terms of these parameters, the mass and width differences of a $B$-meson system are given by
\begin{equation}
   \label{eq:DEDGamma}
    \Delta E \: = \: 2|\mathbf{E}| \Re {\sqrt{1-r^2 -2ir\cos\theta_{\mathbf{e}\boldsymbol{\gamma}}}} \, , \qquad \Delta \Gamma \: = \: -4|\mathbf{E}| \Im{\sqrt{1-r^2 -2ir\cos\theta_{\mathbf{e}\boldsymbol{\gamma}}}} \, .
\end{equation}
Including the CP-violation measure, $|q/p|$, allows one to uniquely determine the oscillation parameters\- of the effective Hamiltonian~${\rm H}_{\rm eff}$ by the relation~\cite{Karamitros:2022oew},
\begin{equation}
   \label{eq:qoverp4}
    \left| \frac{q}{p} \right|^4 \: = \: \frac{1+r^2 - 2r \sin \theta_{\mathbf{e}\boldsymbol{\gamma}}}{1+r^2 + 2r \sin \theta_{\mathbf{e}\boldsymbol{\gamma}}} \ .
\end{equation}
From~\eqref{eq:qoverp4}, it is then easy to see that when either $r = 0$ or $\sin \theta_{\mathbf{e}\boldsymbol{\gamma}} = 0$, one has $|q/p| =1$, so~CP violation through mixing of states will vanish. In this case, ${\rm H}_{\rm eff}$ may be diagonalised by means of unitary transformations, and dispersive and dissipative effects become distinct from one another. In contrast, for a CUQ, for which~$\cos\theta_{\mathbf{e}\boldsymbol{\gamma}}=0$ and~$r<1$, CP violation induced by the mixing of the two quantum states, which manifests itself in a non-zero value of $|(|q/p| -1)|$, is maximised at any given $r$. Moreover, for a CUQ, the expressions given in~\eqref{eq:DEDGamma} and~\eqref{eq:qoverp4} simplify to
\begin{equation}
    \Delta E \: = \: 2|\mathbf{E}|\sqrt{1-r^2} \, , \qquad \Delta \Gamma \: = \: 0 \, , \qquad \left| \frac{q}{p} \right| \: = \: \sqrt{\frac{1 \pm r}{1 \mp r}} \; ,
\end{equation}
where the sign of $\theta_{\mathbf{e}\boldsymbol{\gamma}}$ is determined by the signs contained within the expression for $|q/p|$. 

\begin{table}[t!]
    \centering
    \begin{tabular}{|c||c|c|c|}
        \hline
        {} & $\Delta E \, [{\rm ps}^{-1}]$ &  $\Delta\Gamma \, [{\rm ps}^{-1}]$  & $|q/p|-1$\\[0.1cm]
        \hline\hline
        Experiment&  $0.5069 \pm 0.0019$ & $(0.7 \pm 7)\times 10^{-3}$ & $ (1.0\pm 0.8) \times 10^{-3} $\\
        \hline
        Theory &  $0.535\pm 0.021$ & $\lrb{2.7 \pm 0.4}\times 10^{-3}$ & $(2.6 \pm 0.3) \times 10^{-4}$\\[0.1cm]
        \hline\hline
        {} & $r$ &  $\theta_{\mathbf{e}\boldsymbol{\gamma}} [^\circ]$  & $|\mathbf{E}| [\textrm{ps}^{-1}]$\\[0.1cm]
        \hline\hline
        Experiment&  $(1\pm 4) \times 10^{-3}$ & $-90 \pm 90$ & $ 0.253 \pm 0.001 $\\
        \hline
        Theory &  $(2.5 \pm 0.4) \times 10^{-3}$ & $-5 \pm 3$ & $0.28 \pm 0.01$\\[0.1cm]
        \hline
    \end{tabular}
    \caption{\em 
    The experimental values for the various CP-violating parameters are taken from~\cite{HeavyFlavorAveragingGroupHFLAV:2024ctg} and the most recent theoretical SM predictions from are extracted from~\cite{Albrecht:2024oyn}.
    Also shown are the model parameters, $r$, $\theta_{\mathbf{e}\boldsymbol{\gamma}}$ and~$|\mathbf{E}|$, as deduced from the $B^0$-meson oscillation data, along with their SM predictions.}
    \label{tab:MesonData}
\end{table}

It proves more convenient for our subsequent analysis to invert the above relations and express the Bloch sphere parameters, $r$, $\theta_{\mathbf{e}\boldsymbol{\gamma}}$ and $|{\bf E}|$, as functions of $\Delta E$, $\Delta\Gamma$ and $|q/p|$, for all CUQ scenarios with $r<1$. To do so, we~introduce the two dimensionless shorthand quantities,
\begin{equation}
    \delta_M \: = \: \frac{\left| q/p \right|^4 - 1}{\left|q/p \right|^4 + 1} \; , \qquad \eta \: = \: \frac{(\Delta E)\lrb{\frac{1}{2}\Delta \Gamma}}{(\Delta E)^2 - \lrb{\frac{1}{2}\Delta \Gamma}^2} \; .
\end{equation}
In terms of these quantities, the value of $r$ is found to be
\begin{equation}
    r \: = \: \frac{\sqrt{1+4\eta^2} - \sqrt{1-\delta^2_M\phantom{|}\!}}{\sqrt{4\eta^2 + \delta^2_M}} \ .
\end{equation}
Knowing $r$, we may determine the remaining two Bloch-sphere parameters,  
\begin{equation}
   \label{eq:Etheta}
    |\mathbf{E}|^2 \: = \: \frac{(\Delta E)^2 - \lrb{\frac{1}{2}\Delta \Gamma}^2}{4\,(1-r^2)} \;, \qquad \sin \theta_{\mathbf{e}\boldsymbol{\gamma}} \: = \: -\,\frac{1+r^2}{2r}\delta_M \, .
\end{equation}
The limiting scenario of a CUQ corresponds to vanishing width difference, i.e.~$\Delta \Gamma \to 0$, which implies $\eta \to 0$. Since $|\sin \theta_{\mathbf{e}\boldsymbol{\gamma}}| = 1$ for a CUQ, the expression for $\sin\theta_{\mathbf{e}\boldsymbol{\gamma}}$ in~\eqref{eq:Etheta}
now becomes a constraint which leads to the following self-consistent relations:
\begin{equation}
    r \: = \: \frac{1}{|\delta_M|} - \frac{1}{|\delta_M|}\sqrt{1-\delta_M^2} \ , \qquad |\mathbf{E}| \: = \: \frac{\Delta E}{2\sqrt{1-r^2}} \ .
\end{equation}
If mixing CP violation vanishes in a CUQ, this means that not only $\delta_M$, but also $r$ is zero. Thus, in a CUQ, CP-violating effects can only be sourced from a non-zero value of $r$. This fact motivates our search for anharmonicities in $B$-meson oscillations as a potential signal of CP-violating mixing phenomena.

In Table~\ref{tab:MesonData} we show the central values and bounds at the $1\,\sigma$ Confidence Level (CL) for the frequently quoted quantities $\Delta E$, $\Delta\Gamma$ and $|q/p| -1$, as well as the respective values for the relevant Bloch sphere parameters $r$, $\theta_{\mathbf{e}\boldsymbol{\gamma}}$ and $|{\bf E}|$, as these are estimated from both experi\-mental and theoretical analyses. For definiteness, the experimental values are taken from~\cite{HeavyFlavorAveragingGroupHFLAV:2024ctg} and the most recent theoretical predictions in the Standard Model (SM) can be found in~\cite{Albrecht:2024oyn}. In the SM, the $B^0\bar{B}^0$ system is predicted to be far away from a CUQ realisation. Nevertheless, physics beyond the SM could be responsible for generating the CUQ dynamics in the $B^0_{\rm d}$-meson system.

\subsection{CP-Violating Observables}

To elucidate our Fourier series approach to analysing anharmonicities in CUQ oscillations, we will now provide a concrete and practical study in the $B$-meson system. 
Our~approach\- can be used to complement and refine existing and upcoming analyses by experimental collaborations such as CDF~\cite{CDF:1996kan}, BaBar~\cite{BaBar:2013xng}, and LHCb~\cite{LHCb:2012mhu,LHCb:2016gsk}.

In general, the oscillation of neutral mesons, such as the $K^0\bar{K}^0$- and $B^0\bar{B}^0$-meson systems, is theoretically quantified through some observable of flavour asymmetry~\cite{Kabir:1989dd}, e.g.
\begin{equation}
    \delta(t) \: = \: \frac{N_s(t) - \overline{N}_s(t)}{N_s(t) + \overline{N}_s(t)} 
    \, ,
    \label{eq:delta_def}
\end{equation}
where $N_s$ is the number of particles of the signal in the initial state, and $\overline{N}_s$ the number of particles in the CP-conjugate state. 
This result can be expressed in terms of the co-decaying Bloch vector as follows:
\begin{equation}
    \delta(t)  \: = \: \frac{\mathbb{P}({\rm P}^0) - \mathbb{P}(\bar{\rm P}^0)}{\mathbb{P}({\rm P}^0) + \mathbb{P}(\bar{\rm P}^0)}  \: = \: \frac{\Tr{\sigma^3 \rho(t)}}{{\rm Tr}\, \rho(t)} \: = \: {\rm b}_3(t) \, ,
    \label{eq:delta_b3}
\end{equation}
with ${\rm P}^0 = K^0,\, B^0, D^0$.
In the CP basis, assuming CPT invariance of the effective Hamiltonian leads to the constraints~\cite{Grimus:1988,Kabir_1996,
ParticleDataGroup:2024cfk}
\begin{equation}
    {\rm E}_{11} \: = \: {\rm E}_{22}\,,\qquad 
    \Gamma_{11} \: = \: \Gamma_{22}\, ,    
\end{equation}
which implies ${\rm E}_3=\Gamma_3=0$. For CUQs (where $\mathbf{e}\cdot\boldsymbol{\gamma}=0$), the last constraint is translated to  
\begin{equation}
    \mathbf{e}\times\boldsymbol{\gamma} \: = \: \begin{pmatrix}
        0,  \, 0, \, 1 
    \end{pmatrix} \, .
\end{equation}
Therefore, with the help of~\eqref{eq:delta_b3}, the flavour asymmetry in~\eqref{eq:delta_def} takes on the simple form,
\begin{equation}
    \delta(t)  \: = \: {\rm b}_3(t) \: = \: \vecb(t) \cdot (\mathbf{e}\times\boldsymbol{\gamma})\, .
    \label{eq:delta_b_proj}
\end{equation}

In realistic experimental settings, background events must also be taken into account, since they can spoil the idealized form of the flavour asymmetry $\delta(t)$. To properly deal with such events, one defines the mixing asymmetry~\cite{LHCb:2016gsk}
\begin{equation}
    A(t) \: \equiv \: \frac{\lrb{N_s(t) + N_b(t)} - \lrb{\overline{N}_s(t) + \overline{N}_b(t)}}{\lrb{N_s(t) + N_b(t)} + \lrb{\overline{N}_s(t) + \overline{N}_b(t)}} \: = \: \frac{\lrb{N_s(t) - \overline{N}_s(t)} + \lrb{N_b(t) - \overline{N}_b(t)}}{\lrb{N_s(t) + \overline{N}_s(t)} + \lrb{N_b(t) + \overline{N}_b(t)}} \, ,
    \label{eq:A_def}
\end{equation}
where $N_b(t)$ is the particle number of the background and $\overline{N}_b(t)$ its CP-conjugate quantity. The background asymmetry is expected to vanish at the leading order~\cite{LHCb:2012mhu}, implying $N_b(t) = \overline{N}_b(t)$. Then, the resulting mixing asymmetry is related to the flavour asymmetry via
\begin{align}
   \label{eq:A_delta}
    A(t) \: &= \: \frac{N_s(t) + \overline{N}_s(t)}{\lrb{N_s(t) + \overline{N}_s(t)} + \lrb{N_b(t) + \overline{N}_b(t)}}\:\frac{N_s(t) - \overline{N}_s(t)}{N_s(t) + \overline{N}_s(t)} \nonumber\\
    &= \: \frac{N_s(t) + \overline{N}_s(t)}{\lrb{N_s(t) + \overline{N}_s(t)} + \lrb{N_b(t) + \overline{N}_b(t)}} \: \delta(t) \, .
\end{align}

Assuming a CUQ realisation for the $B^0\bar{B}^0$-meson system and using~\eqref{eq:delta_b_proj}, the mixing asymmetry~\eqref{eq:A_delta} can be further expressed as
\begin{equation}
     A(t) \: = \: \frac{N_s(t) + \overline{N}_s(t)}{\lrb{N_s(t) + \overline{N}_s(t)} + \lrb{N_b(t) + \overline{N}_b(t)}} \ \vecb(t) \cdot (\mathbf{e}\times\boldsymbol{\gamma}) \;.
    \label{eq:A_b_proj}
\end{equation}
Evidently, in order to study the flavour asymmetry alone, we must divide out the additional time dependence that enters through the changing ratio between the signal and the background. 

Restricting our analysis to the $B^0\bar{B}^0$-meson system,  the total number of particles in the source and background are assumed to decay exponentially according to the world average decay rates~\cite{ParticleDataGroup:2024cfk}:
\begin{equation}
    \Gamma_s \: = \: 1/\tau(B^0_{\rm d}) \:= \: 1/1.519 \: {\rm ps}^{-1} \, , \qquad \Gamma_b \: = \: 1/\tau(B^+) \:= \: 1/1.638 \: {\rm ps}^{-1} \, .
\end{equation}
The initial background-to-signal ratio, $\alpha_{B^+}$, is determined by the experimental setup. In~\cite{LHCb:2016gsk}, oscillation data for years 2011 and 2012 were given for two decay channels of the $B_{\rm d}^0$ system: $B^0_{\rm d} \to D^-\mu^+\nu_\mu X$ and $B^0_{\rm d} \to D^{-*}\mu^+\nu_\mu X$. Moreover, it was reported in~\cite{LHCb:2016gsk} that $\alpha_{B^+} =6\% ~(3\%)$ for the 2011 (2012) data set.

\begin{figure}[t!]
    \centering
    \begin{subfigure}{0.49\linewidth}
    \centering
    \includegraphics[width=\linewidth]{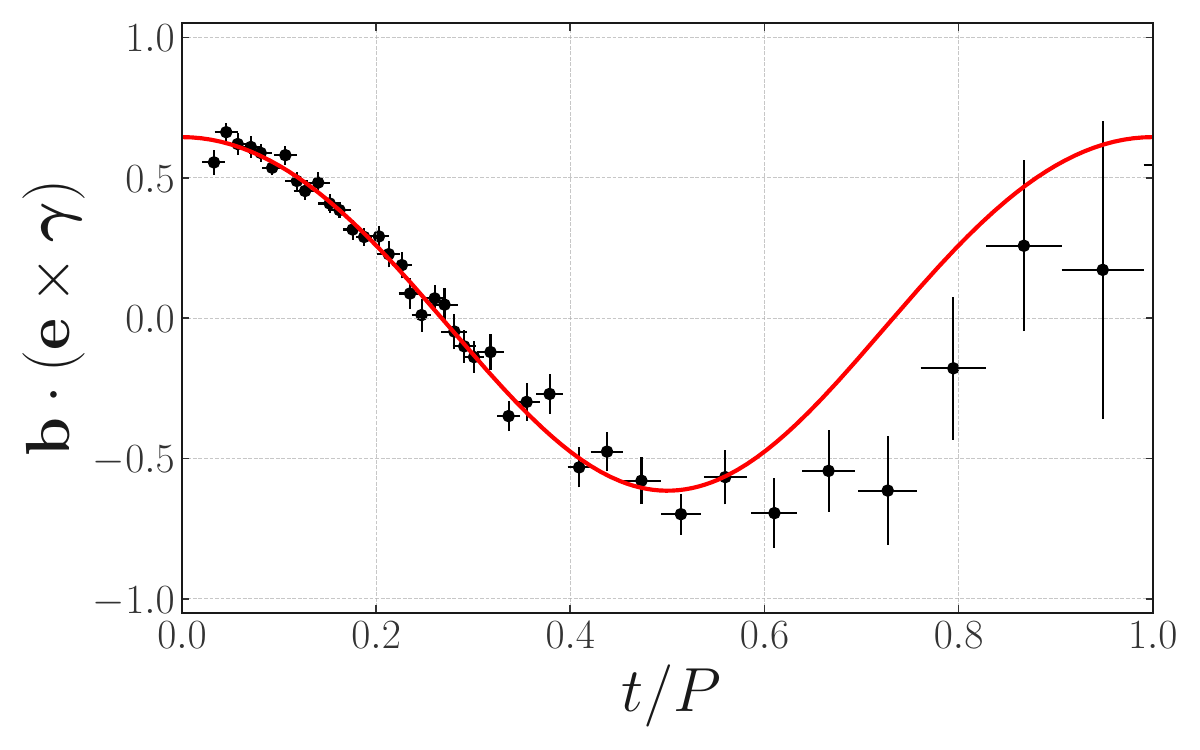}
    \caption{\empty}
    \label{fig:B2D2011}
\end{subfigure}
\begin{subfigure}{0.49\linewidth}
    \centering
    \includegraphics[width=\linewidth]{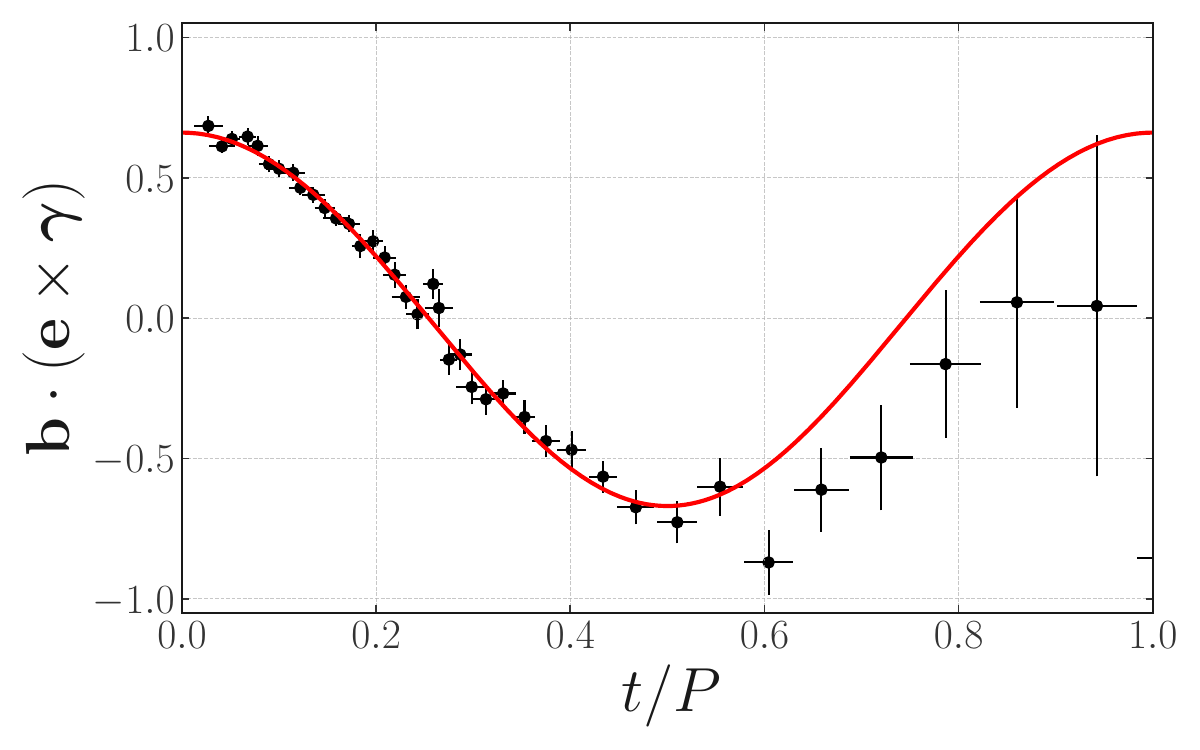}
    \caption{\empty}
    \label{fig:B2DStar2011}
\end{subfigure}
\begin{subfigure}{0.49\linewidth}
    \centering
    \includegraphics[width=\linewidth]{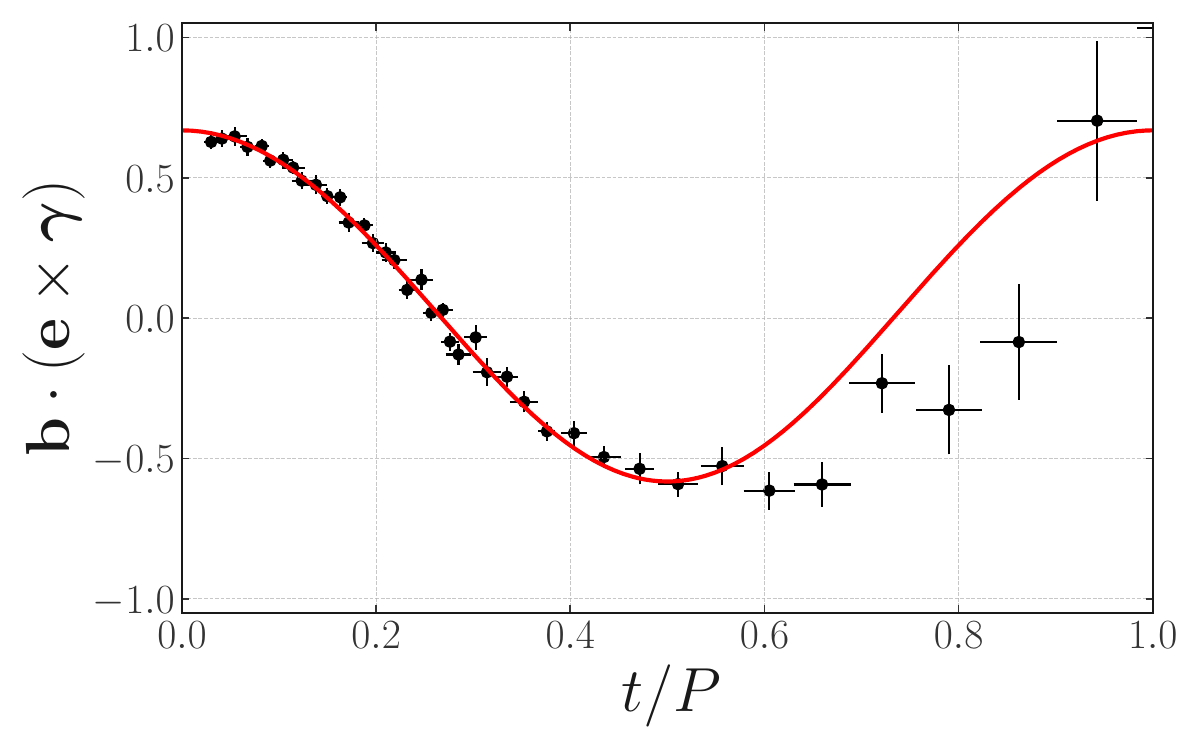}
    \caption{\empty}
    \label{fig:B2D2012}
    \end{subfigure}
    \begin{subfigure}{0.49\linewidth}
    \centering
    \includegraphics[width=\linewidth]{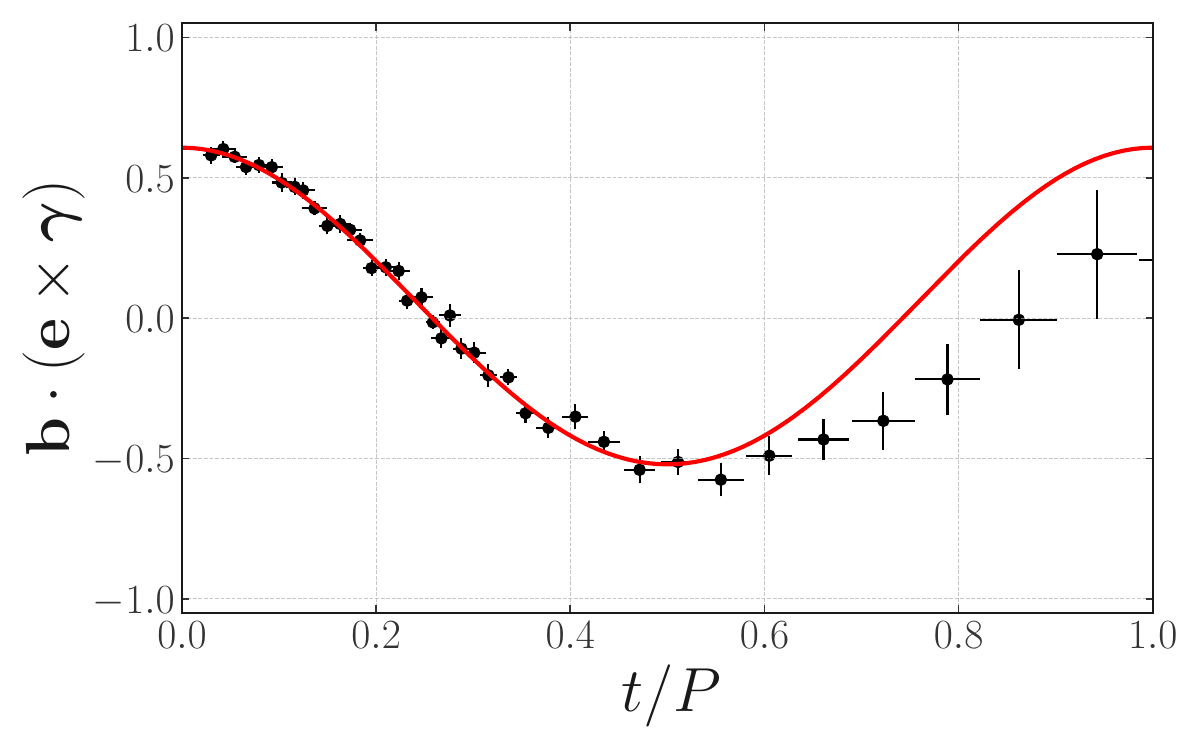}
    \caption{\empty}
    \label{fig:B2DStar2012}
    \end{subfigure}
    \caption{Fourier series fit for the data given in~\cite{LHCb:2016gsk}, for the year 2011: panels ($a$) and ($b$), and the year 2012: panels ($c$) and ($d$). Panels ($a$) and ($c$) show the flavour asymmetry for the channel $B^0_{\rm d}\to D^-\mu^+\nu_\mu X$ and panels ($b$) and ($d$) for the channel $B^0_{\rm d}\to D^{-*}\mu^+\nu_\mu X$. The black data points and their associated error bars have been rewritten from their original form into projections of  along the $\mathbf{e}\times\boldsymbol{\gamma}$ direction. For this fit, we make use of the first two harmonics of the Fourier series, as well as the constant term.}
    \label{fig:Fourier_B_meson}
\end{figure}

In addition, one has to account for the acceptance of the signal in the detector. As mentioned in~\cite{LHCb:2012mhu}, this acceptance is modelled through the inclusion of an {\em empirical efficiency function}, 
\begin{equation}
    \epsilon\lrb{t \left| a_1, a_2  \right.} \: = \: \tan^{-1} \lrb{a_1 e^{a_2 t}} \, ,
\end{equation}
which depends on two unknown parameters, $a_1$ and $a_2$, to be determined below from the data.
With these additional considerations, one can extract the relevant projection of ${\bf b}(t)$ in terms of the mixing asymmetry:
\begin{equation}\label{eq:AsymInv}
    \vecb(t)\cdot\lrb{\mathbf{e}\times\boldsymbol{\gamma}} \: = \: \bigg\{ 1 + \frac{\alpha_{B^+}}{\epsilon(t|a_1,a_2)}\exp\Big[\!-\big(\Gamma_b - \Gamma_s\big)t\Big]\bigg\} \, A(t) \, .
\end{equation}
In the above, the values of $a_1$ and $a_2$ were determined through a $\chi^2$-minimising fit of the experimental data, which yielded
\begin{equation}
    a_1 \: = \: 0.321 \, , \qquad a_2 \: = \: -0.116 \, , \qquad \frac{\chi_{\nu}^2}{\nu} \: = \: 0.92 \, .
\end{equation}

Since the model parameter $r$ is expected to be small, we may assume that the oscillation plane of the co-decaying Bloch vector, ${\bf b}(t)$, is spanned by $\boldsymbol{\gamma}$ and $\mathbf{e}\times\boldsymbol{\gamma}$. The validity of this approximation is discussed in Appendix~\ref{App:Geometry}.
By virtue of~\eqref{eq:AsymInv}, we can extract $\vecb(t)\cdot\lrb{\mathbf{e}\times\boldsymbol{\gamma}}$ from the data, and use this information to evaluate the Fourier coefficients of the oscillation profile. The original analysis of LHCb~\cite{LHCb:2016gsk} identifies the mass splitting of the $B$-meson system, $\Delta m_{\rm d}$, with the measured angular frequency~$\omega$ of the mixing asymmetry oscillation. Consequently, we equate $\omega$ with $\Delta m_{\rm d}$. The Fourier coefficients are calculated by making use of a multiple linear regression of the projected Bloch vector, $\vecb(t)\cdot\lrb{\mathbf{e}\times\boldsymbol{\gamma}}$, against the Fourier modes, $\cos(n \omega t)$, where the Fourier coefficients are the regression values that minimise the $\chi^2$ of the fit. Details of the analysis including estimation of the Fourier coefficients, estimation of $r$, and their associated errors are given in Appendix~\ref{App:Errors}.

Figure~\ref{fig:Fourier_B_meson} shows the mixing asymmetry oscillation data as a projection of $\vecb (t)$ onto ${\mathbf{e}\times\boldsymbol{\gamma}}$. Figures~\ref{fig:B2D2011} and~\ref{fig:B2DStar2011} show the $2011$ data for the decay processes, $B^0_{\rm d} \to D^-\mu^+\nu_\mu X$ and ${B^0_{\rm d} \to D^{-*}\mu^+\nu_\mu X}$, respectively. Likewise, Figures~\ref{fig:B2D2012} and~\ref{fig:B2DStar2012} exhibit the corresponding $2012$ data for the above two decays.
%$B^0_{\rm d} \to D^-\mu^+\nu_\mu X$ and $B^0_{\rm d} \to D^{-*}\mu^+\nu_\mu X$, respectively. 
The red line is a truncated Fourier series fit by taking into account the sum of $n=0,\, 1$ and $2$ harmonics. The Fourier coefficients extracted from this fit, with their respective errors, are given in Table~\ref{tab:Fourier_Coeffs}. Alongside these Fourier coefficients, we check whether each coefficient deviates significantly from zero, where the latter value represents our {\em null hypothesis}. The $p$-value of each coefficient is also stated in Table~\ref{tab:Fourier_Coeffs}, which has been calculated using a standard $t$-distribution with mean $\mu=0$. We have found small $p$-values for the Fourier coefficient, $d_1$, which means the hypothesis of the basic sinusoidal harmonic becomes statistically significant.  

\begin{table}[t!]
    \centering
    \begin{tabular}{|c||c|c|c|c|}
        \hline
         - & \multicolumn{2}{c|}{$B_{\rm d}^0 \to D^-\mu^+\nu_\mu X \, (2011)$} & \multicolumn{2}{c|}{$B_{\rm d}^0 \to D^{-*}\mu^+\nu_\mu X\, (2011)$}  \\[0.1cm]
        \hline\hline
        {} & $d_n \pm \delta d_n$ & $p$-value & $d_n \pm \delta d_n$ & $p$-value \\ 
        \hline
        $d_0$ & $0.04 \pm 0.12$ & $74\%$ & $0.006 \pm 0.137$ & $96\%$ \\
        \hline
        $d_1$ & $0.630 \pm 0.007$ & $< 0.01\%$ & $0.665 \pm 0.007$ & $<0.01\%$ \\
        \hline
        $d_2$ & $-0.03 \pm 0.01$ & $ 6\%$ & $0.01 \pm -0.01$ & $42\%$ \\
        \hline\hline
        - & \multicolumn{2}{c|}{$B_{\rm d}^0 \to D^-\mu^+\nu_\mu X \, (2012)$} & \multicolumn{2}{c|}{$B_{\rm d}^0 \to D^{-*}\mu^+\nu_\mu X\, (2012)$}  \\[0.1cm]
        \hline\hline
        {} & $d_n \pm \delta d_n$ & $p$-value & $d_n \pm \delta d_n$ & $p$-value \\ 
        \hline
        $d_0$ & $0.06 \pm 0.07 $ & $45\%$ & $0.04 \pm 0.06$ & $58\%$\\
        \hline
        $d_1$ & $ 0.625 \pm 0.004$ & $<0.01\%$ & $0.564 \pm 0.003$ & $<0.01\%$ \\
        \hline
        $d_2$ & $ -0.013 \pm 0.009 $ & $18\%$ & $0.007 \pm 0.008$ & $37\%$ \\
        \hline
    \end{tabular}
    \caption{\em Fourier coefficients extracted from the regression fit of the data given in~\cite{LHCb:2016gsk}. The errors are estimated as a combination of the errors associated with individual data points as well as the errors inherent to fitting the Fourier coefficients. For completeness, we additionally give p-values to provide an assessment of the significance of each Fourier coefficient as a departure from zero.}
    \label{tab:Fourier_Coeffs}
\end{table}

As can also be seen in Table~\ref{tab:Fourier_Coeffs}, the other two harmonics, $d_0$ and $d_2$, acquire sizeable $p$-values through our analysis, which point to low statistical significance from the null hypothesis. As~a result, it may be difficult to reach a high quality estimate for the dimensionless parameter~$r$. In~the following subsection, we will explain the reasons for this low significance. As stated in~\cite{LHCb:2012mhu,LHCb:2016gsk}, it is worth drawing attention to the fact that the lifetime of $B^0$ mesons is about $6~{\rm ps}$, whereas the oscillation period is measured to be around $12~{\rm ps}$. Hence, a significant number of $B^0$ mesons have decayed before half an oscillation has been observed. As a consequence, data points after $t\simeq \textrm{P}/2$ are of lower precision, and therefore the error bars widen with time, as this is reflected in Figure~\ref{fig:Fourier_B_meson}.

\begin{table}[t!]
    \centering
    \begin{tabular}{|c||c|c|c|c|}
        \hline
         {} & \multicolumn{2}{c|}{$B_{\rm d}^0 \to D^-\mu^+\nu_\mu X \, (2011)$} & \multicolumn{2}{c|}{$B_{\rm d}^0 \to D^{-*}\mu^+\nu_\mu X\, (2011)$}  \\[0.1cm]
        \hline\hline
        {} & $ \mathcal{D}_n $ & $r$ & $\mathcal{D}_n $ & $r$ \\
        \hline
        $\mathcal{D}_0$ & $15 \pm 44$ & $0.2\pm 0.6$ & $109 \pm 2464$ & $0.03 \pm 0.06$ \\
        \hline
        $\mathcal{D}_1$ & $0.04 \pm 0.02$ & $0.13 \pm 0.06$ & $0.02 \pm 0.02$ & $0.05\pm0.06$ \\
        \hline\hline
        {} & \multicolumn{2}{c|}{$B_{\rm d}^0 \to D^-\mu^+\nu_\mu X \, (2012)$} & \multicolumn{2}{c|}{$B_{\rm d}^0 \to D^{-*}\mu^+\nu_\mu X\, (2012)$}  \\[0.1cm]
        \hline\hline
        {} & $ \mathcal{D}_n $ & $r$ & $\mathcal{D}_n $ & $r$ \\ 
        \hline
        $\mathcal{D}_0$ & $11 \pm 15 $ & $0.3 \pm 0.3$ & $16 \pm 29$ & $0.2 \pm 0.4$\\
        \hline
        $\mathcal{D}_1$ & $ 0.02 \pm 0.01$ & $0.06 \pm 0.05$ & $0.01 \pm 0.02$ & $0.05 \pm 0.06$ \\
        \hline
    \end{tabular}
    \caption{\em Estimations of the anharmonicity factors and values of $r$ for the $2011$ and $2012$ data points provided in~\cite{LHCb:2016gsk}. Alongside these central values, we provide estimates of their associated $1\sigma$ error bounds.}
    \label{tab:r_estimates}
\end{table}

By adopting the procedure outlined in Section~\ref{sec:Fourier}, we may make estimates of the dimensionless parameter $r$. For $N$ Fourier coefficients, $N-1$ estimates for $r$ can be obtained. These results are presented in Table~\ref{tab:r_estimates}. As far as the Fourier coefficients themselves are concerned, the large errors pervade into the estimates of the anharmonicity factors, ${\cal D}_0$ and ${\cal D}_1$, and the model parameter~$r$. From the data analysed here, we can take a weighted average of the values for~$r$ giving the central estimate $r = 0.07 \pm 0.03$. We see that given the present data set and the current analysis, it is difficult to attain an estimate of $r$ which differs significantly from~zero. The estimates we have shown in~Table~\ref{tab:r_estimates} deviate from the Rabi oscillation assumption with very low statistical significance amounting to values of~$p=8\%$. As we clarify in the next subsection, there are well-founded reasons to expect this result.

\subsection{Detector Resolution}

It is interesting to assess whether a given signal may be better described by a typical Rabi oscillation or by a CUQ oscillation. To determine this, we consider $N$ equally time-spaced measurements over the oscillation period~$\text{P}$, defined as $t_k/\text{P} \equiv k/N $, with $k=0,1,\dots, N$. Then, under the working assumption of $r=0$, the $k$-th measurement of the projection of the co-decaying Bloch vector, ${\bf b}(t_k)$, along the fixed unit vector, $\mathbf{e}\times\boldsymbol{\gamma}$, has the expectation: $E_k = \cos\lrb{2 \pi \frac{k}{N}}$. Instead, if we assume that the data are distributed according to a CUQ scenario, the signal would have observations, $O_k$ at each $t_k$, given by
\begin{equation}
    O_k \: = \: \cos \theta(t_k) \: = \: \frac{\cos(\omega t_k)-r}{1-r\cos(\omega t_k)} \: \simeq \: \cos(\omega t_k) - r + r\cos^2(\omega t_k)\: +\: \mathcal{O}(r^2) \, .
\end{equation}
Assuming that each measurement has approximately equal variance, $\sigma^2$, we can calculate a statistical value $\chi^2$ with $N$ degrees of freedom,
\begin{equation}
    \chi^2 \: = \: \sum_{k=1}^N \frac{\lrb{O_k - E_k}^2}{\sigma^2} \: = \: \frac{r^2}{4\sigma^2} \sum_{k=1}^N \sin^4\lrb{2\pi\frac{k}{N}} \, . 
\end{equation}
Making use of double-angle identities and well-known expressions for finite sums, one can evaluate this sum and find the simple expression for the $\chi^2$ value,
\begin{equation}
    \chi^2 \: = \: \frac{3 N r^2}{8\sigma^2} \ .
\end{equation}
For high values of $N$, the value of $\chi^2$ approaches a normal distribution with expectation $N$ and variance $2N$. Under this assumption, we may evaluate the significance of $\chi^2$ using a typical one-tailed $Z$-test:
\begin{equation}
    Z \: = \: \frac{\chi^2 - \mu_{\chi^2}}{\sqrt{{\rm var}[\chi^2]}} \: = \: \lrb{\frac{3r^2}{8\sigma^2}-1}\frac{N}{\sqrt{2N}} \: = \: \frac{\sqrt{N}}{\sqrt{2}}\lrb{\frac{3r^2}{8\sigma^2}-1} \, .
\end{equation}
By choosing an appropriate level of statistical significance, e.g.~$99\%$~CL, one can determine the relevant value of~$Z^*$ and produce a bound on the required resolution of the experimental data. Hence, for a value $Z^* \leq Z$, we have the lower limit for $r^2$,
\begin{equation}
    %Z^* \: \leq \:  \frac{\sqrt{N}}{\sqrt{2}}\lrb{\frac{3r^2}{8\sigma^2}-1} \: \implies \: 
    r^2 \: \geq \: \frac{8\sigma^2}{3} \lrb{1+\frac{\sqrt{2}}{\sqrt{N}} Z^*} \, .
\end{equation}
From this last inequality, we see that regardless of the amount of data taken, or the significance level chosen, there is a lower precision bound at $r \sim 1.6\,\sigma$, below which the data are inferred to be consistent with Rabi oscillations. Recalling that the leading order of the Fourier coefficients is $c_n \sim r^{n}$, we see that below this statistical bound, the anharmonicities will be contained within the error bars of the data, and hence, indistinguishable from statistical variation. 

If the oscillation amplitude, $R$, happens to be less than~1, one must still account for the fact that the estimates are for the effective parameter, $\tilde{r}$. The latter must then be re-mapped onto the true value of~$r$ using the expression given in~\eqref{eq:r_effective}. Repeating the above analysis by assuming an amplitude of $R<1$ results in the following precision bound:
\begin{equation}
    \tilde{r}^2\: \geq \: \frac{8\sigma^2}{3R^2} \lrb{1+\frac{\sqrt{2}}{\sqrt{N}} Z^*}\,, 
\end{equation} 
which may then be translated to the lower limit on $r^2$,
\begin{equation}
    r^2 \: \geq \: \frac{8\sigma^2 \lrb{\sqrt{N}+\sqrt{2} Z^*}}{3R^4 \sqrt{N} + 8\sigma^2 \lrb{1-R^2} \lrb{\sqrt{N}+\sqrt{2} Z^*}} \ .
\end{equation}
From this last expression, we may estimate, for small values of $r\ll 1$, the minimum precision bound:~${r \sim 1.6\,\sigma/R^2}$. 

The data we used to study $B^0_{\rm d}\bar{B}^0_{\rm d}$-meson oscillations have scale errors ${\sigma \simeq 2 \times 10^{-2}}$ and an amplitude~${R \simeq 0.55}$, and use $N=49$ points in each data set. The one-tailed distribution at~$99\%$~CL leads to $Z^* \simeq 2.33$, from which we can estimate a lower limit on $r$ at~$99\%$~CL given by the inequality $r \gtrsim 0.13$. From this lower limit, we infer that the current data set does not have the required precision to probe the anharmonicities of $B$-meson oscillations, and the result we have extracted thus far is likely an overestimate due to the statistical fluctuations rather than a true estimate of the physical anharmonicities in the data. We envisage that experimental analyses more sophisticated than the simplified analysis presented here will be able to extract anharmonicities in the mixing asymmetry of the $B^0_{\rm d}\bar{B}^0_{\rm d}$-meson system. Likewise, we expect that improved precision in future experiments will offer the sensitivity required for conducting dedicated studies into anharmonic oscillation profiles. Therefore, our approach to CUQ oscillations, which employs the novel methodology presented in this section, provides new impetus to search for new physics, as well as opens new avenues to perform precision tests even within~the~SM.

%\bigskip
%\vfill\eject

\section{Conclusions}\label{sec:Concl}
\setcounter{equation}{0}

The present work extends our first study on Critical Unstable Qubits~\cite{Karamitros:2022oew} in several aspects${}$.
In doing so, we have employed the Bloch-sphere representation of the density matrix and the Pauli decomposition of the effective Hamiltonian~${\rm H}_{\rm eff}$, in order to write the latter in terms of the unit energy and decay vectors, $\mathbf{e}$ and $\boldsymbol{\gamma}$, as well as the dimensionless parameter~$r$. Hence, on the basis of the Bloch-sphere formalism, CUQs are unstable qubits characterised by the critical condition: $\mathbf{e}\cdot\boldsymbol{\gamma}=0$, with $r<1$. 
We have elaborated on our previous analysis of CUQs by generalising the derivation of the time evolution of the co-decaying Bloch vector~${\bf b}(t)$, for an arbitrary initial vector ${\bf b}(0)$ at $t=0$. The so-derived analytic form of~${\bf b}(t)$ was then used to obtain exact expressions for the Fourier series of CUQ oscillations.

A distinctive physical feature of CUQs is that these are two-level quantum systems where the two decay widths are equal, and so there is no longest-lived state to which ${\bf b}(t)$ relaxes at large times. In this class of quantum systems, the inclusion of decay effects can still be physically meaningful, since non-vanishing values of $r$ can significantly modify the oscillation period and give rise to {\em anharmonicities} in the oscillation pattern between energy levels. This drastically differs from the current paradigm widely assumed in the literature, where the principal harmonic is sufficient to analyse unstable two-level systems. However, as we have shown in this work, by considering higher harmonics, it is possible to probe deeper into the structure of the effective Hamiltonian~${\rm H}_{\rm eff}$. In particular, by taking ratios of Fourier coefficients, we have found a simple scale-invariant way by which the dimensionless parameter~$r$ can be determined. In line with this observation, we have developed an approach to objectively assessing the degree of matching between the approximate $N$-term Fourier series and the exact analytic form. As was shown in Figure~\ref{fig:Fourier_figs}, the inclusion of additional Fourier modes, beyond the principle harmonic, gives a greater degree of matching. Furthermore, we confirmed our expectation that for values of~$r$ close to zero, one can achieve a high degree of accuracy with relatively few Fourier modes. However, as $r$ grows towards~1, it was necessary to include higher modes in order to accurately describe the variance around the exact oscillation profile of a CUQ.

To make connection with current experiments, we have highlighted a novel method by which current and future experimental data on $B$-mesons may be used to extract the value of~$r$ from their mixing oscillation asymmetry. In general, confirming a non-zero value of $r$ may provide by itself strong evidence for CP violation in meson--antimeson systems. As can be seen in Figure~\ref{fig:Fourier_B_meson} and in Tables~\ref{tab:Fourier_Coeffs} and~\ref{tab:r_estimates}, given the current datasets, only generic estimates for $r$ can be deduced, although these are plagued by large uncertainties. However, future experiments will collect more data and so they are expected to offer a higher level of accuracy in the analyses performed for the $B^0_{\rm d}$-meson system.

The results we have presented in this paper use a particular class of initial conditions for which the co-decaying Bloch vector, ${\bf b}(t)$, lies wholly on the plane spanned by $\boldsymbol{\gamma}$ and $\mathbf{e}\times\boldsymbol{\gamma}$. For~$r\ll 1$, this plane approximately contains the trajectory of the ${\bf b}(t)$ for any choice of the initial vector $\vecb(0)$. But if ${\bf b}(0)$ points to an arbitrary direction and $r < 1$ is not too small, we have shown in Figures~\ref{fig:BlochPlane} and \ref{fig:Arrow_figs} that these properties remain. In this regard, we explicitly demonstrated in Appendix~\ref{App:Geometry} that there always exists a projection of ${\bf b}(t)$ which follows an equivalent trajectory on a plane determined by the model vectors, $\mathbf{e}$ and $\boldsymbol{\gamma}$, and by~${\bf b}(0)$.  This property is general and results from the vanishing of the torsion, ${\cal T}$, for all trajectories of~${\bf b}(t)$, in agreement with earlier findings in~\cite{Karamitros:2022oew} that were made for a restricted set of initial conditions.

The present study largely completes the analysis of CUQs, whose time evolution is governed by an infinite series of non-zero Fourier modes. On the other hand, our work opens the door to even more detailed analyses considering quasi-critical unstable qubit scenarios where $\mathbf{e}$ is not quite orthogonal to $\boldsymbol{\gamma}$. Numerical solutions to the evolution equations seem to suggest that the oscillation period is solely dependent on $r$ and independent of the angle, $\theta_{\mathbf{e}\boldsymbol{\gamma}}$, between $\mathbf{e}$ and $\boldsymbol{\gamma}$. Future studies should consider how a similar Fourier-type decomposition can be performed to extract not only~$r$, but also~$\theta_{\mathbf{e}\boldsymbol{\gamma}}$. This would extend the analysis to include meson--antimeson systems with a decay width splitting, thus providing an alternate method to detect CP violation and search for new physics. Furthermore, since this study captures much of the physics pertinent to isolated quantum systems, a natural extension would be to consider these ideas in the context of interacting quantum systems that happen to be in a state of entanglement. 

In conclusion, our analysis lays the foundation for new ways of probing the structure of the effective Hamiltonian~$\text{H}_{\rm eff}$ with greater precision. In particular, we have explicitly demonstrated how our approach can be used in scalar-meson-mixing systems. Other applications, such as axion-photon mixing, open quantum systems, and quantum computing, are worthy of their own dedicated studies.

\subsection*{Acknowledgements} 
\vspace{-1mm}

\noindent
The work of AP  is supported in part by the Lancaster-Manchester-Sheffield Consortium for Fundamental Physics, under STFC Research Grant ST/T001038/1. TM acknowledges support from the STFC Doctoral Training Partnership under STFC training grant ST/V506898/1. The~authors thank Conor Fitzpatrick for discussions on experimental analyses of $B$-meson oscillation data at the LHCb.

%%%%%%%%%%%%%%%%%%%%%%%%%%%%%%%%%%%%%%%%
% APPENDICES
\newpage
%%%%%%%%%%%%%%%%%%%%%%%%%%%%%%%%%%%%%%%%
\setcounter{section}{0}
%\section*{Appendix}
\appendix

\renewcommand{\theequation}{\Alph{section}.\arabic{equation}}
\setcounter{equation}{0}  % reset counter
%%%%%%%%%%%%%%%%%%%%%%%

\section{Equivalence of Critical Unstable Qubit Scenarios}\label{App:Geometry}
\setcounter{equation}{0}

\begin{figure}[b!]
    \captionsetup[subfigure]{labelformat=empty}
    \centering
    \begin{subfigure}{0.32\linewidth}
        \centering
        \includegraphics[width=\linewidth]{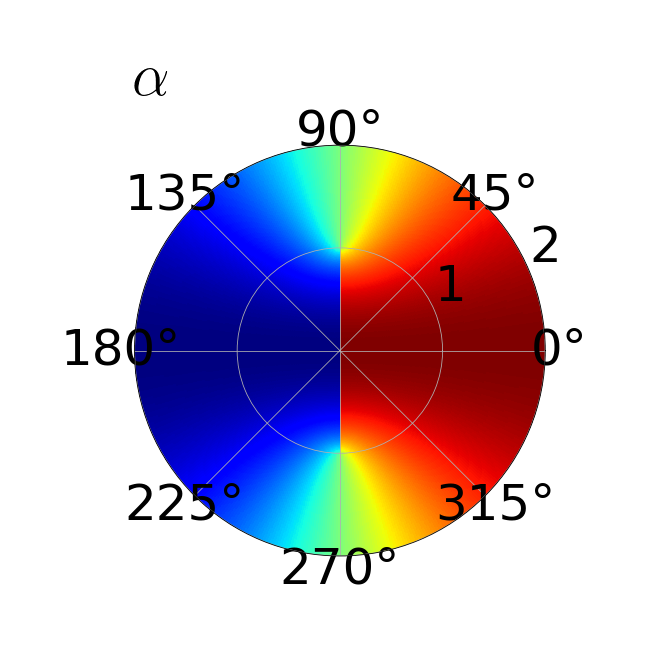}
    \end{subfigure}
    \begin{subfigure}{0.32\linewidth}
        \centering
        \includegraphics[width=\linewidth]{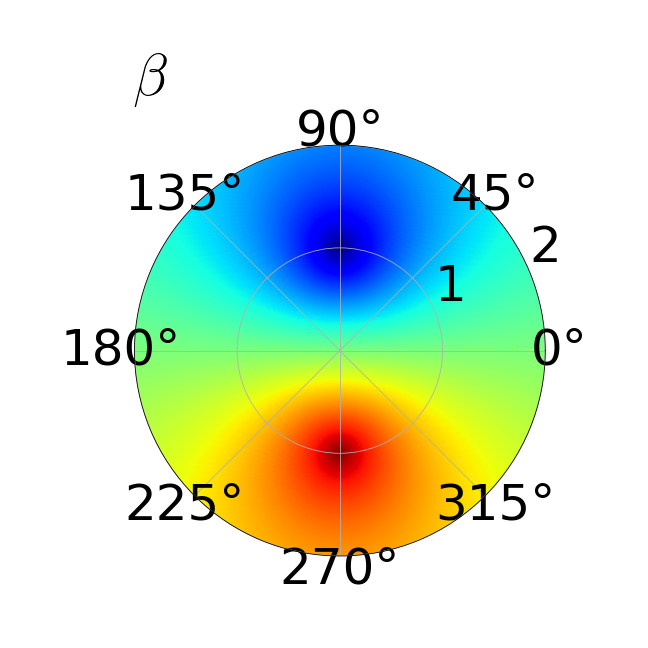}
    \end{subfigure}
    \begin{subfigure}{0.32\linewidth}
        \centering
        \includegraphics[width=\linewidth]{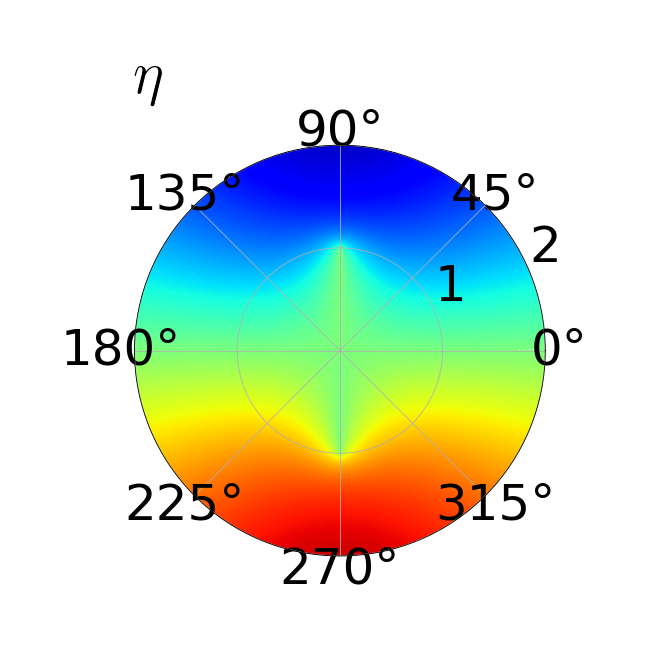}
    \end{subfigure}
    \caption{The stable fixed point solutions of the general 3D
      dynamical flow in a plane polar representation of $r \in (0, 2]$ and
      $\theta_{\mathbf{e}\boldsymbol{\gamma}}\in [0,360^\circ)$. Blue hues indicate
      values of $\alpha$, $\beta$, and $\eta$ close to~$-1$, red hues
      identify their values close to $+1$, and green hues show their regions close to~0.}
    \label{fig:Fixed_Points}
\end{figure}

In~\cite{Karamitros:2022oew}, it was analytically shown that there are stationary solutions for the co-decaying Bloch vector, ${\bf b}(\tau )$, with exact solutions given in terms of $r$ and $\theta_{\mathbf{e}\boldsymbol{\gamma}} = \cos^{-1}(\mathbf{e}\cdot\boldsymbol{\gamma})$. These solutions were identified through the static points in the flow equations of~$\be (\tau)$. In this appendix, we extend our previous analysis to clearly demonstrate the equivalence of all CUQ scenarios. Additionally, we will highlight an alternative approach to identifying fixed points based on separating the coordinate space into different dynamical regions. 

In Figure~\ref{fig:Fixed_Points}, we show the longest-lived states of the Bloch vector. Here, $\be(\tau)$ has the decomposition,
\begin{equation}
    \vecb(\tau) \: = \: \alpha(\tau) \, \mathbf{e} + \beta(\tau) \, \mathbf{e}\times\boldsymbol{\gamma} + \eta (\tau)\, \mathbf{e}\times\lrb{\mathbf{e}\times\boldsymbol{\gamma}} \, .
\end{equation}
The figure represents systems with $0<r<2$ through the radius, and the full range of $\theta_{\mathbf{e}\boldsymbol{\gamma}} \in [0,360^\circ]$ as the polar angle. As can be seen in the figure, the co-ordinate space is largely smooth, and so, changes to $r$ and $\theta_{\mathbf{e}\boldsymbol{\gamma}}$ give well-behaved changes to the longest-lived state. There is, however, a noticeable discontinuity in the space of solutions for $\alpha$. CUQs, characterised by $r<1$ and $\mathbf{e}\cdot\boldsymbol{\gamma}=0$, are contained within this discontinuity. To study this region in more detail, we focus on the geometric structure of CUQs and demonstrate the equivalence of all systems where $\mathbf{e}\cdot\boldsymbol{\gamma} = 0$.

The analysis we have provided on CUQs in~\cite{Karamitros:2022oew}, and followed up on in this work, was done assuming a convenient initial condition for the co-decaying Bloch vector, $\be(0)$. In particular, $\vecb(0)$ was presumed to lie on the plane spanned by $\boldsymbol{\gamma}$ and $\mathbf{e}\times\boldsymbol{\gamma}$. On this plane, the evolution equation indicates that there is no motion along $\mathbf{e}$, since
\begin{equation}
    \frac{\rm d}{{\rm d}\tau} \big(\mathbf{e}\cdot\be(\tau)\big) \: = \: -\lrb{\boldsymbol{\gamma}\cdot\vecb}\lrb{\mathbf{e}\cdot\vecb} \: = \: 0 \, .
\end{equation}
Consequently, $\be(\tau)$ may be parametrised solely by two real-valued coefficients as follows:
\begin{equation}
    \vecb(\tau) \: = \: \beta(\tau)\, \boldsymbol{\gamma}\, +\, \eta(\tau)\, \mathbf{e}\times\boldsymbol{\gamma} \; .
\end{equation}
The two functions, $\beta(\tau)$ and $\eta(\tau)$, are described by a coupled system of evolution equations,
\begin{subequations}\label{eq:CSFlows}
    \begin{align}
    \frac{\de \beta}{\de \tau}\ &=\ \frac{1}{r}\eta\, +\, 1\, -\, \beta^2 \, , \\
    \frac{\de \eta}{\de\tau}\ &=\ -\frac{1}{r}\beta\, -\, \eta \beta \; .
\end{align}
\end{subequations}
When $r>1$, these equations describe a flow which approaches a stationary point as $\tau\to \infty$. However, for a CUQ, where $r<1$, these flows characterise stable cycles.

However, it is not guaranteed that such a special circumstance can always be prepared, nor should it be that the physical phenomena we see in these cases be simply a result of a particular choice of initial condition. Therefore, we wish to verify that the novelties of CUQs are preserved in the most general set-up in which the initial co-decaying Bloch vector ${\bf b}(0)$ was chosen to point to an arbitrary direction. To this end, we will describe the geometric nature of the curve traced out by~$\mathbf{b}(\tau)$, following up on the observation given in~\cite{Karamitros:2022oew} that the torsion~${\cal T}$ of ${\bf b}(\tau )$ vanishes, i.e. 
\begin{equation}
    \mathcal{T} \: = \: \frac{\lrb{\be^\prime\times\be^{\prime\prime}}\cdot\be^{\prime\prime\prime}}{\left|  \be^\prime\times\be^{\prime\prime} \right|^2} \: = \: 0 \, ,
\end{equation}
where primed terms indicate derivatives with respect to the dimensionless time parameter, $\tau$. We will see that the plane which contains the curve traced out by $\be(\tau)$ may be fully determined through the initial condition $\be(0)$, and the two model vectors $\mathbf{e}$ and $\boldsymbol{\gamma}$. Remarkably, we will see that $\vecb(\tau)$ no longer simply oscillates about $\mathbf{e}$, but a shifted vector, $\mathbf{u}$, which contains a correction to $\mathbf{e}$ proportional $\be(0)$. Moreover, and perhaps more surprisingly, the plane containing the trajectory of $\be(\tau)$ is found to be parallel to the decay vector, $\boldsymbol{\gamma}$, for any choice of initial condition provided that $\mathbf{e}\cdot\boldsymbol{\gamma}=0$. On this plane, it is possible to demonstrate the full mathematical equivalence between general CUQs and the special case already detailed, including the surprising result that the dimensionless oscillation period is independent of the initial condition.

We consider the Frenet-Serret construction~\cite{doCarmo:1976} for which we require three orthonormal basis vectors: ($i$) the tangent vector, $\vect$, ($ii$) the normal vector, $\vecn$, and ($iii$) the bi-normal vector, $\vecm$, which form a basis of the space. The evolution equation that we study indicates that for CUQs, the curve traced out by $\be(\tau)$ will be continuous and singularity free, avoiding any complications in the construction of tangent vectors. This tangent unit vector, at some $\tau$, is easily identified through the evolution equation of $\vecb$:
\begin{equation}
    \vect(\tau) \: = \: \frac{1}{|\be^\prime(\tau)|}\be^\prime(\tau)\, ,
\end{equation}
with
\begin{equation}
    |\be^\prime(\tau)|^2 \: = \: \frac{1}{r^2}\lrsb{|\be(\tau)|^2-(\mathbf{e}\cdot\be(\tau))^2} + 1 - \frac{2}{r} (\mathbf{e}\times\boldsymbol{\gamma})\cdot\be(\tau) + \lrb{\boldsymbol{\gamma}\cdot\be(\tau)}^2 \lrsb{|\be(\tau)|^2 - 2} \, .
\end{equation}
As previously, primed terms indicate the derivative with respect to the dimensionless run parameter $\tau=|\boldsymbol{\Gamma}|t$. The normal vector, $\vecn(\tau)$, is similarly identified by taking the derivative of the tangent vector, up to an appropriate normalisation factor:
\begin{align}
    \kappa(\tau) \vecn(\tau) \: &= \: \frac{\vect^\prime(\tau)}{|\be^\prime(\tau)|}   %\nonumber\\
    %&= \: \frac{1}{|\be^\prime(\tau)|^2} \lrsb{\vecb^{\prime\prime}(\tau) - \lrb{\vecb^{\prime\prime}(\tau)\cdot\vect(\tau)}\vect(\tau)}  \nonumber\\
    \: = \: \frac{-\frac{1}{r} \mathbf{e}\times \vect(\tau) + (\boldsymbol{\gamma}\cdot\vect(\tau))(\be(\tau) \cdot \vect(\tau))\vect(\tau) - (\boldsymbol{\gamma}\cdot\vect(\tau)) \be(\tau) }{|\be^\prime(\tau)|}  \, .
\end{align}
It is already known that the torsion vanishes, and so, the plane spanned by the tangent and normal vectors contain the full trajectory traced out by $\vecb(\tau)$. It should be emphasised, however, that this plane does not necessarily contain the origin, nor $\be(\tau)$ itself, but solely the curve traced out by $\vecb(\tau)$, since the plane is not only tilted, but also shifted away from the plane spanned by $\mathbf{e}$ and $\mathbf{e}\times\boldsymbol{\gamma}$. The full \textit{moving-trihedron} basis is completed by including the bi-normal vector, i.e. the vector normal to the plane. This is constructed as the outer product between the tangent and normal vectors, $\vecm(\tau) = \vect(\tau)\times\vecn(\tau)$. A direct calculation of the inner product between $\vecm(\tau)$ and $\boldsymbol{\gamma}$ shows that these two vectors are perpendicular to one another. Consequently, the decay vector, $\boldsymbol{\gamma}$, must be collinear to $\vect(\tau)$ and $\vecn(\tau)$, and so one of the basis vectors for the plane containing the trajectory of $\be(\tau)$. In essence, the vanishing of this inner product indicates that the plane containing the trajectory of $\be(\tau)$ must, surprisingly, be parallel to the decay vector.

The moving-trihedron construction we have used gives a complete basis upon which we could study the dynamics of a CUQ. However, it offers little connection with the model vectors $\mathbf{e}$ and $\boldsymbol{\gamma}$. The true utility of the moving tri-hedron construction is two-fold: ($i$) it demonstrates that $\boldsymbol{\gamma}$ is a basis vector of the plane, and ($ii$) it provides a complete description of the plane. For this latter point, the moving tri-hedron provides two vectors, $\vect$ and $\vecn$. Since the torsion is vanishing, the plane basis is time-independent, and hence we may solely consider the initial condition of $\vecb$ to build an equivalent basis. Given that $\boldsymbol{\gamma}$ is a basis vector of the plane, we assume that there exists some other vector $\boldsymbol{v}$ which is perpendicular to $\boldsymbol{\gamma}$ and completes the basis. Since these vectors are coplanar to $\vect$ and $\vecn$, we know that $\lrcb{\boldsymbol{\gamma}, \boldsymbol{v}}$ are a passive orthogonal transformation of the basis $\vect$ and $\vecn$, i.e.~$\mathbb{O}(2)$ rotations. One can find an analytic expression for $\boldsymbol{v}$ by considering projections along each basis vector. In this way, we obtain 
\begin{equation}\label{eq:PerpVec}
    \boldsymbol{v} \: = \: \frac{1}{N} \lrBigcb{\mathbf{e}\times\boldsymbol{\gamma} + r \, \lrBigb{\boldsymbol{\gamma}\times\be(0)} \times \boldsymbol{\gamma}} \, ,
\end{equation}
which is manifestly perpendicular to $\boldsymbol{\gamma}$. The normalisation constant, $N$, is chosen to ensure $\boldsymbol{v}$ is of unit magnitude, such that
\begin{equation}
    N^2 \: = \: r^2\lrsb{\mathbf{e}\cdot\be(0)}^2 + \lrsb{1+r(\mathbf{e}\times\boldsymbol{\gamma}\cdot\be(0))}^2 \, .
\end{equation}
One then completes the space and forms a right handed basis by including the vector perpendicular to the plane,
\begin{equation}\label{eq:Plane_Norm}
    \boldsymbol{u} \: = \: \boldsymbol{\gamma}\times\boldsymbol{v} \: = \: \frac{1}{N} \lrBigcb{\mathbf{e} + r\, \boldsymbol{\gamma}\times\be(0)} \, .
\end{equation}
Note that in the particular case that $\be(0)$ is contained to the plane spanned by $\mathbf{e}$ and $\mathbf{e}\times\boldsymbol{\gamma}$, including $\mathbf{e}\cdot\vecb(0)=0$, these expressions return the previously known basis of CUQ scenarios, i.e. $\lrcb{\mathbf{e}, \, \boldsymbol{\gamma} , \, \mathbf{e} \times\boldsymbol{\gamma}}$. Further to this, we see that the correction for general CUQs is proportional to $r$. Therefore, for small values of $r$, where oscillations are expected to dominate over decay effects, the basis vectors approximately take the well-known form: $\boldsymbol{v} \simeq \mathbf{e}\times\boldsymbol{\gamma}$ and $\boldsymbol{u}\simeq \mathbf{e}$. However, the plane which contains the trajectory of $\be(\tau)$ may not contain the origin, cf. Figure~\ref{fig:BlochPlane}.

With analytic expressions found for each of the basis vectors, we can turn to proving the equivalency of all CUQs. This can be done by clearly demonstrating that the evolution equations given in~\eqref{eq:CSFlows} can be reproduced for any $\be(0)$. We represent $\be(\tau)$ on the $\tau$-independent basis $\lrcb{\boldsymbol{u}, \boldsymbol{\gamma}, \boldsymbol{v}}$:
\begin{equation}
    \vecb(\tau) \: = \: \alpha(\tau) \, \boldsymbol{u} + \beta(\tau) \, \boldsymbol{\gamma} + \eta(\tau) \, \boldsymbol{v}  \, .
\end{equation}
These coefficients then follow a set of three flow equations:
\begin{subequations}
    \begin{align}
        \frac{{\rm d} \alpha}{{\rm d}\tau} \: &= \: \frac{1}{r}\beta(\boldsymbol{v}\cdot\mathbf{e}) - \beta\alpha \, , \\
        \frac{{\rm d} \beta}{{\rm d}\tau} \: &= \: -\frac{1}{r}\lrsb{\alpha(\boldsymbol{v}\cdot\mathbf{e}) - \eta(\boldsymbol{u}\cdot\mathbf{e})}-\beta^2+1 \, , \\
        \frac{{\rm d} \eta}{{\rm d}\tau} \: &= \: -\frac{1}{r} \beta(\boldsymbol{u}\cdot\mathbf{e})-\beta\eta \, ,
    \end{align}
\end{subequations}
which can be found by projecting $\vecb^\prime$ along each of the basis vectors. Due to the vanishing torsion, we know that the motion along $\boldsymbol{u}$ vanishes, and so ${\rm d}\alpha/{\rm d\tau} = 0$. In general, we may take $\beta \neq 0$ as an initial condition, and therefore, we infer that $\boldsymbol{v}\cdot\mathbf{e} = r\alpha$. Furthermore, since $\alpha(\tau)=\alpha(0)=\boldsymbol{u} \cdot \be(0)$, one can verify that the replacement of $\boldsymbol{v}\cdot\mathbf{e}$ with $r\alpha$ is consistent with the pertinent projections along the basis vectors. As before, the 3D system is reduced to a 2D flow described by the two differential equations,
\begin{subequations}
    \begin{align}
        \frac{{\rm d} \beta}{{\rm d}\tau} \: &= \:\frac{\boldsymbol{u}\cdot\mathbf{e}}{r}\eta - \alpha^2 
        - \beta^2+1\,, \\
        \frac{{\rm d} \eta}{{\rm d}\tau} \: &= \: -\frac{\boldsymbol{u}\cdot\mathbf{e}}{r} \beta-\beta\eta\,.
    \end{align}
\end{subequations}
These equations contain a number of additional factors which one must consider when compared with the flow equations presented in~\eqref{eq:CSFlows}, such as the factor $\boldsymbol{u}\cdot\mathbf{e}$ in the oscillation terms, and the previously absent $\alpha^2$ contribution to $\beta^\prime$. To resolve this issue, observe that in the pure state case, $|\vecb|=1$, the flow equations expectedly return that $\beta^2+\eta^2$ is constant since for CUQs, pure states remain pure. Hence, $\vecb(\tau)$ traces out a circle of radius $\sqrt{1-\alpha^2}$. In line with this, we may rescale the flow equations to be bounded by the unit circle on the plane spanned by $\boldsymbol{\gamma}$ and $\boldsymbol{v}$ through the definitions:
\begin{equation}\label{eq:ReMaps}
    R \: = \: \sqrt{1-\alpha^2} \, , \qquad \tilde{\beta} \: = \: \frac{\beta}{R} \, , \qquad \tilde{\eta} \: = \: \frac{\eta}{R} \, , \qquad \tilde{\tau} \: = \: R\tau\,, \qquad \tilde{r} \: = \: \frac{r R}{\boldsymbol{u}\cdot\mathbf{e}}  \, .
\end{equation}
Importantly, even with $|\vecb(0)| < 1$, $R$ remains constant even as $\vecb(\tau)$ changes in direction and magnitude, however, loses its interpretation as the radius of a circle. By construction, it is clear that $R$ is bounded at unity from above. However, careful consideration of the parameter space shows that $R$ must additionally be bounded from below. Taking $x=\lrb{\mathbf{e}\cdot\be(0)}^2$, we obtain the inequality,
\begin{equation}
    \alpha^2 \: \leq \: \frac{x}{1 + r^2 - 2r \sqrt{1-x}}\ .
\end{equation}
Maximising the RHS of this inequality with respect to $x$ gives an upper limit on $\alpha^2$ for any given~$r$. This upper limit can then be turned into a lower limit on $R$, given by
\begin{equation}
    R \: \geq\: \frac{r}{\sqrt{1+r^2}} \ .
\end{equation}
This expression crucially bounds $R$ above zero, with $R=0$ only possible when $r\to 0$, i.e. Rabi oscillations. In contrast, as $r\to \infty$, we see that the interval which contains $R$ narrows, and $R$ must approach unity. Thus, as $r\to \infty$, the plane must contain both the origin and $\be(\tau)$. This is mirrored by the plane vectors, where we see in the limit $r\to \infty$, the binormal vector, $\vecm$, approaches $\boldsymbol{\gamma}\times\be(0)$ provided $\be(0) \not\propto \boldsymbol{\gamma}$. Hence, the normal to the plane is orthogonal to $\be(0)$.

By means of the transformations stated in~\eqref{eq:ReMaps}, one may recast the 2D flow equations into an identical form with those given in~\eqref{eq:CSFlows}, i.e.
\begin{subequations}\label{eq:S2Flows}
    \begin{align}
    \frac{\de \tilde{\beta}}{\de \tilde{\tau}}\ &=\ \frac{1}{\tilde{r}}\tilde{\eta}\, +\, 1\, -\, \tilde{\beta}^2 \, , \\
    \frac{\de \tilde{\eta}}{\de\tilde{\tau}}\ &=\ -\frac{1}{\tilde{r}}\tilde{\beta}\, -\, \tilde{\eta} \tilde{\beta} \; .
\end{align}
\end{subequations}
Because of this form invariance of the flow equations, we may rescale the two coefficients of~$\be(\tau)$ into the unit circle and so map the general CUQ back onto the particular case already studied on the $\lrcb{\boldsymbol{\gamma}\, , \, \mathbf{e}\times\boldsymbol{\gamma}}$ basis. Hence, previous observations, such as anharmonic oscillations and coherence-decoherence oscillations, are preserved in general CUQ scenarios, although one must be careful with the use of rescaled parameters rather than the original parameters. In particular, we must pay attention to the use of $\tilde{r}$ rather than $r$.

The results and expressions shown in this work are clearly intrinsically dependent on the dimensionless model parameter $r$. Specifically, the evolution of $\theta$ [cf.~\eqref{eq:AngSol}], and consequently the Fourier coefficients [cf.~\eqref{eq:FourierCoeffC}] now take $\tilde{r}$ as an input rather than $r$, whenever the initial condition does not lie on the plane spanned by $\boldsymbol{\gamma}$ and $\mathbf{e}\times\boldsymbol{\gamma}$. Hence, different initial conditions are expected to have different trajectories for the same value of $r$. In terms of the observable quantities $\mathcal{C}_n$, the changes in the amplitude cancel out as expected.
But, the dependence on $\tilde{r}$ remains, because
\begin{equation}
    \tilde{c}_n(\tilde{r}) \: = \: R \, c_n(\tilde{r}) \quad \implies\quad   \mathcal{C}_n \: = \: \frac{c_{n+1}(\tilde{r})}{c_n(\tilde{r})} \, .
\end{equation}
One can then extract the underlying value of $r$ from $\tilde{r}$ using the relation,
\begin{equation}
    r \: = \: \frac{\tilde{r}}{\sqrt{R^2 + \tilde{r}^2 \lrb{1-R^2}}} \ . \tag{\ref{eq:r_effective}}\nonumber
\end{equation}

With these differences in the dynamical properties of CUQs, one may expect that the use of~$\tilde{\tau}$ rather than $\tau$ [cf.~\eqref{eq:ReMaps}] would result in a different oscillation period for a pure-state unstable qubit. But, surprisingly enough, this is not the case. After performing the appropriate transformations, we~find that the factors of $R$ and $\boldsymbol{u}\cdot\mathbf{e}$ conveniently cancel out. To make this explicit, we use the expression for the period given in~\eqref{eq:DLPeriod}. Consequently, we know that the oscillation period of a pure quantum-state system, in which $\tilde{\eta}^2+\tilde{\beta}^2=1$, is given, in terms of~$\tilde{r}$, by
\begin{equation}
    \tilde{\rm P} \: = \: \frac{2\pi \tilde{r}}{\sqrt{1-\tilde{r}^2}} \ .
\end{equation}
We may relate the period of the true system $\widehat{\rm P}$ to the period of the remapped system, $\tilde{\rm P}$, by appropriately rescaling the latter with $R$, by means of~(\ref{eq:ReMaps}). In this way, we have
\begin{equation}
    \widehat{\rm P} \: = \: \frac{1}{R}\, \frac{2\pi \tilde{r}}{\sqrt{1-\tilde{r}^2}} \: = \: \frac{2\pi r}{\sqrt{(\boldsymbol{u}\cdot\mathbf{e})^2 - R^2 r^2}} \: = \: \frac{2\pi r}{\sqrt{1-r^2}} \, .
\end{equation}
In the final step, we have used the completeness of the space, the orthogonality of $\mathbf{e}$ and $\boldsymbol{\gamma}$, the pure-state assumption, and $\boldsymbol{v}\cdot\mathbf{e}=r\alpha$, in order to make the replacement: $(\boldsymbol{u}\cdot\mathbf{e})^2 = 1-r^2+r^2R^2$. Consequently, we find that the oscillation period of a general CUQ is identical to the period of the particular case studied in the main body.

\begin{figure}[t!]
    \centering
    \begin{subfigure}{0.49\linewidth}
    \centering
    \includegraphics[width=\linewidth]{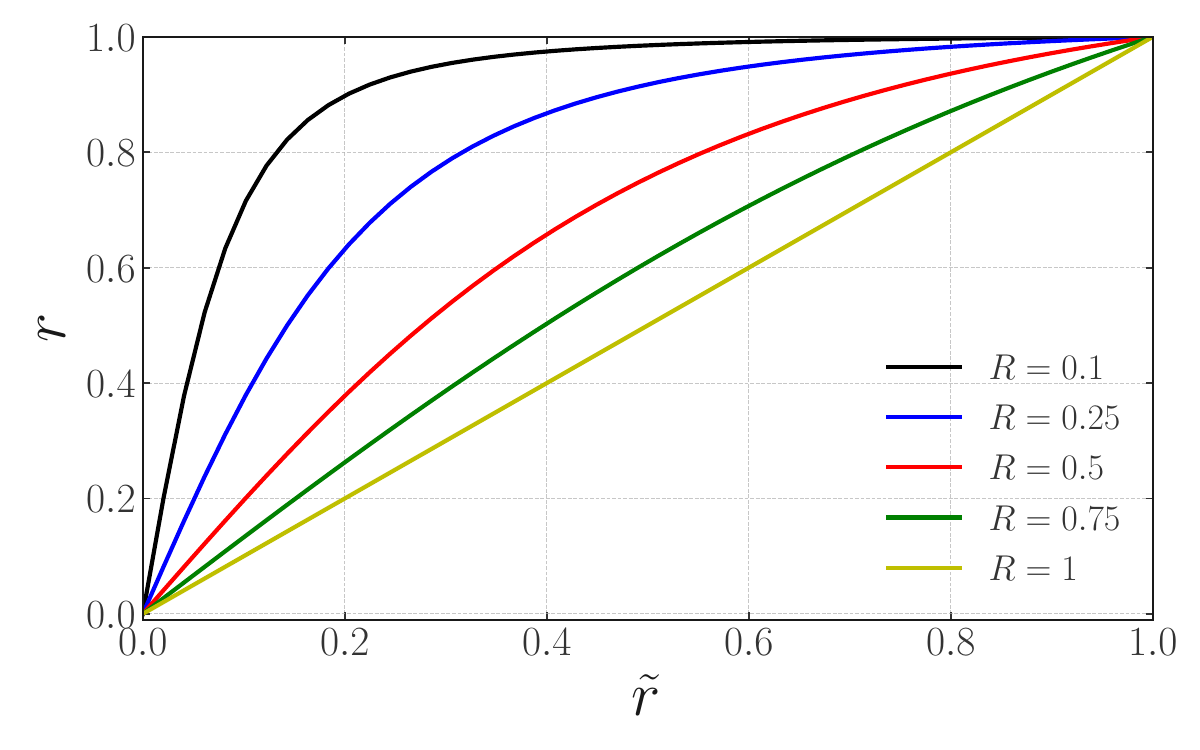}
    \caption{\empty}
    \label{fig:r_tilde}
    \end{subfigure}
    \begin{subfigure}{0.49\linewidth}
    \centering
    \includegraphics[width=\linewidth]{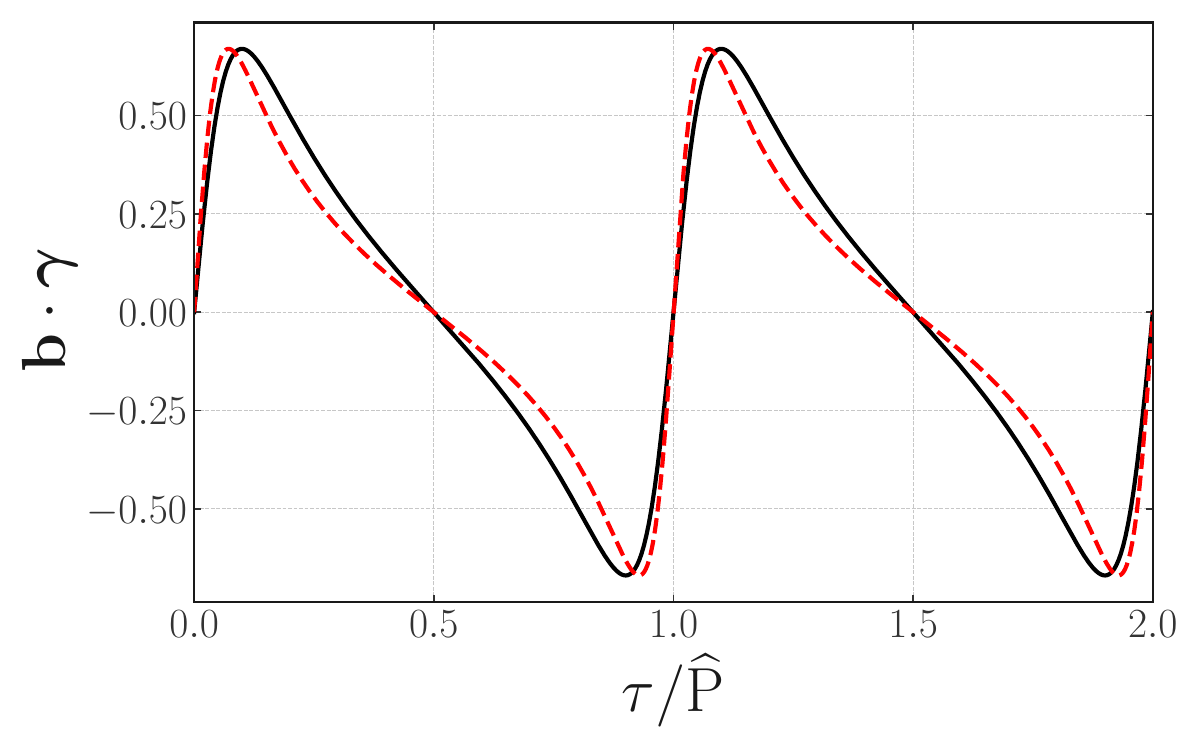}
    \caption{\empty}
    \label{fig:ReScale}
    \end{subfigure}
    \caption{The importance of properly accounting for the impact of different initial conditions in the estimations of $r$. The left panel ($a$) shows the true value of $r$ as a function of the effective value, $\tilde{r}$, for various values of $R$, which enters through the initial condition.The right panel ($b$) shows how this feeds through to the oscillation profile of the CUQ. Both the black and red lines consider CUQ scenarios with $r=0.9$. The black line is the true evolution with $\vecb(0)=\mathbf{e}$. The red dashed line has initial condition $\mathbf{e}\times\boldsymbol{\gamma}$ but has the amplitude scaled down to match the black line.}
    \label{fig:Rescale_Figs}
\end{figure}

Figure~\ref{fig:Rescale_Figs} highlights the importance of 
extracting the true value of $r$. Figure~\ref{fig:r_tilde} shows the adjusted value of $r$ as a function of $\tilde{r}$ for five inputs of the radius, $R$. As can be seen in the same figure, $\tilde{r}$ and $r$ are equal when $R=1$, or at the extreme values $\tilde{r}=0$ and $\tilde{r}=0$. Otherwise, we expect to find a significant difference between $\tilde{r}$ and $r$, particularly when $R\ll 1$. Notice that the mapping between $\tilde{r}$ and $r$ is bijective and is confined to the unit square $r(\tilde{r}): [0,1] \mapsto [0,1]$. Thus, we see that even in the general construction, the critical conditions are upheld even after the remapping stated in~\eqref{eq:ReMaps}. Most importantly, this means that one cannot change physical regimes simply through particular choices of the initial condition of the CUQ.

In Figure~\ref{fig:ReScale}, we can see the difference between $\tilde{r}$ and $r$ on full display. Here, the black line shows the projection of $\be(\tau)$ along $\boldsymbol{\gamma}$ for a CUQ scenario with initial condition $\vecb(0)=\mathbf{e}$, while setting $r=0.9$. In contrast, the red dashed line shows the trajectory of $\be(\tau)$ with $\vecb(0)=\mathbf{e}\times\boldsymbol{\gamma}$ and $r=0.9$, rescaled by the oscillation radius. As can be seen in Figure~\ref{fig:ReScale}, these two lines have identical oscillation frequencies. However, the true evolution of $\be(\tau)$ evolves differently from the system which is simply rescaled by the oscillation radius $R$. It is clear that both cases largely capture the gross features of the anharmonic oscillation. Nevertheless, it is equally apparent that the erroneous use of $r$ rather than $\tilde{r}$ will lead to an overestimation of the underlying parameters and will cause difficulties in fitting theoretical expectations to the oscillation~data.

The results presented in this Appendix show that there is a clear correspondence between the particular case we have considered previously and the general case, with features such as the attractor solution at $\tilde{\beta}^2 + \tilde{\eta}^2 = 1$ reproduced for $r>1$, and the anharmonic trajectory of $\be(\tau)$ manifest on the plane when $r<1$. This result strengthens the results presented in~\cite{Karamitros:2022oew} and removes any potential ambiguities that may otherwise have been covered by a convenient choice of parameterisation. Also, we have shown that there are notable nuances present in the trajectory, and as such, simply rescaling the coordinates does not provide a straightforward equivalence between different CUQ scenarios.

Following on from this equivalency, we may assume that the trajectory of $\be(\tau)$ is restricted to some plane and follows the 2D flow equations given in~\eqref{eq:S2Flows}. In what follows, we will now present an alternative method for the identification of a CUQ scenario, focusing on dynamical features of flows rather than algebraic expressions.

Isoclines of the 2D flow are defined to be trajectories in $(\tilde{\beta},\tilde{\eta})$ for which the gradient is constant~\cite{guckenheimer2013nonlinear}. The isoclines are, therefore, determined as 
\begin{equation}
    \frac{\de\tilde{\beta}}{\de\tilde{\eta}} \: = \: \frac{\de\tilde{\beta}}{\de \tilde{\tau}}\frac{\de \tilde{\tau}}{\de\tilde{\eta}} \: = \: \frac{\frac{1}{\tilde{r}}\tilde{\eta} + 1 - \tilde{\beta}^2}{ -\frac{1}{\tilde{r}}\tilde{\beta} - \tilde{\eta} \tilde{\beta}} \: = \: \frac{\tilde{\eta} + \tilde{r}(1 -\tilde{\beta}^2)}{ -\tilde{\beta} (\tilde{r} + \tilde{\eta})} \: = \: k \, .
\end{equation}
Isolating $\tilde{\eta}$ identifies a characteristic function for $\tilde{\eta}$ in terms of $\tilde{\beta}$, across which, the trajectories are of identical gradient. In terms of the parameter, $k$, this characteristic function is given by
\begin{equation}
    \tilde{\eta} \: = \: \frac{1}{1+k\tilde{\beta}}\lrsb{-k\tilde{r}\tilde{\beta} - \tilde{r}(1-\tilde{\beta}^2)} \, .
\end{equation}

\begin{figure}[t!]
    \centering
    \begin{subfigure}{0.49\linewidth}
    \centering
    \includegraphics[width=\linewidth]{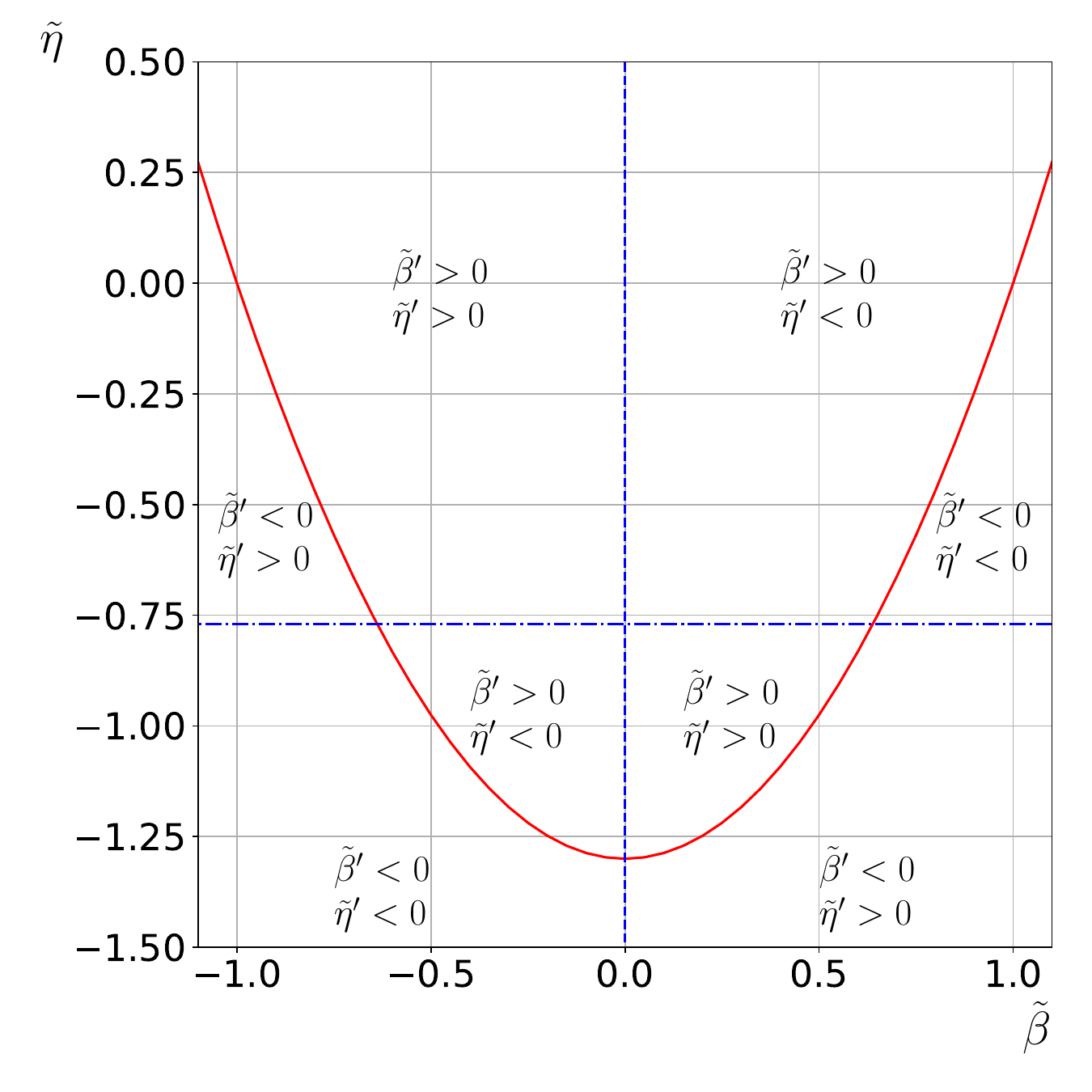}
    \caption{\empty}
    \label{fig:AttRegions}
    \end{subfigure}
    \begin{subfigure}{0.49\linewidth}
    \centering
    \includegraphics[width=\linewidth]{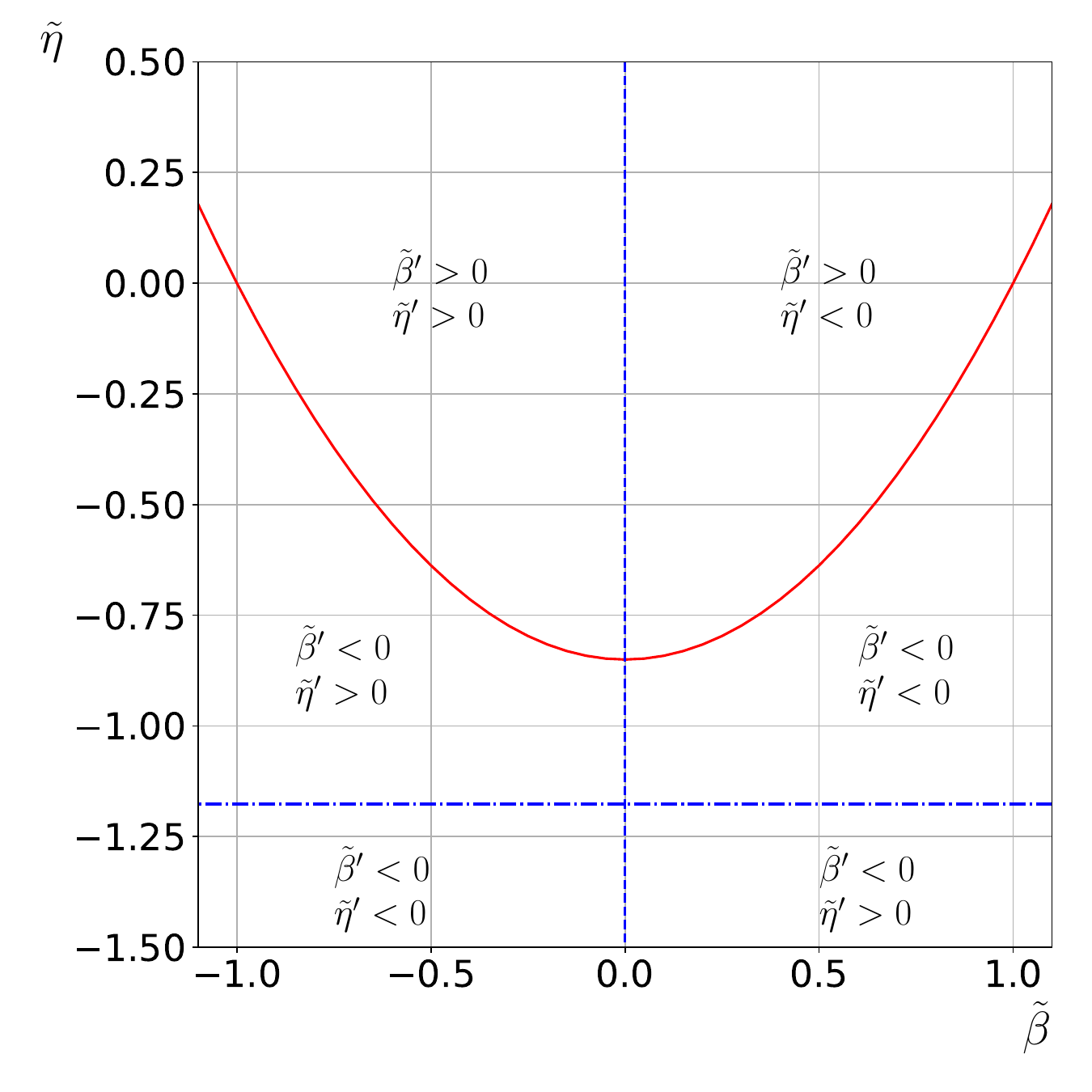}
    \caption{\empty}
    \label{fig:CSRegions}
    \end{subfigure}
    \caption{Phase space plot separated out into different kinematic regions using nullclines. Red lines indicate $\tilde{\beta}$-nullclines and blue lines show the $\tilde{\eta}$-nullclines. The left figure ($a$) is a model where $\tilde{r}>1$ and the right figure ($b$) is a model where $\tilde{r}<1$.}
    \label{fig:Regions}
\end{figure}

There are particular values of $k$ which are of particular importance. The value $k=0$ and the limit $k\to\infty$ identify the nullclines of the system. For $k=0$, trajectories are defined by constant~$\beta$, and so appropriately referred to as $\tilde{\beta}$-nullclines. Instead, $\tilde{\eta}$-nullclines are obtained by taking $k\to\infty$, but holding $\eta$ constant. These nullclines segment the phase space into different dynamical regions from which we can infer the properties of the flow. In~detail, 
we~have
\begin{subequations}
\begin{eqnarray}
    \tilde{\beta}-{\rm nullcline}:&&\ \tilde{\eta}(\tilde{\beta}) \: = \: -\,\tilde{r}(1-\tilde{\beta}^2) \, ,\\
    \tilde{\eta}-{\rm nullcline}:&&\  \tilde{\eta}(\tilde{\beta})\: =\: -\frac{1}{\tilde{r}} \ , \quad \tilde{\beta}(\tilde{\eta})\: =\: 0 \, .
\end{eqnarray}
\end{subequations}
Note that the case $\tilde{\beta}=0$ is not a true nullcline since it exists for any value of $k$, but should be highlighted as it contributes a change in the sign of the flow. By studying the phase space in Figure~\ref{fig:Regions}, one can see that when $\tilde{r}>1$, there are two relevant stationary points, identified by the intersection of nullclines. This gives the two stationary solutions,
\begin{equation}
    \tilde{\beta} \: = \: \pm \frac{\sqrt{\tilde{r}^2-1}}{\tilde{r}} \ , \qquad \tilde{\eta} \: = \: -\frac{1}{\tilde{r}} \ .
\end{equation}

\begin{figure}[t!]
    \centering
    \begin{subfigure}{0.49\linewidth}
    \centering
    \includegraphics[width=\linewidth]{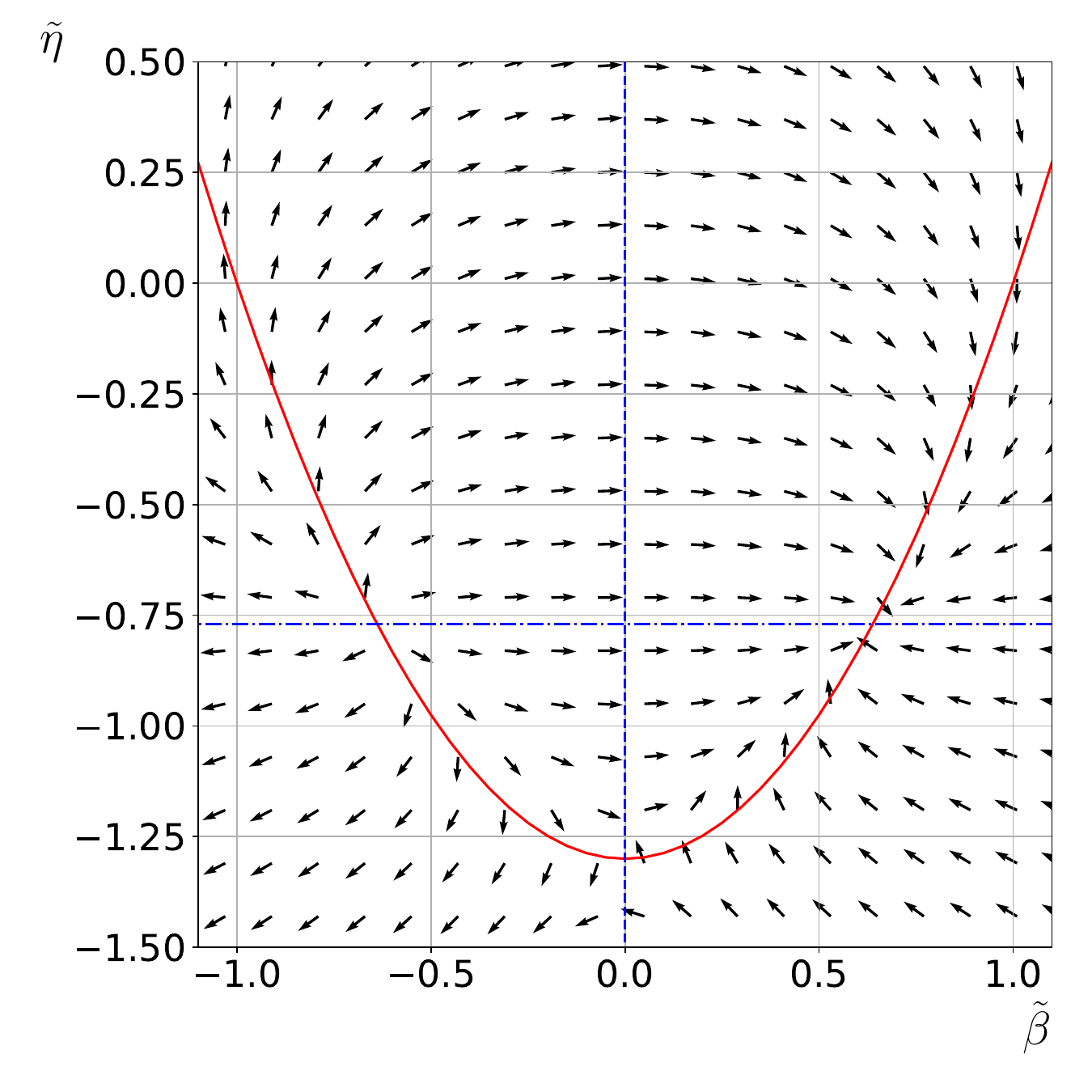}
    \caption{\empty}
    \label{fig:AttVect}
    \end{subfigure}
    \begin{subfigure}{0.49\linewidth}
    \centering
    \includegraphics[width=\linewidth]{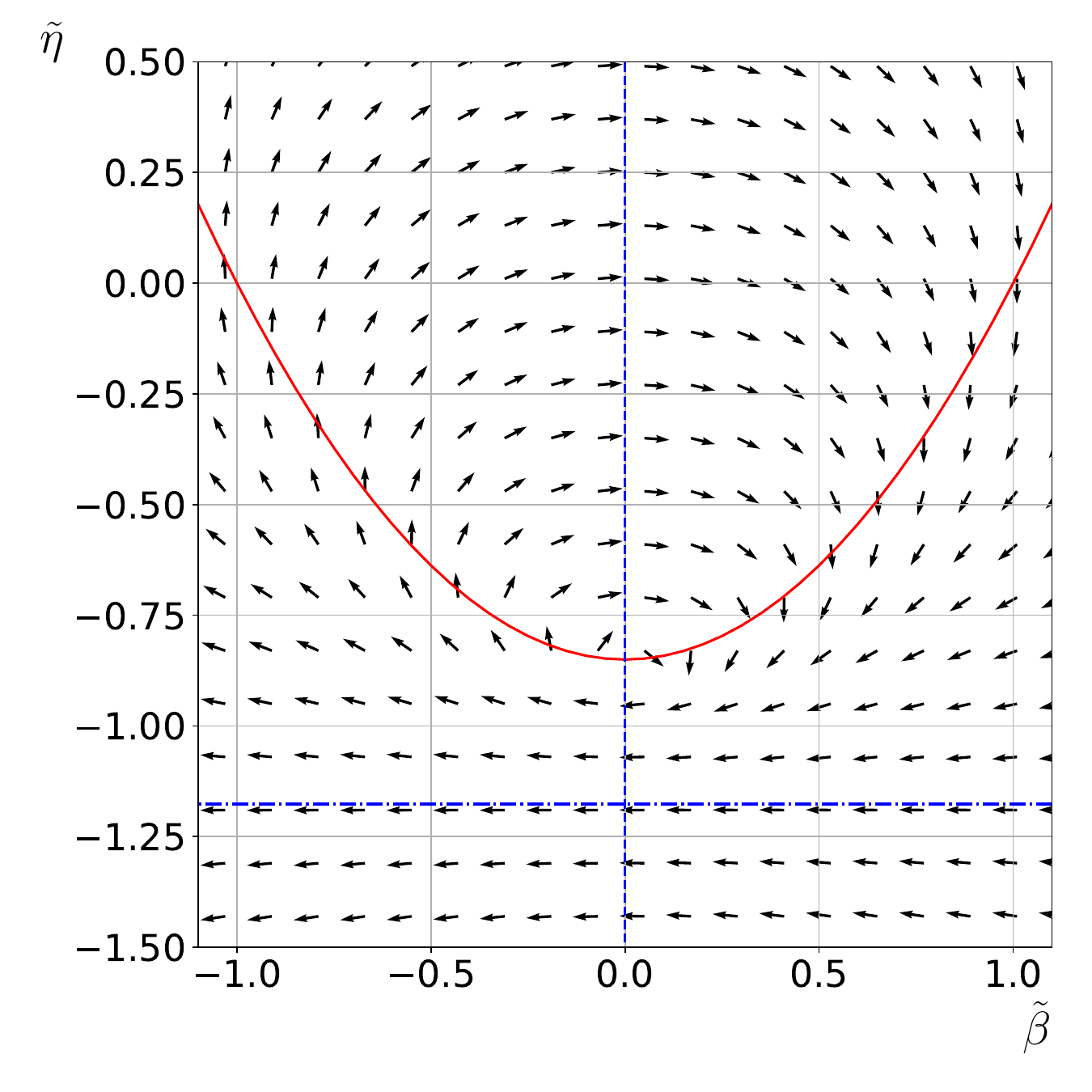}
    \caption{\empty}
    \label{fig:CSVect}
    \end{subfigure}
    \caption{Vector fields for the phase space spanned by $\tilde{\beta}$ and $\tilde{\eta}$. The left figure ($a$) shows the vector field for a CUQ scenario with $\tilde{r}>1$, and the right figure ($b$) shows a vector field for a CUQ scenario with $\tilde{r}<1$.}
    \label{fig:Flows}
\end{figure}
By studying the flows around these stationary points in Figure~\ref{fig:AttRegions}, it can be seen that any stationary point with $\tilde{\beta}<0$ is unstable, with perturbations about this point diverging. In contrast, any stationary point with $\tilde{\beta}>0$ is stable, with perturbations sinking into some stationary point. As a result, the gradient vectors and flows all point away from a negative stationary point and towards a positive stationary point. This is illustrated in Figure~\ref{fig:AttVect}. In contrast, when we consider CUQ scenarios, where $\tilde{r}<1$, there are no intersections of the nullclines. Consequently, the stationary points vanish. By examining Figure~\ref{fig:CSRegions}, we observe the presence of the so-called Lyapunov-stable cycles~\cite{guckenheimer2013nonlinear}, indicating that one should expect to find oscillatory solutions. Indeed, inspecting Figure~\ref{fig:CSVect}, we see the existence of closed loops in the flows indicating indefinite oscillations of the co-decaying Bloch vector, $\be (\tau )$.

\section{Estimation of Fourier Coefficients}\label{App:Errors}
\setcounter{equation}{0}  % reset counter

The statistical methods used to fit the neutral oscillation data delineated in Section~\ref{sec:ApplBmeson} may be found in a number of references~\cite{casella2002statistical,alma9912184053502466, wasserman2004all}. In this appendix, we describe the particular techniques used to fit the signal to the Fourier series, as well as the estimation of $r$ and its associated error. The data we consider come in the form of the $N_d$ ordered pairs $\lrb{A_\alpha, t_\alpha}$. The asymmetry $A(t)$ is a function that depends on two unknown acceptance parameters $a_1$ and $a_2$, as well as the projection of the co-decaying Bloch vector, ${\bf b}(t)$, along $\mathbf{e}\times\boldsymbol{\gamma}$. Previous studies on $B^0\bar{B}^0$-meson  oscillations indicate that $r$ is very close to zero. Therefore, for the purpose of estimating $a_1$ and $a_2$, we take $\vecb(t)\cdot \lrb{\mathbf{e}\times\boldsymbol{\gamma}} \simeq b \cos(\Delta m_{\rm d} t)$. This gives three unknown constants, i.e.~$a_1$, $a_2$, and $b$. Since $\epsilon(t|a_1,a_2)$ is an efficiency function, we assume that $a_1>0$ in general. Equation~\eqref{eq:AsymInv} may then be rewritten as the linear equation,
\begin{equation}\label{eq:ALinFit}
    \ln a_1 + a_2 t \: = \: \ln \lrcb{ \tan \lrsb{ \frac{\alpha_{B^+} A(t) }{ b \cos(\Delta m_{\rm d} t) - A(t)} \cdot \exp \lrsb{-(\Gamma_b - \Gamma_s) t} } } \, .
\end{equation}
For any consistent value of the parameter $b$, one can define the data points of observations,~$O_\alpha$, and expectations,~$E_\alpha$, as follows:
\begin{equation}
    O_\alpha \: = \: \ln \lrcb{ \tan \lrsb{ \frac{\alpha_{B^+} A_\alpha }{ b \cos(\Delta m_{\rm d} t_\alpha) - A_\alpha} \cdot \exp \lrsb{-(\Gamma_b - \Gamma_s) t_\alpha} } } \, , \qquad E_\alpha \: = \: \ln a_1 + a_2 t_\alpha \, ,
\end{equation}
through which we can perform a linear regression by minimising the $\chi^2$ parameter,
\begin{equation}
    \chi^2\lrb{a_1, a_2 | b} \: = \: \sum_{\alpha=1}^{N_d} \, \frac{\lrb{ O_\alpha - E_\alpha}^2}{\sigma^2_{\alpha}} \ .
\end{equation}
In the above expression, $\sigma^2_{\alpha} = {\rm var}\lrsb{A_{\alpha}}$, is the variance of each mixing asymmetry data point,~$A_\alpha$. According to our procedure, the regression gradient enables us to extract $a_2$ and the intercept is used to estimate~$\ln a_1$.

To ensure the most accurate estimates for $a_1$ and $a_2$, we additionally scan over the values of~$b$ for which~\eqref{eq:ALinFit} takes real values, i.e.~$b\in [{\rm max} \, A_\alpha \, , \, 1]$ and calculate $\chi^2$ for each $b$. The value of $b$, $a_1$, and $a_2$ which give the minimal $\chi^2$ on this scan is then taken as the best fit. With these parameters estimated, we can then use~\eqref{eq:AsymInv} and the fitted parameters $a_1$ and $a_2$ to infer the full projection $\vecb(t)\cdot \lrb{\mathbf{e}\times\boldsymbol{\gamma}}$. Note that the fitted parameter $b$ is not used, since we only wish to remove the change in the amplitude of the signal due to variations in the ratio between signal and background events.

Once we have extracted $\vecb(t)\cdot \lrb{\mathbf{e}\times\boldsymbol{\gamma}}$, we need to fit the amplitude-corrected oscillation to the Fourier series. For notational convenience, in the following, Roman subscripts refer to Fourier modes, whereas Greek indices refer to data points. With this in mind, we define the observed and expected values as follows:
\begin{equation}
    O_\alpha \: = \: \vecb(t_\alpha)\cdot \lrb{\mathbf{e}\times\boldsymbol{\gamma}}\, , \qquad  E_\alpha \: = \: d_0 + \sum_n \, d_n \cos(n\omega t_\alpha) \, .
\end{equation}
As before, our approach to this fit is to minimise the $\chi^2$ statistic for $N$ Fourier modes. With this aim, we create datasets of the deviations, $\mathcal{O}_\alpha$, from the mean value of the oscillation data and the mean value of the Fourier modes of the oscillation, i.e.
%\begin{subequations}
\begin{equation}
    \mathcal{O}_\alpha \: = \: O_\alpha - \overline{O}\;.
\end{equation}
For convenience, we also define the quantities,
\begin{equation}
    x_{n,\alpha} \: = \: \cos \lrb{n\omega t_\alpha}\,,\qquad X_{n,\alpha}\: =\: x_{n,\alpha} -\, \overline{x}_{n} \, .
\end{equation}
%\end{subequations}
In the above, $\overline{O}$ and $\overline{x}_{n}$ are weighted averages, with weights $\sigma_\alpha^{-2} = {\rm var}[O_\alpha]^{-1}$. The expected value at each $t_\alpha$ of the Fourier series may then be concisely expressed as
\begin{equation}
    E_\alpha \: = \: \mathbf{d} \cdot \mathbf{X}_\alpha
\end{equation}
where $\mathbf{d} = \lrb{d_1 , \dots , d_n}$, and the $d_0$ term has been absorbed into the definition of $\mathbf{X}$. After each $\mathbf{d}$ is determined, one can easily re-derive $d_0$ going through the consistency condition of the fit,
\begin{equation}
    d_0 \: = \: \overline{O} - \mathbf{d}\cdot \overline{\boldsymbol{x}} \, .
\end{equation}
Minimising $\chi^2$ for the fit gives a set of coupled equations which may be solved numerically to find the unknown coefficients represented by the vector, $\mathbf{d} = \mathcal{M}^{-1} \, \mathbf{u}$, with
\begin{equation}
    \mathcal{M}_{ij} \: = \: \sum_{\alpha=1}^{N_d} \frac{1}{\sigma^2_\alpha} X_{i,\alpha} \, X_{j,\alpha}\, , \quad u_i \: = \: \sum_{\alpha=1}^{N_d} \frac{1}{\sigma^2_\alpha} X_{i,\alpha} \, \mathcal{O}_\alpha \, .
\end{equation}
Additionally, assuming independence of each observation, $\mathcal{O}_k$, gives a way to find the variance of the Fourier coefficients
\begin{equation}
     {\rm var}[d_i] \: = \: \sum_{k=1}^N \, \mathcal{M}^{-1}_{ik} \mathcal{M}_{kk} \mathcal{M}^{-1}_{ki} \, , \quad {\rm var}[d_0] \: = \: \sum_k x_k^2 {\rm var}[d_k] + \lrb{\sum_{\alpha=1}^{N_d} \frac{1}{\sigma^2_\alpha}}^{-1} \, .
\end{equation}

Once the Fourier coefficients are extracted from the data, one can follow the prescription laid out in the main body to estimate the anharmonicity factors, $\mathcal{D}_n$. These may then be used to infer the value of $\tilde{r}$, and finally the true value of $r$. Analytic expressions for $\tilde{r}$ in terms of~$\mathcal{D}_n$, as well as for $r$ in terms of $\tilde{r}$ are given in~\eqref{eq:Fourier_Ratios} and~\eqref{eq:r_effective}, respectively. Here, we simply state relevant error estimates, which we obtain through partial derivatives. The error on each anharmonicity factor may be found as:
\begin{equation}
    \delta \mathcal{D}_n \: \simeq \: \mathcal{D}_n\, \sqrt{\lrb{\frac{\delta d_n}{d_n}}^2 + \lrb{\frac{\delta d_{n-1}}{d_{n-1}}}^2} \ .
\end{equation}
One may then estimate the error on the effective value of $\tilde{r}$ using the simple expressions:
\begin{equation}
    \delta \tilde{r}(\mathcal{D}_0) \: \simeq \: \frac{\mathcal{D}_0 r^3}{4}\, \delta \mathcal{D}_0 \, , \qquad \delta \tilde{r}(\mathcal{D}_{n \geq 1}) \: \simeq \: \frac{r^2}{2\mathcal{D}_n^2}\,\big(1-\mathcal{D}_n^2\big)\,\delta \mathcal{D}_n \, .
\end{equation}
Finally, the true value of $r$, calculated using $\tilde{r}$, has the error band,
\begin{equation}
    \delta r \: \simeq \: \lrb{\frac{r}{\tilde{r}}}^3 R^2 \delta \tilde{r} \, .
\end{equation}
Analytic results giving estimations of the errors on $\mathcal{C}_n$ are identical, aside from the absence of the $\mathcal{C}_0$ term.

We end this appendix by noting that the actual value of $r$ quoted is the weighted sum of all extracted values, where the weighting is done as usual, according to the estimated variance of each value, i.e.
\begin{equation}
    \overline{r} \: = \: \frac{1}{\sum_k \lrb{\delta r_k}^{-2}} \sum_k \frac{r_k}{\lrb{\delta r_k}^2} \, , \quad {\rm var}[\overline{r}] \: = \: \bigg[\sum_k \frac{1}{\lrb{\delta r_k}^2}\bigg]^{-1} \, .
\end{equation}

\vfill\eject
\bibliography{bibs-refs}{}
\bibliographystyle{JHEP}
\end{document}